\documentclass[aps,amsmath, amssymb, prd, twocolumn, superscriptaddress,floatfix,showpacs,bibliography,10pt]{revtex4-1}

\usepackage{graphicx}
\usepackage{hyperref}
\usepackage{tensor}

\usepackage{color}
\usepackage[normalem]{ulem}

\usepackage{sistyle}
\usepackage{mathtools}
\newcommand{\eqcolon}{\mathrel{\resizebox{\widthof{$\mathord{=}$}}{\height}{ 
$\!\!=\!\!\resizebox{1.2\width}{0.8\height}{\raisebox{0.23ex}{$\mathop{:}$}}\!\!$ }}}
\newcommand{\coloneq}{\mathrel{\resizebox{\widthof{$\mathord{=}$}}{\height}{ 
$\!\!\resizebox{1.2\width}{0.8\height}{\raisebox{0.23ex}{$\mathop{:}$}}\!\!=\!\!$ }}}

\graphicspath{{images/}}

\begin{document}

\title{Gravitational wave polarization from combined Earth-space detectors}

\author{Lionel \surname{Philippoz}}
\email[]{plionel@physik.uzh.ch}
\affiliation{Physik-Institut, Universit\"at Z\"urich, Winterthurerstrasse 190, 8057 Z\"urich}

\author{Adrian \surname{Bo\^itier}}
\email[]{boitier@physik.uzh.ch}
\affiliation{Physik-Institut, Universit\"at Z\"urich, Winterthurerstrasse 190, 8057 Z\"urich}

\author{Philippe \surname{Jetzer}}
\email[]{jetzer@physik.uzh.ch}
\affiliation{Physik-Institut, Universit\"at Z\"urich, Winterthurerstrasse 190, 8057 Z\"urich}

\date{\today}

\begin{abstract}

In this paper, we investigate the sensitivity to additional gravitational wave polarization modes of future detectors. We first look at the 
upcoming Einstein Telescope and its combination with existing or planned Earth-based detectors in the case of a stochastic gravitational wave 
background. We then study its correlation with a possible future space-borne detector sensitive to high-frequencies, like DECIGO. Finally, we adapt 
those results for a single GW source and establish the sensitivity of the modes, as well as the localization on the sky.
\end{abstract}

%\pacs{04.30.-w, 04.80.Nn}

\maketitle

\section{Introduction}\label{ET_section:introduction}
Since the first detection of gravitational waves (GW) by the LIGO collaboration, a total of 5 black hole mergers \cite{DetectionGW150914, 
DetectionGW151226, DetectionGW170104, DetectionGW170608, DetectionGW170814} has been observed so far\footnote{The current state of the detections 
by the LIGO/VIRGO Collaboration can be found on: \url{https://www.ligo.caltech.edu/page/detection-companion-papers}}, as well as a neutron star 
merger visible through gravitational waves and all accessible frequency bands of the electromagnetic spectrum \cite{BNS2017, MultiMessenger2017}. 
The results matched with Einstein's theory of general relativity (GR) up to measurement precision.

Until now, many tests of general relativity have been performed (e.g. the perihelion precession of Mercury, the geodetic precession and the 
Lense-Thirring effect by Gravity Probe B \cite{GProbeB2008} or the weak equivalence principle by MICROSCOPE \cite{MICROSCOPE2017} to name only a 
few) and so far, they all agree with general relativity. Modifications to GR have been constrained by experiments, but there are still some 
possibilities which cannot be excluded, see Will \cite{Will2014} to get an overview. As we will discuss in the next part, one could for example 
modify GR by adding a scalar or a vector field which only couple to the metric and therefore act as correction to GR. These fields would allow 
additional polarizations to the two tensor polarizations, + (plus) and $\times$ (cross), predicted by GR. The scalar field would create the 
breathing ($b$) and the longitudinal ($l$) and the vector field the $x$ and $y$ polarizations.

The standard model of cosmology describes the creation of the universe as an exponentially fast expansion of a quantum state. In quantum mechanics, 
no field or degree of freedom can be zero. If one now expands the universe, the quantum fluctuations of the fields get macroscopic and create a 
homogeneous and isotropic background where all polarizations are excited equally.

Using the electromagnetic spectrum we can only observe events as far back as the cosmic microwave background (CMB). The neutrinos decouple a bit 
earlier and would allow us to see further back in time, given we would figure out how to measure low energy particles, which almost never 
interact.~If we could however measure a gravitational wave background (GWB), then we could test cosmological models way further back in time. One 
expects that the gravitational waves decouple at the Planck time due to the weak coupling of the metric to the other fields. This might allow us to 
get information about quantum gravity and thus an energy scale which is far out of reach of modern particle colliders. Since one expects all 
polarizations to be excited in the GWB, this would serve as a test for GR or allow to put constrains on alternative theories of gravity by 
checking the presence or absence of additional polarization modes in a given signal.\\

The second-generation ground-based detectors advanced LIGO and advanced VIRGO can detect GW from binary black holes (BBH) and binary neutron 
stars (BNS) \cite{BNS2017}. A similar detector is being built in Japan (KAGRA) \cite{KAGRA2017} and another advanced LIGO is planned in the 
near future in India (IndIGO) \cite{IndIGO2011}. With the Einstein Telescope (ET) \cite{ET}, a cluster of three detectors arranged in an 
equilateral triangle with an arm length of \SI{10}{km}, one plans to build a third-generation detector in Europe which is supposed to be about 10 
times more sensitive to a GW signal than the current generation.

Space-borne detectors are also on their way. LISA pathfinder was a success \cite{LISApathfinder2018}, which is very promising for LISA 
\cite{LISA2017}, a cluster of three satellites planned to be launched as the next ESA L3 mission. LISA will be put on a heliocentric orbit, 
at about 20$^\circ$ behind the Earth. DECIGO \cite{Kawamura2006, YagiSeto2011} was originally planned to consist of 4 clusters distributed in
Earth-orbit around the Sun, each forming a \SI{1000}{km} equilateral triangle with three satellites. A scaled down version with arms of only \SI{100}{km}, B-DECIGO \cite{BDECIGO2018}, initially presented as Pre-DECIGO \cite{PreDECIGO}, has recently been proposed as a first 
generation of deci-Hz detectors. It is planned to revolve around the Earth at an altitude of \SI{2000}{km}.

The ground-based detectors of the second generation are not capable of detecting the gravitational wave background on their own and it is unlikely
that an improvement of about one order of magnitude in sensitivity would be sufficient. But if we combine the signals of all the detectors which 
are built to measure BBH and BNS anyway, then one could enhance the sensitivity by three to four orders of magnitude and thus get more restrictive 
constraints on the GWB or even detect it.

With its high and low frequency interferometers, ET is designed to measure in a frequency range from \SI{1.5}{Hz} to \SI{10}{kHz}. It therefore  
makes perfect sense to cross correlate its signals with the ones of any second-generation ground-based detector, or even DECIGO. The correlation 
between ET and DECIGO has the advantage that their noises are very different, since ET is Earth-based and DECIGO is in space and therefore does 
not have any seismic noise for instance. A correlation of ET or DECIGO with LISA would however be difficult since the designed sensitivity of the 
latter lies within the range \SI{10^{-4}}{Hz} to about \SI{1}{Hz}, out of the frequency band considered for ET/DECIGO.\\

Testing GR by using the gravitational wave background can be of interest due to its constant and isotropic nature. One does not have to 
extract complex waveforms in a combination of all the 6 possible polarizations from the strain, which may differ according to the modified 
theory considered. Note that because of the isotropy of the GWB, one cannot distinguish between + and $\times$ tensor polarizations, or $x$ and
$y$ vector polarizations, but it is nevertheless possible to separate the three modes (tensor T, vector V and scalar S), which can already 
give us information on the involved fields.

In the case of point sources, one can additionally determine the direction of the incoming GW on the sky, as well as distinguishing
between the polarizations. However, this makes the calculation more complicated since we have to deal with 8 degrees of freedom instead of 3.\\

This paper is outlined as follows: in section~\ref{ET_section:theory} we recall some theoretical basics about polarization and summarize the 
derivation of the signal to noise ratio ($SNR$) for a polarization mode when combining multiple detectors in the case of a GWB, as done 
in Nishizawa et al. \cite{Nishizawa2009, Nishizawa2010}. We also derive the general expression of the power spectral density and the overlap 
reduction functions for detectors having arbitrary opening angles. In section~\ref{ET_section:ET}, we apply those results in the case of ET and 
consider its correlation with ground-based detectors. In section~\ref{ET_section:DECIGO}, we introduce DECIGO in the detector network and 
investigate how the time dependent sensitivity of a cross correlation between DECIGO and Earth detectors can be used to distinguish the three 
polarization modes, as an alternative method to the maximum likelihood method on all detector pairs. Finally, we consider the case of point 
sources and derive the $SNR$ for a single polarization and the variance on the incoming direction of the GW in 
section~\ref{ET_section:pointsources}.

\section{Theory and Methods}\label{ET_section:theory}
In this section, we introduce the techniques used to calculate the sensitivities to GW-polarizations of various combinations of detectors. We 
first give a short overview of GW and the notion of polarization in GR or alternative theories of gravitation, as well as the detection 
principle. We continue by extracting the signal of a correlation between two detectors. Then, we take multiple detector pairs and combine their 
signals in an optimal way to distinguish the polarizations and enhance the sensitivity.

The sensitivity is dependent on the noise power spectrum of the detector and geometry factors, which in the case of a gravitational wave 
background are the overlap reduction functions (ORFs). To calculate the sensitivity for a collection of detectors including ET or DECIGO we need 
to generalize the formula for the noise power spectral density to arbitrary opening angles and we can simplify the expression for the ORFs for 
ground-based detectors which comes in handy since many of the detectors we consider here are ground based.

\subsection{Polarizations of Gravitational Waves}\label{ET_section:polarization}
The linearization of the Einstein field equations leads to a linear wave equation for perturbations in the metric. Since the metric is required to 
be symmetric, the degrees of freedom of a 4-dimensional tensor of rank 2 are reduced from 16 to 10. The Einstein equations are invariant under a 
change of reference frame, while the linearized version is only invariant under an infinitesimal change of coordinates, which reduces the degrees 
of freedom to 6. By choosing an orthonormal basis $(\hat{m}, \hat{n}, \hat{\Omega})$, where $\hat{\Omega} \parallel \vec{k}$ is the direction 
of travel \cite{Nishizawa2009}, we can write a general solution as:
\begin{align}
h_{ij}(t,\vec{x}) &= \begin{pmatrix}
h_{11}	&	h_{12}	&	h_{13}	\\
h_{12}	&	h_{22}	&	h_{23} \\
h_{13}	&	h_{23}	&	h_{33}
\end{pmatrix} e^{2\pi i f\left(t-\frac{\hat{\Omega}\cdot\vec{x}}{c}\right)} + c.c. \notag\\ &= \sum_A  h_A(t,\vec{x}) e^A_{ij} e^{2\pi i 
f\left(t-\frac{\hat{\Omega}\cdot\vec{x}_I}{c}\right)} + c.c.,
\end{align}
where the $e^A$ are the basis tensors of the possible polarizations we describe afterwards, and $h_A$ is the amplitude of the GW in the 
polarization $A$.

Therefore, we can have at most 6 polarizations. Since this is a vacuum equation in the case of unmodified GR, the equation is invariant under a 
gauge transformation on the fields $h_{\mu\nu} \mapsto h'_{\mu\nu} = h_{\mu\nu} - \epsilon_{\mu,\nu} - \epsilon_{\nu,\mu}$, with 
$\square\epsilon^\mu = 0$. This further reduces the degrees of freedom to the 2 tensor polarizations + and $\times$. They are purely transversal 
waves, which enlarge distances in one direction and squeeze space in the orthogonal direction. The basis tensors of the tensor mode are given by:
\begin{align}
e^+ = \hat{m}\otimes\hat{m} - \hat{n}\otimes\hat{n},	 &&	e^{\times} = \hat{m}\otimes\hat{n} + \hat{n}\otimes\hat{m}.
\end{align}

We will now look at two representative examples of modifications of GR and their consequences on gravitational waves. Adding a scalar field to the 
Lagrangian is one possibility to modify GR \cite{Will2014}. A general scalar-tensor action can be written as:
\begin{equation}
S[g,\phi] = \frac{1}{16\pi G}\int \left[R - 2g^{\mu\nu}\partial_{\mu}\phi\partial_{\nu}\phi - U(\phi) \right]\sqrt{-g} d^4x.
\end{equation}
This leads to the two scalar polarizations called the breathing mode $b$, since it stretches and squeezes space simultaneously in all transversal 
directions, and the longitudinal mode $l$, which is a purely longitudinal wave. Their basis tensors are given by:
\begin{align}
e^b = \hat{m}\otimes\hat{m} + \hat{n}\otimes\hat{n},	 &&	e^l = \sqrt{2}\hat{\Omega}\otimes\hat{\Omega}.
\end{align}

Another possibility would be to add a vector field Lagrangian as follows:
\begin{align}\notag
S[g,V] &= \frac{1}{16\pi G}\int \left[ (1 + \omega V_\mu V^\mu)R - K^{\mu\nu}_{\rho\sigma}\nabla_\mu V^\rho \nabla_\nu V\sigma \right.\\
&\left.+ \lambda(V_\mu V^\mu + 1) \right]\sqrt{-g} \ d^4x
\end{align}
with
\begin{equation}
K^{\mu\nu}_{\rho\sigma} = c_1g^{\mu\nu}g_{\rho\sigma} + c_2\delta^\mu_\rho\delta^\nu_\sigma + c_3\delta^\mu_\sigma\delta^\nu_\rho - c_4V^\mu V^\nu 
g_{\rho\sigma},
\end{equation}
where the $c_i$ are coefficients which would have to be determined by experiments. This modification generates the two vector polarizations 
$x$ and $y$ which oscillate in direction of travel and in one orthogonal to it. Their respective basis tensors are given by:
\begin{align}
e^x = \hat{m}\otimes\hat{\Omega} + \hat{\Omega}\otimes\hat{m},	&&	e^y = \hat{n}\otimes\hat{\Omega} + \hat{\Omega}\otimes\hat{n}.
\end{align}

We can finally express a general solution in terms of all six polarizations:
\begin{equation}
h_{ij}(t,\vec{x}) = \begin{pmatrix}
h_b + h_+	&	h_{\times}	&	h_x	\\
h_{\times}	&	h_b - h_+	&	h_y \\
h_x			&	h_y			&	h_l
\end{pmatrix} e^{2\pi i f\left(t-\frac{\hat{\Omega}\cdot\vec{x}}{c}\right)} + c.c.
\end{equation}

When a gravitational wave stretches or squeezes an arm of a Michelson interferometer, then one can observe a phase shift. This phase shift is 
larger if the amplitude of the wave is larger and if the detector arms are optimally aligned given an incoming wave with a certain polarization. 
So 
the signal in the detector can be written as:
\begin{equation}
h_{ij}(t,\vec{x}) = D^{ij} \sum_A  h_A(t,\vec{x}) e^A_{ij} e^{2\pi i f\left(t-\frac{\hat{\Omega}\cdot\vec{x}}{c}\right)}.
\end{equation}
The detector tensor $D = \frac{1}{2}(\hat{u}\otimes\hat{u} - \hat{v}\otimes\hat{v})$ describes the orientations of the interferometer arms, given 
by the unit vectors $\hat{u}$, $\hat{v}$, and the polarization of the wave can be written as a linear combination of the basis tensors $e^A$ 
described above. If we contract the two tensors, we get a scalar quantity, the angular pattern function, which describes the geometric dependence 
of the signal:
\begin{equation}
F^A \coloneq D^{ij}e^A_{ij}.
\end{equation}
A GW thus produces one scalar signal in each detector, which means that we need to combine at least 6 detectors to distinguish them. In the 
case of a gravitational background however, we expect a direction independent signal. Therefore, one can only distinguish between the three modes: 
tensor, vector and scalar; three independent signals are thus sufficient, but more signals would of course improve the sensitivity.

By using the matched filtering method, one can calculate the signal to noise ratio of a given signal. Since the ET project has declared a signal 
to noise ratio of at least 8 as their condition to accept an event as an actual signal \cite{ET}, we set the $SNR$ to 8 and calculate the minimal 
amplitude 
a gravitational wave needs to have to be recognized as a true signal by a certain collection of detectors and use this as a measure of their 
combined sensitivity.\\
The $SNR$ is related to the false alarm rate $\alpha$ and detection rate $\gamma$ by \cite{AllenRomano1999}:
\begin{equation}
SNR \geqslant \sqrt{\frac{2}{n}}\left( \text{erfc}^{-1}(2\alpha) - \text{erfc}^{-1}(2\gamma) \right)
\end{equation}

Once one has chosen a minimal signal to noise ratio, one has to choose either a false alarm or a detection rate. If we would split our observation 
time ($T = \SI{1}{yr}$ would be a realistic choice for a GWB observation) into small time intervals of for example \SI{4}{s} and do statistical 
tests on them, then one false alarm in \SI{27000}{yr} would be equivalent to a false alarm rate of $\alpha = \frac{\SI{4}{s}}{\SI{27000}{yr}} = 
4.8\cdot10^{-12}$. This would give us about $n = 7.8\cdot10^6$ time splits and result in a detection rate $\gamma \approx 1$ under the assumption 
of a $SNR$ of 8. The detection rate is related to the false dismissal rate $\beta$ by $\gamma = 1 - \beta$ which gives us a false dismissal rate 
of $\beta = 3.3\cdot10^{-18}$.\\
\\
We will now derive an expression for the $SNR$ in terms of the GW signal and the detector noise.

\subsection{Combined Sensitivity of Multiple Detectors}
Since the signal of a GW is usually smaller than the noise, one can rely on two different techniques in order to get rid of the noise. First, we 
can multiply the Fourier transform of the signal with a suitable filter function, which turns out to be proportional to the signal, and integrate 
over all frequencies. This method is called matched filtering. Secondly, we can cross correlate the strains $s_{I,J} = h_{I,J} + n_{I,J}$ of two 
detectors $I$ and $J$. Since the noises $n_{I,J}$ of the two detectors are not correlated between them and also not correlated to the signals 
$h_{I,J}$, we can get rid of the noise by taking the expectation of the Fourier transform (FT) of the complex conjugated strain $\tilde{s}_I^*$ of 
detector $I$ multiplied with the FT of the strain $\tilde{s}_J$ of detector $J$:
\begin{equation}\label{Eq:E[sI*sJ]}
\mathbb{E}[\tilde{s}_I^*\tilde{s}_J] = \mathbb{E}[\tilde{h}_I^*\tilde{h}_J] + \underset{= 0}{\underbrace{\mathbb{E}[\tilde{h}_I^*\tilde{n}_J]}} + 
\underset{= 0}{\underbrace{\mathbb{E}[\tilde{n}_I^*\tilde{h}_J]}} + \underset{= 0}{\underbrace{\mathbb{E}[\tilde{n}_I^*\tilde{n}_J]}}.
\end{equation}
Nishizawa et al. \cite{Nishizawa2009} used the matched filtering method on a cross correlated signal and derived the $SNR$ for a detector pair 
$(I,J)$, and we now shortly remind the result.

The energy density parameter $\Omega_{GW}$ of the GWB can be written as a sum over all modes $M$, where each mode has two polarizations $M_1$ 
and $M_2$ as discussed previously:
\begin{align}
&\Omega_{GW}(f) = \sum_M \Omega_{GW}^M = \sum_M \left(\Omega_{GW}^{M_1}+\Omega_{GW}^{M_2}\right), \notag\\
&M=\begin{pmatrix} M_1 \\ M_2 \end{pmatrix} \in \left\lbrace T= \begin{pmatrix} + \\ \times \end{pmatrix}, \ V= \begin{pmatrix} x \\ y 
\end{pmatrix}, \ S= \begin{pmatrix} b \\ l \end{pmatrix} \right\rbrace.
\end{align}
The power spectral density $S_h^{M_i}$ of the polarization $M_i$ is related to its energy density parameter by:
\begin{equation}
\Omega_{GW}^{M_i}(f) = \frac{2\pi^2}{3H_0^2}f^3 S_h^{M_i}(f).
\end{equation}
By assuming that only one mode $M$ is excited and using the ansatz $S_h^{M_i}(f)=h_{0,M_i}^2\delta(f'-f)$, where $h_{0,M_i}$ is the amplitude of 
the polarization $M_i$, we get the sensitivity of the detector pair $(I,J)$ to the specific mode $M$:
\begin{align}\label{eq:snr_pair}
(SNR_{IJ}^M)^2 &= \frac{3H_0^2}{10\pi^2} \sqrt{T \int_{-\infty}^{\infty} \frac{(\Omega_{GW}^M(|f'|)\gamma_{IJ}^M(|f'|))^2}{f^6P_I(|f'|)P_J(|f'|)} 
df'} \notag \\
&= \frac{1}{5} \sqrt{T \int_{-\infty}^{\infty} \frac{(S_h^M(|f'|)\gamma_{IJ}^M(|f'|))^2}{P_I(|f'|)P_J(|f'|)} df'} \notag \\
&= \frac{T}{5} \frac{(h_{0,M_1}^2+h_{0,M_2}^2)\gamma_{IJ}^M(f)}{\sqrt{P_I(f)P_J(f)}},
\end{align}
where $H_0$ is the Hubble constant, $T$ the observation time, $P_{I,J}$ are the noise power spectral densities of the detectors $I$ and $J$ and 
$\gamma_{IJ}^M$ is the overlap reduction function defined by:
\begin{align}
\gamma_{IJ}^M(f) &\coloneq \frac{5}{2}\int_{\mathbb{S}^2} (F_I^{M_1}F_J^{M_1} + F_I^{M_2}F_J^{M_2}) e^{\frac{2\pi i 
f}{c}\hat{\Omega}_0\cdot\Delta\vec{x}_{IJ}} \frac{d\hat{\Omega}}{4\pi} \notag \\
&= \rho_1^M(\alpha)D_I^{ij}D^J_{ij} + \rho_2^M(\alpha)D_{I,k}^iD_J^{kj}\hat{d}_i\hat{d}_j \notag\\
&+ \rho_3^M(\alpha)D_I^{ij}D_J^{kl}\hat{d}_i\hat{d}_j\hat{d}_k\hat{d}_l,
\end{align}
with $\alpha(f) \coloneq \frac{2\pi f|\Delta\vec{x}_{IJ}|}{c}$ and where the $\rho_i^M$ are linear combinations of the zeroth, second and fourth 
spherical Bessel functions.

By requiring again $SNR \overset{!}{\geqslant} 8$ for a GW with mode $M$ to be considered a true signal, we can rewrite \eqref{eq:snr_pair} to get 
the minimal amplitude a GW would need to be detected as such:
\begin{equation}
|h_0^M(f)|_{min} = 8\sqrt{\frac{5}{T}} \left(\frac{\sqrt{P_I(f)P_J(f)}}{|\gamma_{IJ}^M(f)|}\right)^{1/2}.
\end{equation}

If we have more than two detectors we can use the maximum likelihood method to distinguish the polarizations. We then get a $SNR$ with which we 
recognize a mode $M$, as derived by Nishizawa et al. \cite{Nishizawa2010}:
\begin{align}\label{Eq:SNR_M}
(SNR^M)^2 &= \frac{3H_0^2}{10\pi^2} \sqrt{T \int_{-\infty}^{\infty} \frac{(\Omega_{GW}^M(f))^2 \det\textbf{F}(f)}{f^6\mathcal{F}_M(f)} df} \notag 
\\
&= \frac{1}{5}\sqrt{T \int_{-\infty}^{\infty} \frac{S_h^M(f)^2 \det\textbf{F}(f)}{\mathcal{F}_M(f)} df}.
\end{align}
Using the same ansatz as above, we get the minimal amplitude we require to not only detect a GW with mode $M$, but also distinguish its 
polarization, with a $SNR$ of at least 8:
\begin{equation}
|h_0^M(f)|_{min} = 8\sqrt{\frac{5}{T}} \left( \frac{\mathcal{F}_M}{\det\textbf{F}} \right)^{1/4},
\end{equation}
where the Fisher matrix $\mathbf{F}$ is obtained by summing over the Fisher matrices of all detector pairs $(I,J)$
\begin{equation}
F_{MM'}(f) = \sum_{(I,J)}\int_0^{T_{obs}} \frac{\gamma_{IJ}^M(t,f)\gamma_{IJ}^{M'}(t,f)}{P_I(f)P_J(f)} dt
\end{equation}
and $\mathcal{F}_M$ is the determinant of the minor one gets by removing the $M$-th row and column from $\textbf{F}$.

\subsection{Optical Read-out Noise}\label{ET_section:noiseps}
The quantum fluctuations of the laser cause a fundamental noise source in each detector which is statistically independent from the other 
detectors. The fluctuation in the number density of photons arriving at the detector causes a random fluctuation in the measured power and a 
fluctuation in the light pressure on the mirror which causes the mirror to vibrate randomly. By increasing the laser power, the fluctuation in the 
number density increase in total but is less compared to the average, which causes the relative fluctuations in the measured laser power to 
decrease, but the pressure and therefore the fluctuations in the position of the mirror increases. One therefore needs to balance one effect 
against the other, which causes an uncertainty relation similar to the one arising from quantum mechanics. We are now going to derive the optical 
read-out noise based on \cite{Maggiore}, but for an arbitrary opening angle between the detector arms.

A Michelson interferometer with a Fabry-Perot cavity catches an additional term dependant on the frequency $f$ of the measured gravitational wave 
and on a pole frequency $f_p$ which is a characteristic of the cavity. The power recycling $C$ appears as a higher effective power, and the 
detector efficiency $\eta$ as a lower one, and we modifiy the input power $P_0$ as $P_0 \mapsto \eta C P_0$.

\begin{figure}[h!]
\includegraphics[width=0.7\linewidth]{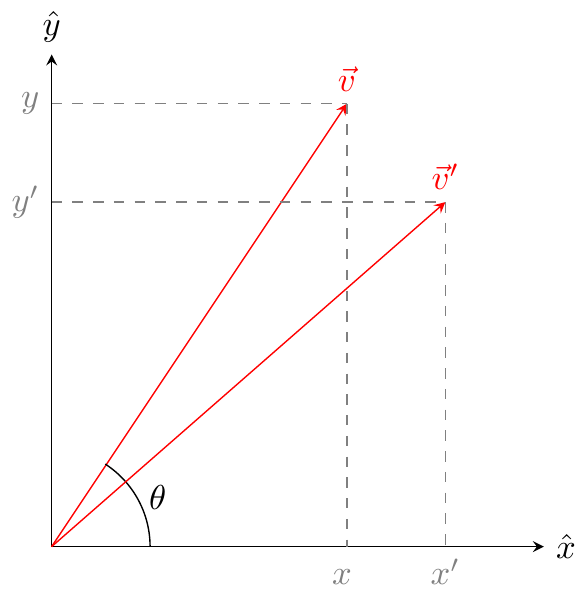}
\caption{\label{ET_fig:h+} The detector arm $\vec{v}$ of a detector with opening angle $\theta$ gets deformed to $\vec{v}'$ under the influence of 
a gravitational wave with + polarization. The other detector arm $\vec{u}$ lies on the $x$-axis.}
\end{figure}

The phase shift of a Fabry-Perot interferometer $\Delta\phi_{FP}$ is related to the one of a Michelson interferometer without cavity 
$\Delta\phi_{Mich}$ by: 
\begin{align}
|\Delta\phi_{FP}| &= \frac{2\mathcal{F}}{\pi} \frac{|\Delta\phi_{Mich}|}{\sqrt{1 + \left(\frac{f}{f_p} \right)^2 }},
\\
|\Delta\phi_{Mich}| &\coloneqq \Delta\phi_u - \Delta\phi_v,
\end{align}
where $\mathcal{F}$ is the finesse of the Fabry-Perot cavity, $L$ the arm length of the detector, $\Delta\phi_u$ and $\Delta\phi_v$ are the phase 
shifts in the arms $u$ and $v$ respectively, and $f_p$ the pole frequency of the cavity is given by:
\begin{equation}
f_p \approx \frac{c}{4\mathcal{F}L}.
\end{equation}
To calculate the phase shift of a Michelson interferometer with opening angle $\theta$, we consider an incoming GW with a + polarization:
\begin{equation}
h_{\mu\nu}^+ = h_+ \begin{pmatrix}
0 & 0 & 0 & 0 \\
0 & 1 & 0 & 0 \\
0 & 0 & -1 & 0 \\
0 & 0 & 0 & 0
\end{pmatrix} \cos(\omega_{GW}t).
\end{equation}
The GW effectively stretches space in $x$-direction and squeezes it in $y$-direction, as depicted in Fig.~\ref{ET_fig:h+}, by a factor 
$h_+(t-\frac{L}{c})$, using the approximation $\frac{\omega_{GW}L}{c}<<1$.
In this choice of reference frame we can write $\vec{v}$ as:
\begin{equation}
\vec{v} = \begin{pmatrix} x \\ y \end{pmatrix} = L \begin{pmatrix} \cos\theta \\ \sin\theta \end{pmatrix}	\ , \quad |\vec{v}| = L.
\end{equation}
Using the previous approximation, we can write down the components of the deformed arm $\vec{v}'$ and express the change in the coordinates as:
\begin{align}
x' &= \sqrt{x^2 + h_+x^2} \approx \left[1 + \frac{1}{2}h_+\right]x \ \Rightarrow \ \Delta x = \frac{1}{2}h_+x, \\
y' &= \sqrt{y^2 - h_+y^2} \approx \left[1 - \frac{1}{2}h_+\right]y \ \Rightarrow \ \Delta y = -\frac{1}{2}h_+y,
\end{align}
where we used the short notation $h_+$ for $h_+(t-\frac{L}{c})$ and expanded to first order. The total change in the length of the detector arm is 
then given by:
\begin{align}
\Delta v &= |\vec{v}'| - |\vec{v}| = \sqrt{x'^2 + y'^2} - \sqrt{x^2 + y^2} \notag \\
&= \frac{1}{2}\frac{2x}{\sqrt{x^2 + y^2}}\Delta x + \frac{1}{2}\frac{2y}{\sqrt{x^2 + y^2}}\Delta y + \mathcal{O}(\Delta^2) \notag \\ 
&\approx \frac{\cos\theta L}{L}\Delta x + \frac{\sin\theta L}{L}\Delta y \notag \\
&= \frac{1}{2}h_+L(\cos^2\theta - \sin^2\theta)
\end{align}
Since the light bounces back and forth, the phase shift catches a factor of two: $\Delta\phi_u = 2k_L\Delta u$ and $\Delta\phi_v = 2k_L\Delta v$. 

With those results, we can calculate the amplitude of the Michelson phase shift:
\begin{align}
|\Delta\phi_{Mich}| &= |\Delta\phi_u - \Delta\phi_v| \notag \\
&= \left|k_Lh_+\left(t-\frac{L}{c}\right)L \right. \notag \\
&\left. - k_Lh_+\left(t-\frac{L}{c}\right)L(\cos^2\theta - \sin^2\theta)\right| \notag \\
&= \frac{4\pi}{\lambda_L} \sin^2\theta L h_+,
\end{align}
with the wave number $k_L$ and wave length $\lambda_L$ of the laser: $k_L = \frac{2\pi}{\lambda_L}$. The change in the pathlength of a photon due 
to the incoming GW is given by:
\begin{equation}
\Delta L = 2(\Delta u - \Delta v) = \sin^2\theta L h_+
\end{equation}
Therefore the transfer function (change in pathlength per GW amplitude) is $\sin^2\theta L$. \\

By inserting $|\Delta\phi_{Mich}|$ and the transfer function for a general opening angle $\theta$ of the detector arms into the equations (9.220), 
(9.234) and (9.122) of \cite{Maggiore} and neglecting the efficiency of the photodetector $\eta \approx 1$ we get the shot-noise
\begin{equation}
\left.\sqrt{S_n(f)}\right|_{shot} = \frac{1}{4\mathcal{F}\sin^2\theta L}\sqrt{\frac{\pi \hbar \lambda_L 
c}{CP_0}}\sqrt{1+\left(\frac{f}{f_p}\right)^2},
\end{equation}
the radiation pressure
\begin{equation}
\left.\sqrt{S_n(f)}\right|_{rad} = \frac{16\mathcal{F}}{M\sin^2\theta L}\sqrt{\frac{\hbar CP_0}{\pi \lambda_Lc}}\frac{1}{(2\pi 
f)^2\sqrt{1+\left(\frac{f}{f_p}\right)^2}},
\end{equation}
and the optical read-out noise is thus given by:
\begin{equation}
S_n(f)\vert_{opt} = S_n(f)\vert_{shot} + S_n(f)\vert_{rad}.
\end{equation}
We will use this last result as the main component of the total noise, for an opening angle $\theta$ ($\pi/3$ for ET and $\pi/2$ for LIGO-like 
detectors).

Each ET detector consists of a high- (HF) and a low- (LF) frequency detector which are then used as one to broaden the frequency range. The 
detector characteristics of these two detectors are listed in Tab.~\ref{ET_tab:ET_characteristics} and will be used throughout this paper. The 
values we are using 
for advanced LIGO are summarized in Tab.~\ref{ET_tab:LIGO_characteristics}.

\begin{table}[h!]
\caption{\label{ET_tab:ET_characteristics} Detector characteristics of the high- (HF) and the low- (LF) frequency detectors, taken from the 
Einstein Telescope 
proposal \cite{ET}, section 5.1}
\begin{ruledtabular}
\begin{tabular}{lll}
\textbf{Quantity} & \textbf{ET-HF} & \textbf{ET-LF} \\ \hline
Input power (after IMC) $P_0$ & \SI{500}{W} & \SI{3}{W} \\
Laser wavelength $\lambda_L$ & \SI{1064}{nm} & \SI{1550}{nm} \\
Arm length $L$ & \SI{10}{km} & \SI{10}{km} \\
Mirror mass $M$ & \SI{200}{kg} & \SI{211}{kg} \\
Finesse $\mathcal{F}$ & 880 & 880 \\
Recycling gain $C$ & 21.6 & 21.6
\end{tabular}
\end{ruledtabular}
\end{table}

\begin{table}[h!]
\caption{\label{ET_tab:LIGO_characteristics} Detector characteristics of the aLIGO detectors, taken from \cite{aLIGO15}, and $C$ from 
\cite{aLIGO16}}
\begin{ruledtabular}
\begin{tabular}{ll}
\textbf{Quantity} & \textbf{aLIGO} \\ \hline
Input power (at PRM) $P_0$ & up to \SI{125}{W} \\
Laser wavelength $\lambda_L$ & \SI{1064}{nm} \\
Arm length $L$ & \SI{4}{km} \\
Mirror mass $M$ & \SI{40}{kg} \\
Finesse $\mathcal{F}$ & 450\\
Recycling gain $C$ & 38
\end{tabular}
\end{ruledtabular}
\end{table}

\subsection{Overlap Reduction Functions $\gamma_{I,J}^M(f)$}\label{ET_section:orf}
\begin{figure}[h!]
\includegraphics[width=0.8\linewidth]{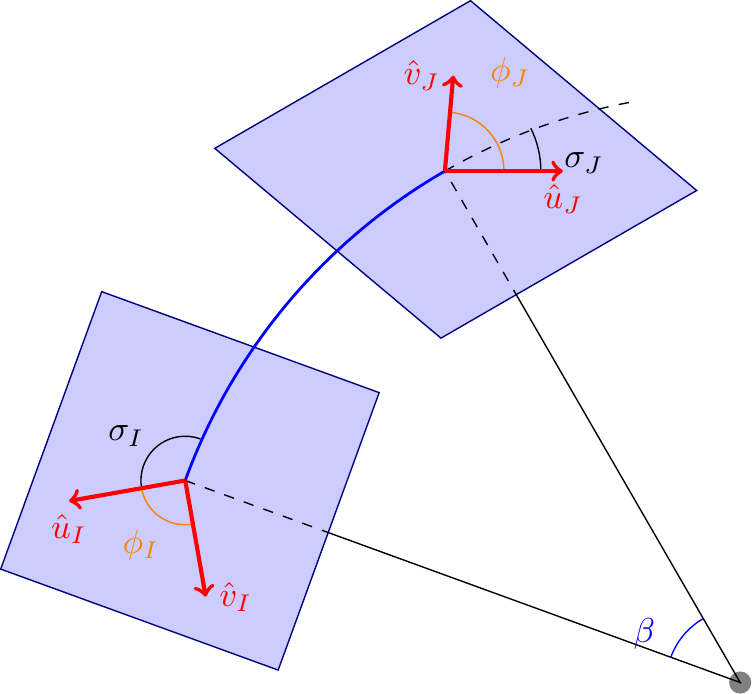}
\caption{\label{ET_fig:Detectors_on_Earth} Depiction of the unit vectors of the detector arms $\hat{u}_I, \hat{v}_I, \hat{u}_J, \hat{v}_J$ (red), 
the opening angles $\phi_I$, $\phi_J$ (orange) and the angles $\sigma_I$, $\sigma_J$ between $\hat{u}_{I,J}$ and the great circle (blue) between 
the detectors $I$ and $J$}
\end{figure}

The angular dependence of the pattern functions $F_A(\hat{\Omega})$ can be split into the relative orientation of the detectors towards each other 
and the orientation of an incoming GW with respect to the two-detector cluster. The overlap reduction functions (ORF) account for the 
relative orientation of the two detectors. \\
We consider a pair $(I,J)$ of Michelson interferometers on Earth with opening angles $\phi_I$ and $\phi_J$. We denote the direction vectors of the 
detector arms as $\hat{u}_{I,J}, \hat{v}_{I,J}$ such that $(\hat{u}_{I,J}, \hat{v}_{I,J}, \hat{z}_{I,J})$, with $\hat{z}_{I,J}$ being the 
direction pointing to the sky, forms a positively oriented frame, as shown in Fig.~\ref{ET_fig:Detectors_on_Earth}. The relative orientation of 
the detectors can be described by the angles $\sigma_{I,J}$ between the detector arms $\hat{u}_{I,J}$ and the separation vector $\Delta \vec{x}$, 
which points from detector $I$ to $J$. \\
The direction vectors of the detector arms in the cluster frame are given by:
\begin{align}
&\hat{u}_I = \begin{pmatrix} \cos\sigma_I \\ \sin\sigma_I \\ 0 \end{pmatrix} \ ,\ 
 \hat{v}_I = \begin{pmatrix} \cos(\sigma_I + \phi_I) \\ \sin(\sigma_I + \phi_I) \\ 0 \end{pmatrix} \ , \ \notag\\
& \hat{d} = \frac{1}{\sqrt{2(1-\cos\beta)}}\begin{pmatrix} \sin\beta \\ 0 \\ \cos\beta - 1 \end{pmatrix} \ , \ \notag \\
&\hat{u}_J = \begin{pmatrix} \cos\beta \cos\sigma_J \\ \sin\sigma_J \\ -\sin\beta \cos\sigma_J \end{pmatrix} \ , \
 \hat{v}_J = \begin{pmatrix}\cos\beta \cos(\sigma_J + \phi_J) \\ \sin(\sigma_J + \phi_J) \\ -\sin\beta \cos(\sigma_J + \phi_J) \end{pmatrix}
\end{align}

The contractions of the two detector tensors are then given by:
\begin{widetext}
\begin{align}
D_I^{ij}D^J_{ij} &= \frac{1}{2}(\hat{u}_I^i\hat{u}_I^j - \hat{v}_I^i\hat{v}_I^j)\frac{1}{2}(\hat{u}^J_i\hat{u}^J_j - \hat{v}^J_i\hat{v}^J_j) 
\notag \\
&= \frac{1}{4}\left[(\sin^2\sigma_{1+}-\sin^2\sigma_{1-})(\sin^2\sigma_{2+}-\sin^2\sigma_{2-})\right. \notag \\
&\qquad +\frac{1}{2}\cos\beta(\sin(2\sigma_{1+})-\sin(2\sigma_{1-}))(\sin(2\sigma_{2+})-\sin(2\sigma_{2-})) \notag \\
&\qquad \left.+\cos^2\beta(\cos^2\sigma_{1+}-\cos^2\sigma_{1-})(\cos^2\sigma_{1+}-\cos^2\sigma_{1-})\right], \label{Eq:DD}\\
\notag \\
D_{I,k}^iD_J^{kj}\hat{d}_i\hat{d}_j &= \frac{1}{4}\left((\hat{u}_I\cdot\hat{d})\hat{u}_I - 
(\hat{v}_I\cdot\hat{d})\hat{v}_I\right)\cdot\left((\hat{u}_J\cdot\hat{d})\hat{u}_J - (\hat{v}_J\cdot\hat{d})\hat{v}_J\right) \notag \\
&= \frac{1+\cos\beta}{8}\left[\frac{1}{4}(\sin(2\sigma_{1+})-\sin(2\sigma_{1-}))(\sin(2\sigma_{2+})-\sin(2\sigma_{2-}))\right. \notag \\
&\qquad \qquad \qquad \left.+\cos\beta(\cos^2\sigma_{1+}-\cos^2\sigma_{1-})(\cos^2\sigma_{2+}-\cos^2\sigma_{2-})\right], \label{Eq:DDdd}\\
\notag \\
D_I^{ij}D_J^{kl}\hat{d}_i\hat{d}_j\hat{d}_k\hat{d}_l &= \frac{1}{4}\left((\hat{u}_I\cdot\hat{d})^2 - 
(\hat{v}_I\cdot\hat{d})^2\right)\left((\hat{u}_J\cdot\hat{d})^2 - (\hat{v}_J\cdot\hat{d})^2\right) \notag \\
&= \frac{(1+\cos\beta)^2}{16}(\cos^2\sigma_{1+}-\cos^2\sigma_{1-})(\cos^2\sigma_{2+}-\cos^2\sigma_{2-}), \label{Eq:DDdddd}
\end{align}
\end{widetext}
with $\sigma_{1+} \coloneq \sigma_I + \phi_I$, $\sigma_{1-} \coloneq \sigma_I$, $\sigma_{2+} \coloneq \sigma_J + \phi_J$ and $\sigma_{2-} \coloneq 
\sigma_J$.\\

Nishizawa et al. \cite{Nishizawa2009} have used a different definition of the angles $\sigma_{1,2}$, which is related to our notation by: 
$\sigma_{1+} = \sigma_1 + \frac{\phi_I}{2}$, $\sigma_{1-} = \sigma_2 - \frac{\phi_I}{2}$, $\sigma_{2+} = \sigma_2 + \frac{\phi_J}{2}$ and 
$\sigma_{2-} = \sigma_2 
- \frac{\phi_J}{2}$.\\

Finally we get the following expression for the ORF $\gamma_{IJ}^M$ of the detectors $I$ and $J$ for the polarization $M$:
\begin{widetext}
\begin{align}
\gamma_{IJ}^M(f) &= \rho_1^M(\alpha)D_I^{ij}D^J_{ij} + \rho_2^M(\alpha)D_{I,k}^iD_J^{kj}\hat{d}_i\hat{d}_j + 
\rho_3^M(\alpha)D_I^{ij}D_J^{kl}\hat{d}_i\hat{d}_j\hat{d}_k\hat{d}_l \notag \\
&= \frac{1}{16}\left\lbrace4\rho_1^M(\sin^2\sigma_{1+}-\sin^2\sigma_{1-})(\sin^2\sigma_{2+}-\sin^2\sigma_{2-})\right. \notag \\
&+ \left(2\rho_1^M\cos\beta + \rho_2^M\frac{1+\cos\beta}{2}\right) \cdot(\sin(2\sigma_{1+})-\sin(2\sigma_{1-})) 
(\sin(2\sigma_{2+})-\sin(2\sigma_{2-})) \notag \\
&+ \left(4\rho_1^M\cos^2\beta + 2\rho_2^M(1+\cos\beta)\cos\beta + \rho_3^M(1+\cos\beta)^2\right) 
\}\cdot\left.(\cos^2\sigma_{1+}-\cos^2\sigma_{1-}) (\cos^2\sigma_{2+}-\cos^2\sigma_{2-})\right\rbrace, \label{Eq:ORFIJ}
\end{align}
\end{widetext}
where we defined the argument $\alpha$ and the relation between the arclength $\beta$ and the distance $|\vec{d}|$ by:
\begin{equation}
\alpha(f) \coloneq \frac{2\pi f|\vec{d}|}{c} \textit{ , $ $ } |\vec{d}| = 2R_E\sin\frac{\beta}{2}.
\end{equation}

\section{Einstein Telescope and Earth-based Detectors}\label{ET_section:ET}

As mentionned in the introduction, the Einstein Telescope is going to be part of the third generation of Earth-based detectors, and we thus want 
to consider several ground based networks involving ET, in order to figure out how ET can affect the overall sensitivity. The estimation of the 
maximal achievable sensitivity could be of use for future detector designs and expectations in the constraint of cosmological parameters. In 
particular, we want to investigate the polarizations of the gravitational background and ET's capability of measuring it.\\

\subsection{Symmetry of the Einstein Telescope}

Since ET consists of three detectors, one can form three detector pairs which can be used to cross correlate the signal. With the resulting three 
noise-free signals, one could in principle (as we will see below, for ET those three signals are not independent) solve for the fraction of the 
power in each polarization mode (tensor $T$, vector $V$, scalar $S$) by using the ORFs.\\
\\
The fraction in Eq.~\eqref{Eq:SNR_M} can be rewritten as:
% \begin{align}\label{Eq:detF/FT}
% \frac{\det\textbf{F}}{\mathcal{F}_T} &= \frac{
% \begin{vmatrix}
% F_{TT} & F_{TV} & F_{TS} \\
% F_{VT} & F_{VV} & F_{VS} \\
% F_{ST} & F_{SV} & F_{SS}
% \end{vmatrix}}
% {\begin{vmatrix}
% F_{VV} & F_{VS} \\
% F_{SV} & F_{SS}
% \end{vmatrix}} \notag \\
% &= F_{TT} - \frac{F_{VV}F_{TS}^2-2F_{VS}F_{TS}F_{TV}+F_{SS}F_{TV}^2}{F_{VV}F_{SS}-F_{VS}^2}.
% \end{align}\\

\begin{widetext}
\begin{equation}\label{Eq:detF/FT}
\frac{\det\textbf{F}}{\mathcal{F}_T} = \frac{
\begin{vmatrix}
F_{TT} & F_{TV} & F_{TS} \\
F_{VT} & F_{VV} & F_{VS} \\
F_{ST} & F_{SV} & F_{SS}
\end{vmatrix}}
{\begin{vmatrix}
F_{VV} & F_{VS} \\
F_{SV} & F_{SS}
\end{vmatrix}} 
= F_{TT} - \frac{F_{VV}F_{TS}^2-2F_{VS}F_{TS}F_{TV}+F_{SS}F_{TV}^2}{F_{VV}F_{SS}-F_{VS}^2}.
\end{equation}
\end{widetext}

This formula was derived via a maximum likelihood method for more than $3$ detectors to find 3 modes and is therefore not well defined for 2 
detectors, which can be seen by writing out the expression for $\mathcal{F}_T$:
\begin{widetext}
\begin{align}
\mathcal{F}_T &= \sum_{(I,J)}\sum_{(I',J')} \int_0^{T_{obs}} \frac{\gamma_{IJ}^V(t)^2\gamma_{I'J'}^S(t')^2 - 
\gamma_{IJ}^V(t)\gamma_{IJ}^S(t)\gamma_{I'J'}^V(t')\gamma_{I'J'}^S(t')}{P_IP_JP_{I'}P_{J'}} dt' dt \notag \\
&= \int_0^{T_{obs}} \frac{\gamma_{IJ}^V(t)^2\gamma_{IJ}^S(t')^2 - \gamma_{IJ}^V(t)\gamma_{IJ}^S(t)\gamma_{IJ}^V(t')\gamma_{IJ}^S(t')}{P_I^2P_J^2} 
dt' dt \textit{ $ $ } = 0,
\end{align}
\end{widetext}
where we used that in our case the ORFs are time independent: $\gamma_{IJ}^M(t) = \gamma_{IJ}^M(0)$. \\

Neglecting for the moment factors of $\frac{T_{obs}}{P_IP_J}$, we find:
\begin{align}
\sqrt{F_{VV}F_{SS}} &\sim \sqrt{\sum_{(I,J)}\sum_{(I',J')} (\gamma_{IJ}^V)^2(\gamma_{I'J'}^S)^2} \notag\\ &= \sqrt{(\gamma_{IJ}^V\gamma_{IJ}^S)^2} 
= \gamma_{IJ}^V\gamma_{IJ}^S \sim F_{VS} \notag \\
\Rightarrow \frac{\det\textbf{F}}{\mathcal{F}_T} &= F_{TT} - \frac{(\sqrt{F_{VV}}F_{TS} - \sqrt{F_{SS}}F_{TV})^2}{F_{VV}F_{SS}-F_{VS}^2} \notag \\
&= \frac{T_{obs}}{P_IP_J} \left[ (\gamma_{IJ}^T)^2  \phantom{\frac{\gamma^2}{\gamma^2}} \right. \notag \\
&\left. - \frac{(\gamma_{IJ}^V\gamma_{IJ}^T\gamma_{IJ}^S - 
\gamma_{IJ}^S\gamma_{IJ}^T\gamma_{IJ}^V)^2}{(\gamma_{IJ}^V)^2(\gamma_{IJ}^S)^2 - (\gamma_{IJ}^V\gamma_{IJ}^S)^2} \right].
\end{align}
This expression is not well defined since the denominator is zero. In order to see whether the vanishing numerator helps, one has to carefully 
take the limit of a slightly non-degenerate case. Even if one uses the formula for more than $3$ detectors, one should be careful, since the 
fraction is ill defined as soon as the $\sum_{(I,J)}\gamma_{I,J}^M$ commute, which happens for the 3 ET-detectors due to the fact that they are 
three identical detectors and their symmetric arrangement leads to ($P_I = P_J \eqcolon P$ $ $ for every pair $(I,J)$):
\begin{align}
&\gamma_{12}^M = \gamma_{23}^M = \gamma_{31}^M \quad  \forall M \in \lbrace T, V, S \rbrace \notag \\
&\Rightarrow \ F_{MM'} = T_{obs}\frac{\gamma_M\gamma_{M'}}{P^2} \\
&\Rightarrow \ \sum_{(I,J)}(\gamma_{IJ}^M)^2\sum_{(I,J)}(\gamma_{IJ}^{M'})^2 = 9(\gamma_{IJ}^M)^2(\gamma_{IJ}^{M'})^2 \notag\\
&= (\sum_{(I,J)}\gamma_{IJ}^M\gamma_{IJ}^{M'})^2.
\end{align}

To use the formula for the ET-detector we have to break the symmetry by changing the ORF of one detector pair by a small 
amount $\epsilon(f)$ and then take the limit:
\begin{align}
\epsilon(f)\rightarrow 0 \textit{ $ $ } \forall f. \notag
\end{align}
Without loss of generality, we can for example consider the case $M = T$, and we perturb one of the ORFs:
\begin{align}
\gamma_{12}^M &= \gamma_{23}^M = \gamma_{31}^M - \epsilon_M \eqcolon \gamma_M \textit{ $ $ } \forall M \in \lbrace T, V, S \rbrace; \notag
\end{align}
When we plug this into the denominator and numerator of the fraction in the right-hand side of Eq.~\eqref{Eq:detF/FT} we get:
\begin{align}
\mathcal{F}_T &= F_{VV}F_{SS} - F_{VS}^2 \notag \\ &= \left(\frac{T_{obs}}{P^2}\right)^2 
\left(\sum_{(I,J)}(\gamma_{IJ}^V)^2\sum_{(I,J)}(\gamma_{IJ}^{S})^2 - (\sum_{(I,J)}\gamma_{IJ}^V\gamma_{IJ}^{S})^2\right) \notag \\
&= 2\left(\frac{T_{obs}}{P^2}\right)^2(\epsilon_V\gamma_S - \epsilon_S\gamma_V)^2
\end{align}
and similarly for the numerator.
Plugging these expressions into Eq.~\eqref{Eq:detF/FT} and taking the limit we arrive at:
\begin{widetext}
\begin{align}
\frac{\det\mathbf{F}}{\mathcal{F}_T} &=  F_{TT} - \frac{F_{VV}F_{TS}^2-2F_{VS}F_{TS}F_{TV}+F_{SS}F_{TV}^2}{F_{VV}F_{SS}-F_{VS}^2} \notag \\
&= \frac{T_{obs}}{P^2}\left(\frac{3}{2}\gamma_T\epsilon_T + 
\frac{3}{2}\gamma_V\gamma_S\epsilon_T\frac{\epsilon_T(\gamma_V\epsilon_S+\gamma_S\epsilon_V)}{(\gamma_V\epsilon_S-\gamma_S\epsilon_V)^2} + 
3\gamma_T\gamma_V\gamma_S\epsilon_T\frac{\epsilon_V\epsilon_S}{(\gamma_V\epsilon_S-\gamma_S\epsilon_V)^2} \right) \notag \\
&\textit{ $ $ $ $ }+ \mathcal{O}(\epsilon^2) \textit{ $ $ } \overset{\epsilon\rightarrow 0}{\longrightarrow} \textit{ $ $ } 0.
\end{align}
\end{widetext}
Due to the symmetry of the Einstein Telescope it is thus impossible to separate the modes out of the signal. The ORFs of each 
detector pair are the same, since they only depend on their relative orientation. Therefore, the detector correlation matrix $\Pi$ has a vanishing 
determinant and the relation between the cross-correlated signals and the modes of the gravitational wave background cannot be inverted:
\begin{equation}
\det\Pi = \begin{vmatrix}
\gamma_{12}^T & \gamma_{12}^V & \gamma_{12}^S \\
\gamma_{23}^T & \gamma_{23}^V & \gamma_{23}^S \\
\gamma_{31}^T & \gamma_{31}^V & \gamma_{31}^S
\end{vmatrix} = \begin{vmatrix}
\gamma_T & \gamma_V & \gamma_S \\
\gamma_T & \gamma_V & \gamma_S \\
\gamma_T & \gamma_V & \gamma_S
\end{vmatrix} = 0.
\end{equation}

Note that even by perturbing the symmetry of ET (slightly changing the arm length or tilting the detector plane), the induced changes are 
negligible and do not allow the use of ET alone to distinguish between the polarization modes. A detailed calculation of the symmetry breakings 
can be found in the appendix~\ref{ET_appendix:ET_perturbation}.

\subsection{Cross Correlation of Future Earth-based Detectors}
We now turn our attention to combinations of ET with other detectors, which are already existing (LIGO, Virgo) or under construction (KAGRA). By 
adding two additional signals to the ET-cluster we break the symmetry and the problem mentionned in the previous section is solved. We add the 
two advanced LIGO detectors in Livingston (LL) and Hanford (LH) to our set of detectors, which results in 6 correlation signals out of which 4 
(ET-ET, ET-LL, ET-LH, LL-LH) are independent. This allows us to distinguish the polarization modes, even if one of the detectors could not be used 
for some reason.

\begin{figure}[h!]
\begin{minipage}{1.0\linewidth}
	\includegraphics[width=1\linewidth]{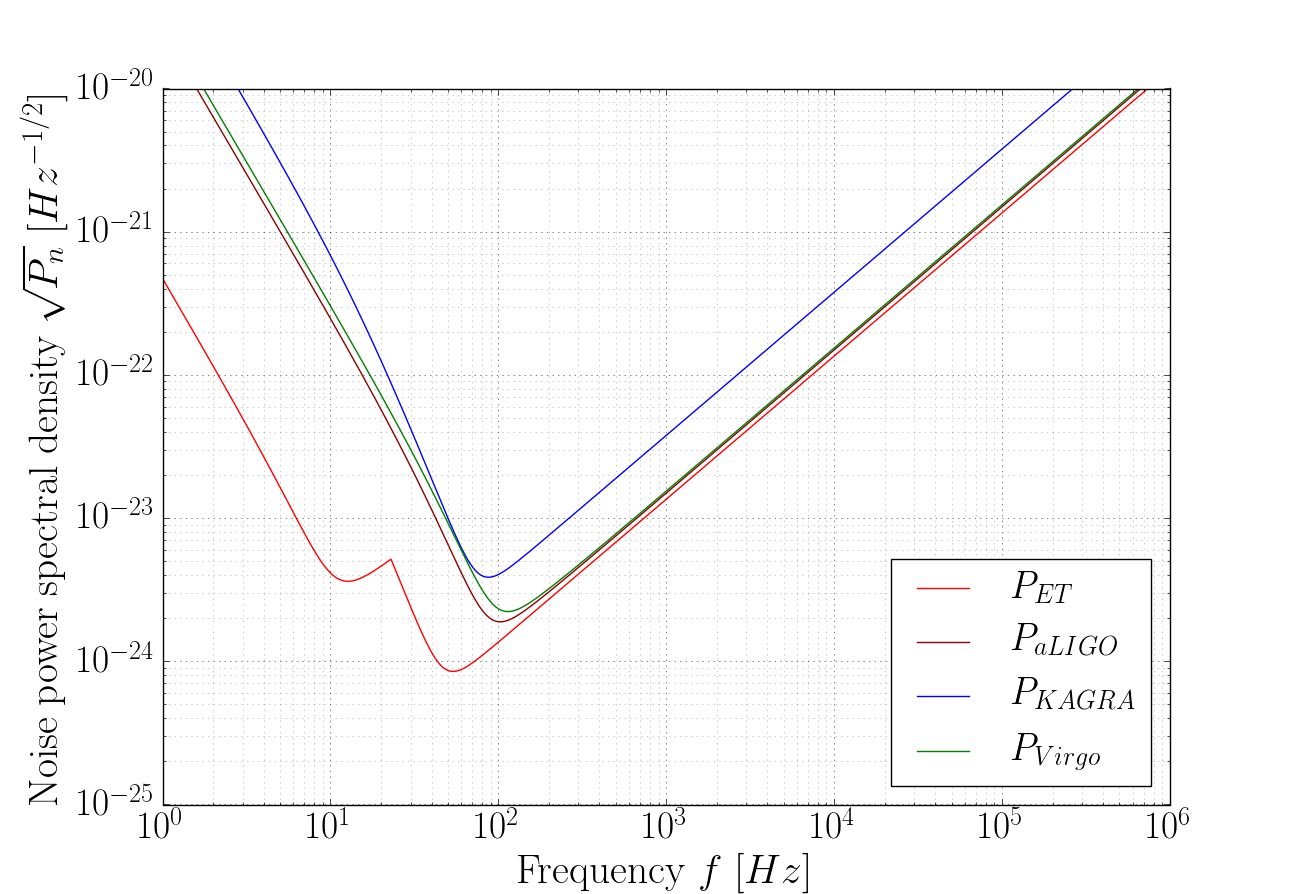}
\end{minipage}
\hfill
\begin{minipage}{1.0\linewidth}
	\includegraphics[width=1\linewidth]{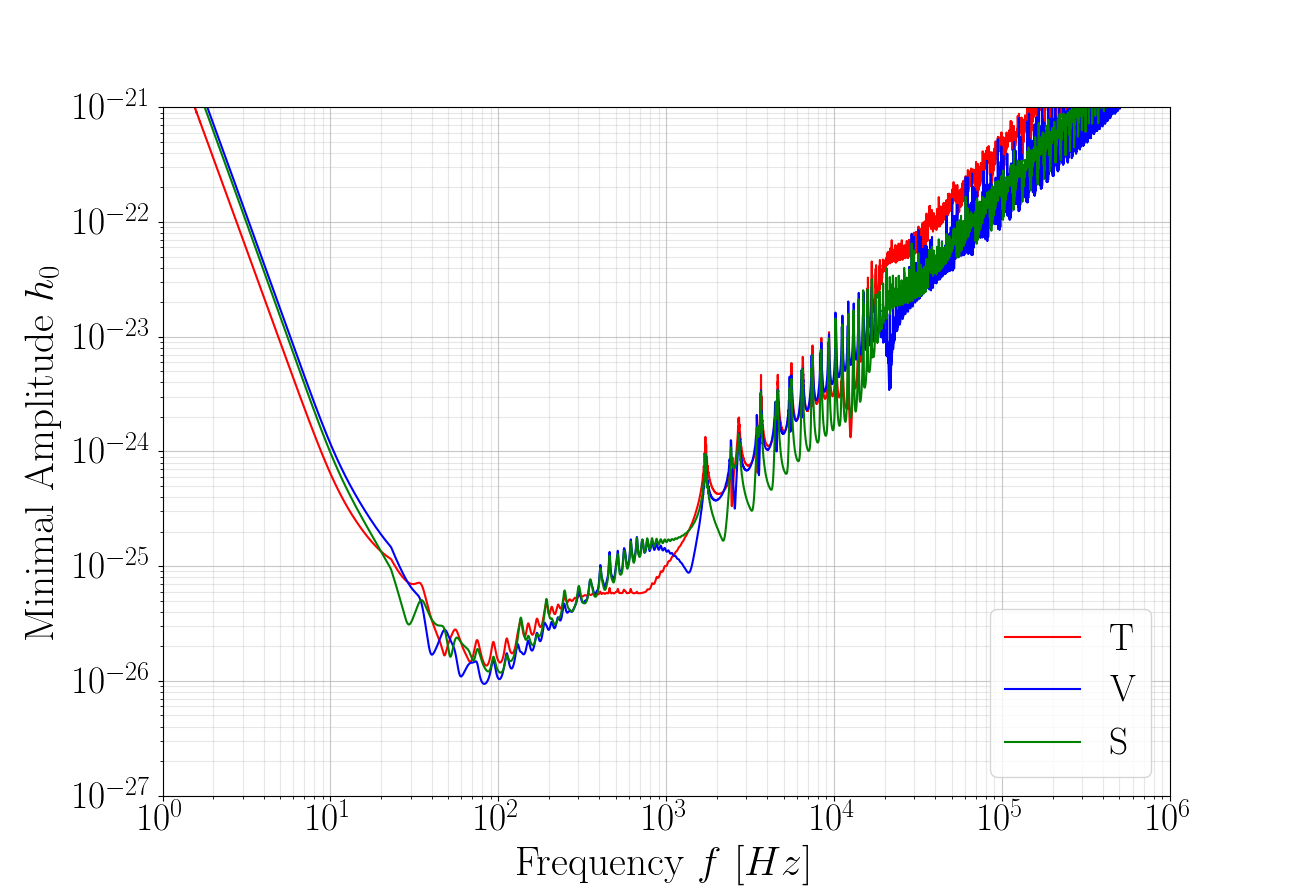}
\end{minipage}
\caption{\label{ET_fig:Earth_detectors} The noise power spectral densities of all involved detectors (above) and the sensitivity of all existing 
and near future Earth detectors combined (ET, LIGO, Virgo and KAGRA). The minimal achievable sensitivity is considered for a $SNR$ of 8.}
\end{figure}

In Fig.~\ref{ET_fig:Earth_detectors} we compare the noise power spectral densities of ET, LIGO, Virgo and KAGRA and show their combined 
sensitivity for the polarization modes of a GW signal.

Note that all ET detectors lie in the same plane and are on the scale of Earth at the same position. Should one of the three ET detectors be 
taken out of the network for any reason, the directions in which its arms were pointing are still covered by the neighbouring detectors. This is 
why the sensitivity would not be significantly affected.

Moreover, by considering the addition of Virgo and KAGRA to the network, beside ET and LIGO, we gain slightly gain sensitivity for frequencies 
above \SI{100}{Hz}.

\section{DECIGO and Correlation with Earth Detectors}\label{ET_section:DECIGO}
After having considered the combined sensitivity of a strictly Earth-based network of detectors, we can now investigate the consequences of a 
future space-borne detector. As already mentionned in the introduction, we focus on the DECIGO project, since the LISA sensitivity lies in a lower 
frequency range than the Earth detectors, and not overlap of their respective frequency bands would be possible.

\subsection{Earth-space Network Sensitivity}

DECIGO is a space-based experiment, and therefore there is no noise due to vibrations of the ground. Since its sensitive region and the one of ET 
and LIGO overlap in the frequency range between \SI{10}{Hz} and \SI{100}{Hz}, it makes sense to cross correlate their signal to get a higher 
precision and confidence for the separation of the signal into the three different polarization modes.

\begin{figure}[h!]
\begin{minipage}{1.0\linewidth}
	\includegraphics[width=1\linewidth]{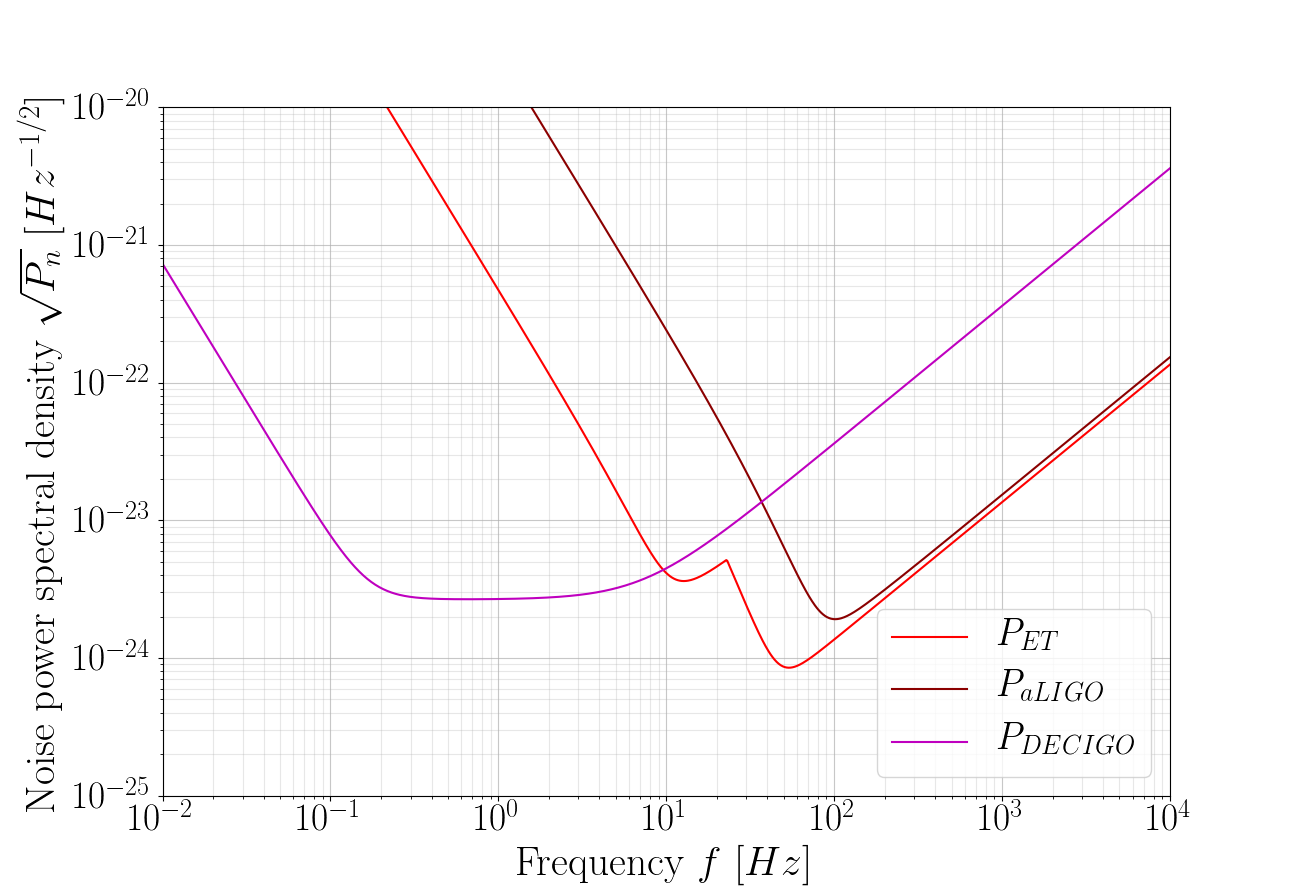}
\end{minipage}
\hfill
\begin{minipage}{1.0\linewidth}
	\includegraphics[width=1\linewidth]{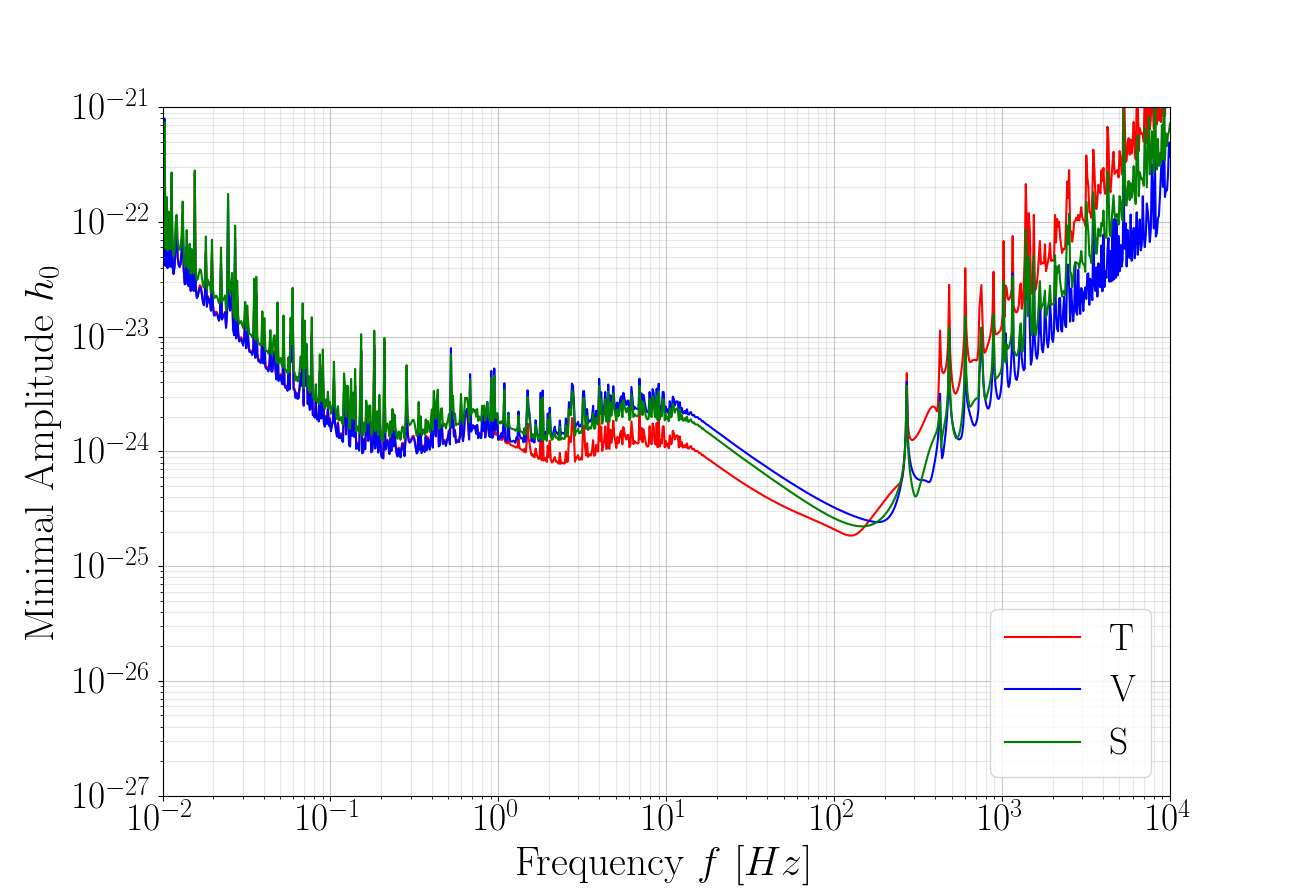}
\end{minipage}
\caption{\label{ET_fig:DECIGO} Noise power spectra of a DECIGO, ET and advanced LIGO detector (above) and the sensitivity of DECIGO alone in its 
C3 configuration.}
\end{figure}

The DECIGO experiment consists of four detector clusters. Each cluster is made up by three satellites which form three independent identical 
Michelson interferometers. One can for example arrange the four clusters in the C3 configuration \cite{YagiSeto2011, Kawamura2006} where two 
clusters are located at the same position near the Earth (about 1 AU behind the Earth, on the same orbit around the Sun) and form a star shape, 
and the remaining two form a triangle together with the star-cluster, which has the Sun at its centre. In Fig.~\ref{ET_fig:DECIGO} we compare the 
noise power spectrum of DECIGO to the ones of ET and LIGO and plot the sensitivity of DECIGO in the C3 configuration.

\begin{figure}[h!]
\begin{minipage}{1.0\linewidth}
	\includegraphics[width=1\linewidth]{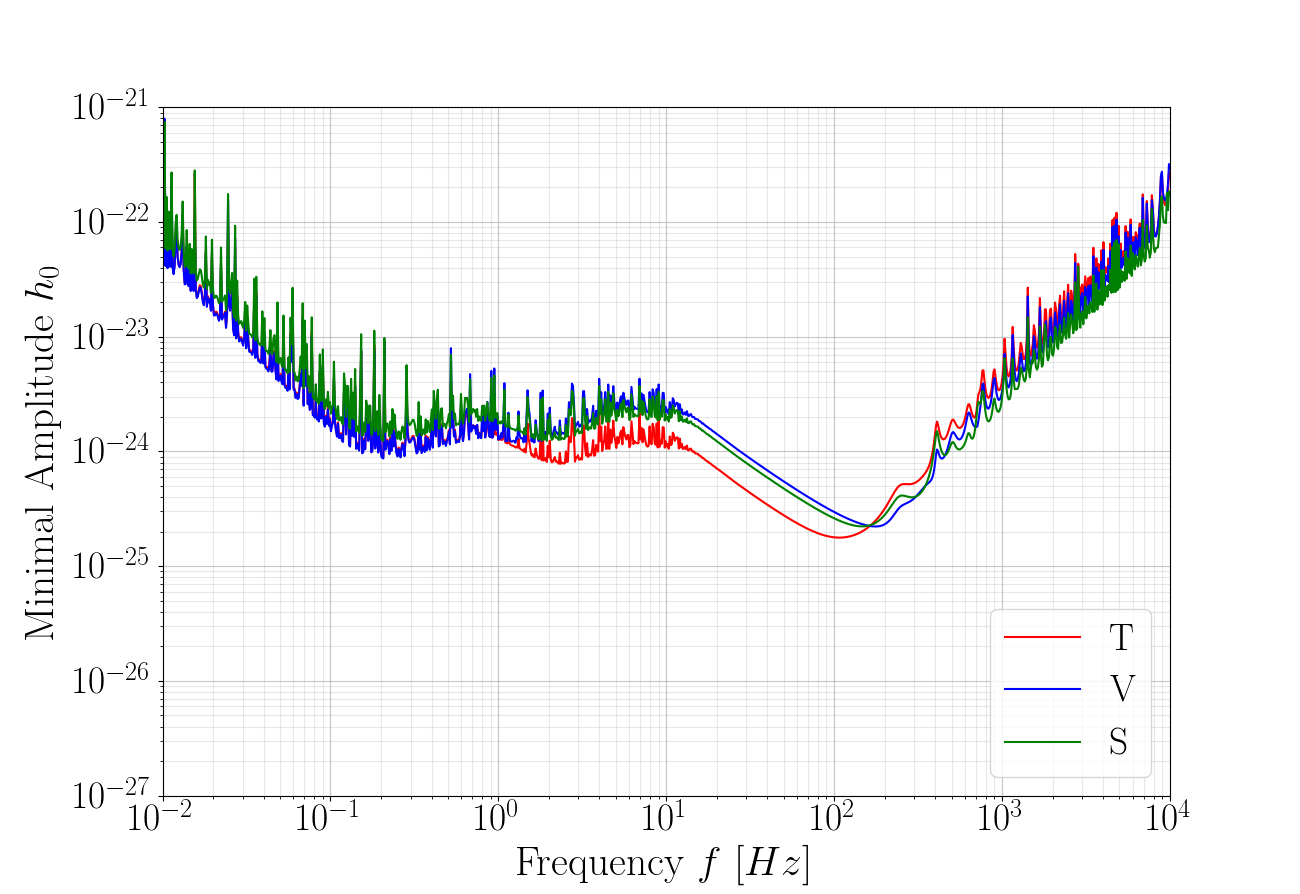}
\end{minipage}
\hfill
\begin{minipage}{1.0\linewidth}
	\includegraphics[width=1\linewidth]{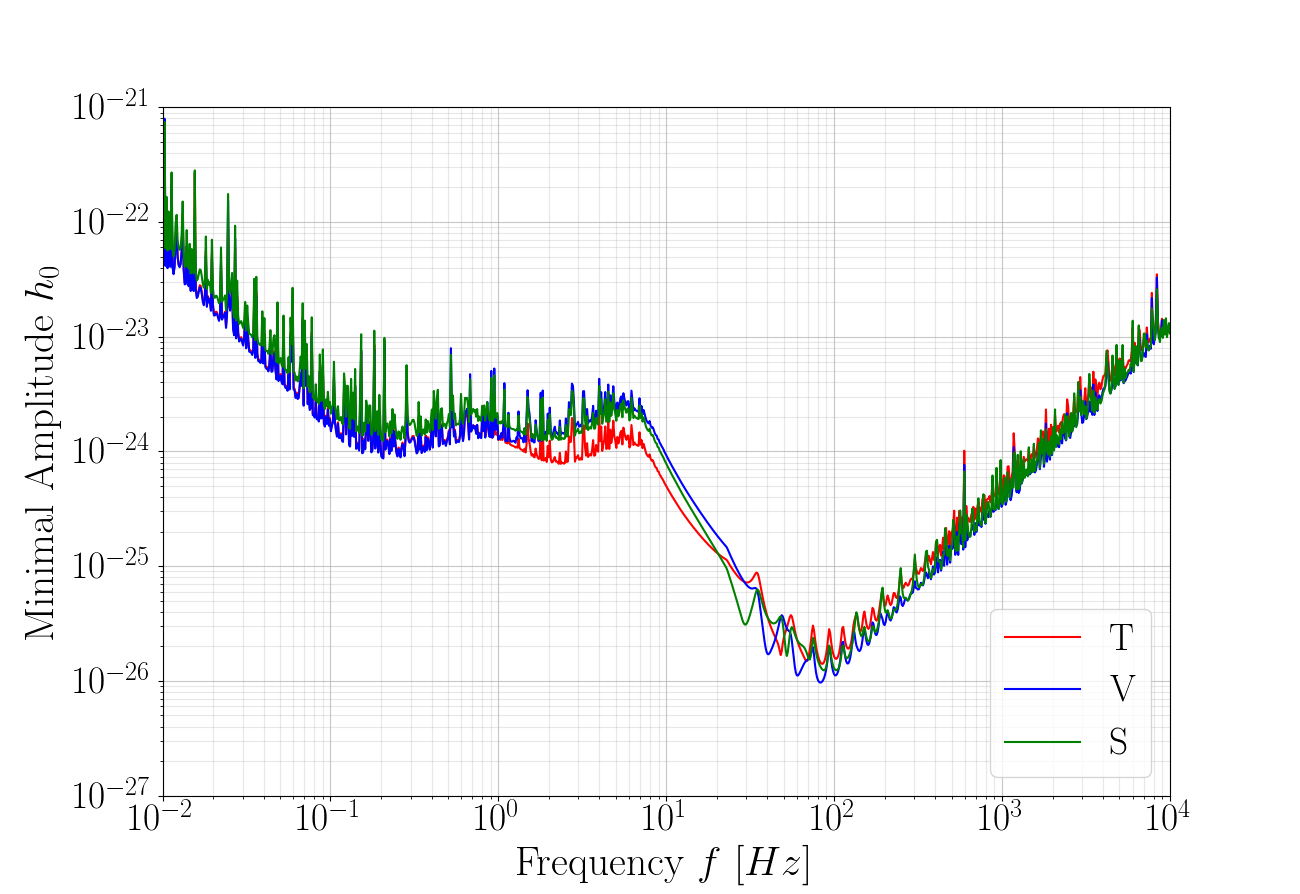}
\end{minipage}
\caption{\label{ET_fig:ELD} ET and all DECIGO detectors in C3 configuration (above) and together with both LIGO detectors (below), averaged over a 
total measurement time of one year. The addition of the LIGO detectors significantly improve the sensitivity around \SI{100}{Hz}. Virgo and KAGRA 
were not included, since the effect is negligible.}
\end{figure}

DECIGO is much more sensitive in the low frequencies than all detectors on Earth combined and is even slightly more sensitive around \SI{10}{Hz}, 
which comes in handy when we combine it with Earth detectors. When we add ET and then LIGO to the set of detectors and sum over all combinations 
of cross-correlations, we get the plots shown in Fig.~\ref{ET_fig:ELD}.

As we can see, ET drags the curves down around \SI{10}{Hz} and  mostly above \SI{100}{Hz}. In particular, the tensor and scalar modes are 
affected and become about as sensitive as the vector mode. Together with LIGO the sensitivity is enhanced by one order of magnitude at LIGO's most 
sensitive frequency range around \SI{100}{Hz}.

\subsection{Time-dependent Sensitivity}
\begin{figure}[h!]
\begin{minipage}{1.0\linewidth}
	\includegraphics[width=1\linewidth]{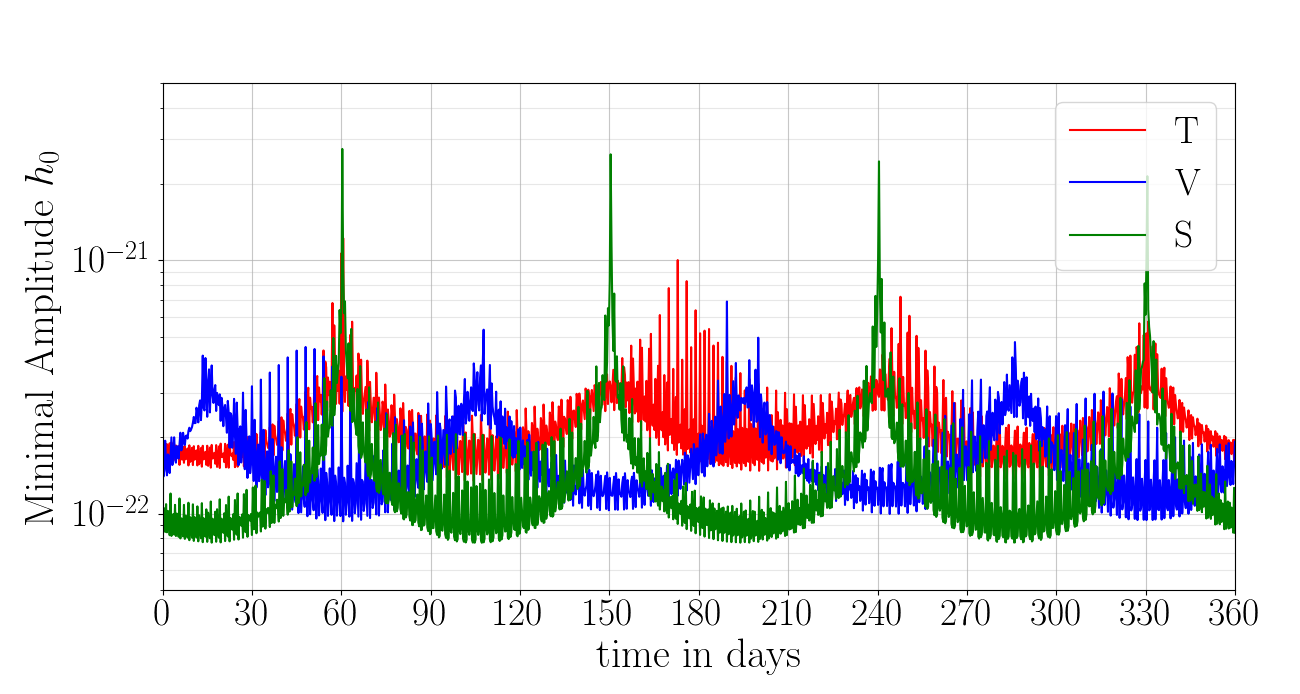}
\end{minipage}
\hfill
\begin{minipage}{1.0\linewidth}
	\includegraphics[width=1\linewidth]{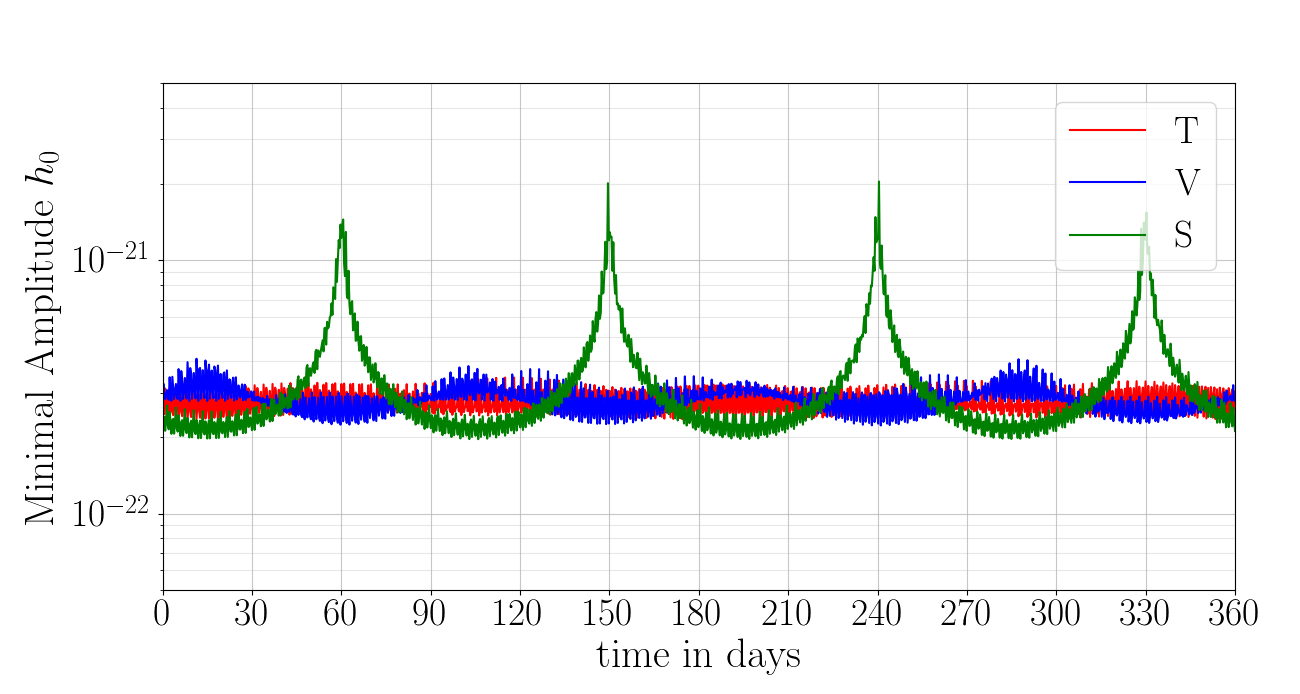}
\end{minipage}
\caption{\label{ET_fig:Tdep} Above: There are two DECIGO detectors in the star cluster, which show the same time dependence when cross correlated 
with an Earth detector. We take all correlations between those two and all ET and LIGO detectors. Below: In a square configuration, where the four 
DECIGO clusters are put on a square around the Sun, there is one detector in each of the three clusters which are far away from the Earth, which 
have the same time dependence when we correlate them with Earth detectors. We form all pairs of these three with ET and LIGO. Both plots 
correspond to a frequency of \SI{100}{Hz}.}
\end{figure}

In the planned C3-configuration of DECIGO we used previously, each cluster rotates around its own axis perpendicular to the detector plane as it 
rotates around the Sun, such that it returns to its original position after one year. A detector on Earth follows Earth rotation and therefore a 
relatively quick oscillation of one day superposed to a slow oscillation of one year. This combined change in the orientations of the detectors in 
a DECIGO-cluster relative to detectors on Earth leads to a time varying sensitivity, which is different for each mode, as can be seen in 
Fig.~\ref{ET_fig:Tdep}. The time dependence of the sensitivity is independent of the frequency. We plot the sensitivities at \SI{100}{Hz}, where 
the DECIGO-Earth detector pairs are most sensitive.

The variation of the sensitivity with time is different if we form the pair with a DECIGO detector close to the Earth or one far away from it. 
The location of the peaks is also different for DECIGO-Earth pairs, formed with different DECIGO detectors. Changing the detector on Earth however 
does not matter, since they are almost at the same place viewed on the solar system scale and oscillate much faster, and therefore do not 
influence the trend on a monthly scale.

\begin{figure}[h!]
\includegraphics[width=1.0\linewidth]{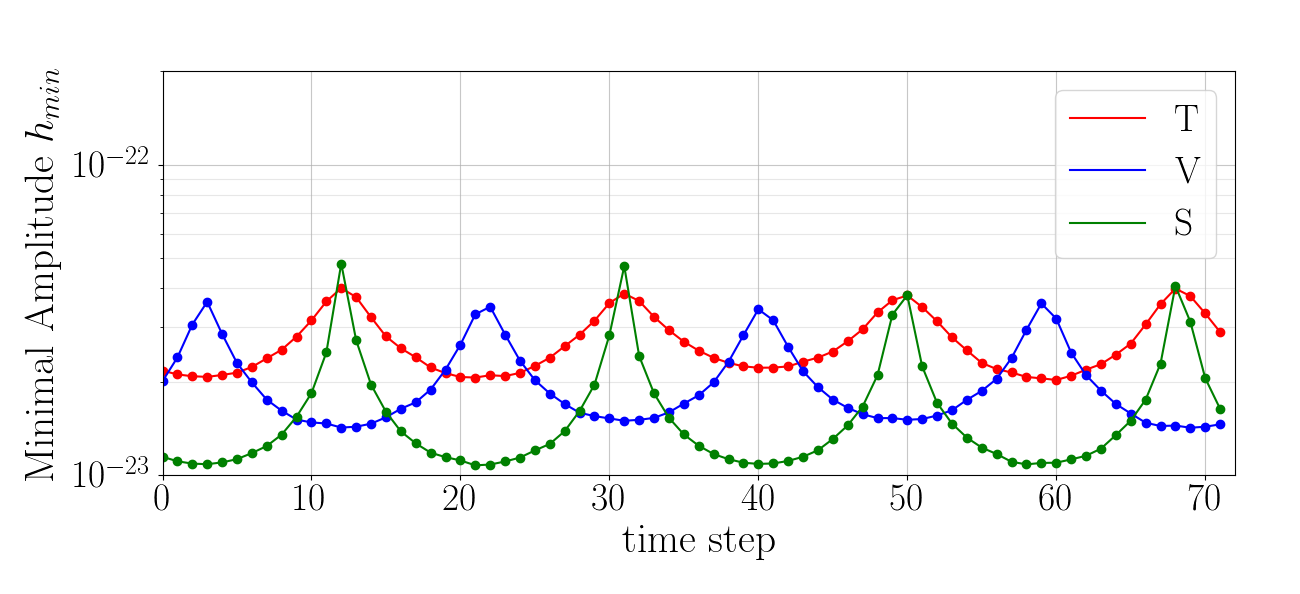}
\caption{\label{ET_fig:TdepClose} Combined sensitivity of the detector pairs with one of two neighbouring detectors in the star cluster and all 
ground-based detectors with an integration time of 5 days at a frequency of \SI{100}{Hz} for one year.}
\end{figure}

We now optimize the sensitivity by combining all detector pairs with a similar time-dependence and average over an integration time of 
5 days in order not to lose to much of the variation. For this purpose, we can always form all pairs with the Earth detectors, but we have to be 
careful which space detectors we pick. If we want to be able to clearly separate the vector mode from the other two, then it makes sense to pick 
the C3 configuration, because it has many detectors close to Earth. For two of the detectors in the star cluster the correlation with any Earth 
detector has almost the same time-dependence since their orientation only differs by $30^\circ$. In Fig.~\ref{ET_fig:TdepClose} we use all those 
pairs and integrate over 5 days to increase the sensitivity. Due to the fact that the vector modes time-dependence is phase-shifted with respect 
to the other two modes, we can easily separate it from the other two in this case.

\begin{figure}[h!]
\includegraphics[width=1.0\linewidth]{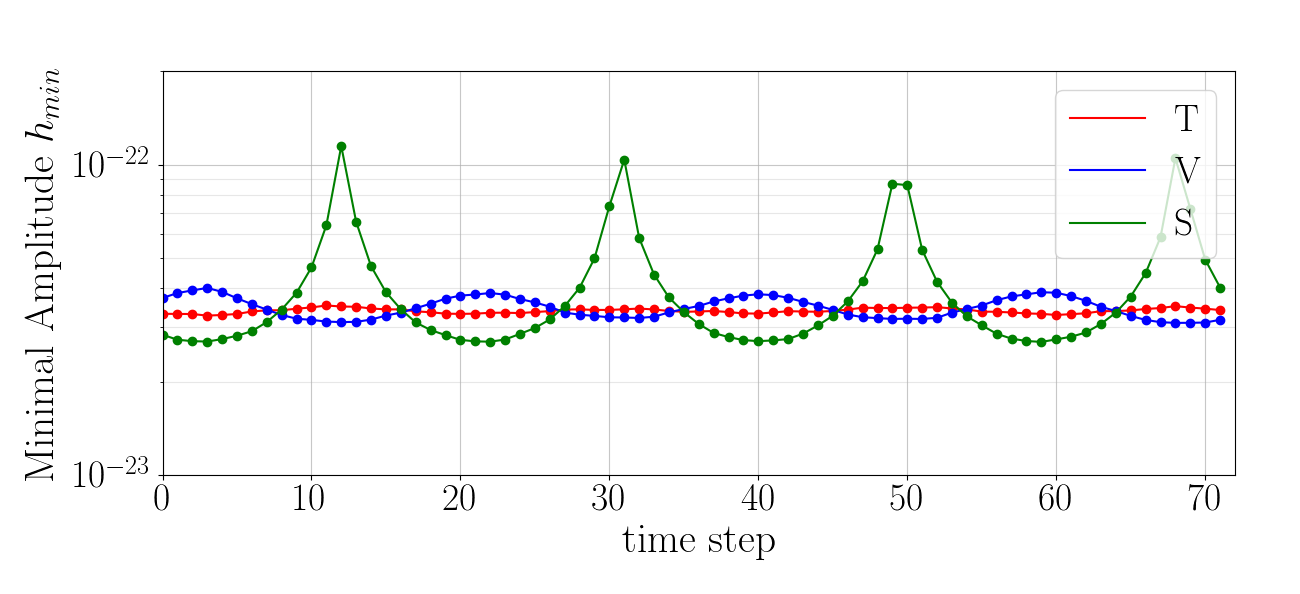}
\caption{\label{ET_fig:TdepFar} Time dependent sensitivity of detector pairs with one of each of the three clusters far from Earth in the square 
configuration with all Earth detectors, binned in time with steps of 5 days at a frequency of \SI{100}{Hz} for one year.}
\end{figure}

However, if we rather want to identify the scalar mode, then it makes more sense to move more clusters further away from Earth and at best on the 
opposite side of the orbit around the Sun, because for detectors which are far away from Earth the scalar mode has large peaks which correspond to 
blind spots. In this case we could arrange the four DECIGO clusters in a square around the Sun, such that only one cluster is close to Earth, and 
one is on the opposite side of Earth orbit. We can then arrange the initial orientation of the clusters, such that the peaks for one detector of 
each of the three clusters far from Earth coincide. Their combined time dependent sensitivity is shown in Fig.~\ref{ET_fig:TdepFar}.

By using the same detectors as previously and combining the data differently, one gets an alternative method to the maximum likelihood method 
for distinguishing the polarization modes. This can help to check the results and gives a higher confidence on a test of GR, without having to 
build another experiment.

Note that we performed the samed analysis for a scaled-down version of DECIGO closer to the Earth, B-DECIGO \cite{BDECIGO2018}. Although the 
sensitivity curves are a bit similar as the DECIGO ones, the results of the time dependent sensitivities does not appear 
to provide good enough differences between the modes. All the details regarding B-DECIGO can be found in the Appendix~\ref{ET_appendix:B-DECIGO}.

\section{Gravitational Waves from Point-sources}\label{ET_section:pointsources}
Until now we have calculated the sensitivity of various combinations of GW-detectors to an isotropic gravitational wave background. Now we attempt 
to do the same for point sources. Since the signal of a point source is coming from a specific direction, we do not average over all solid angles 
and our sensitivity becomes direction dependent.

\subsection{Sensitivity}
The derivation of the expression for the signal to noise ratio works in analogy to what we have done in section~\ref{ET_section:theory}. The 
metric perturbation field at the location of the detector $\vec{x}_I$ can be described as the sum of all 
gravitational waves, incident on the detector $I$, coming from all directions:
\begin{align}
h_{ij}(t,\vec{x}_I) &= \sum_A \int_{\mathbb{S}^2} h_A(t,\vec{x}_I,\hat{\Omega})e^A_{ij}(\hat{\Omega}) d\hat{\Omega} \notag \\
&= \sum_A \int\int_{\mathbb{S}^2} \tilde{h}_A(f,\hat{\Omega}) e^{2\pi i f\left(t-\frac{\hat{\Omega}\cdot\vec{x}_I}{c}\right)} 
e^A_{ij}(\hat{\Omega}) d\hat{\Omega}df.
\end{align}
For a gravitational wave coming from a point source located at $\hat{\Omega}_0$ in the sky, the frequency-space amplitude takes the form:
\begin{equation}
\tilde{h}_A(f,\hat{\Omega}) = \hat{h}_A(f)\delta(\hat{\Omega}-\hat{\Omega}_0).
\end{equation}

The response of the detector $I$ to an incoming gravitational wave is described by the so called pattern functions $F_I^A$, which are defined by 
contracting the basis tensors $e^A$ of the metric perturbations due to GW's for the polarizations $A \in \lbrace +, \times, x, y, b, l\rbrace$ 
with the detector tensor  $D_I$:
\begin{equation}
F_I^A(\hat{\Omega}) \coloneq e^A_{ij}(\hat{\Omega})D_I^{ij}.
\end{equation}
Therefore, the Fourier transform of the signal is given by:
\begin{equation}
\tilde{h}_I(f) = \sum_A \tilde{h}_A(f) e^{-2\pi i f\frac{\hat{\Omega}_0\cdot\vec{x}_I}{c}} F_I^A(\hat{\Omega}_0).
\end{equation}

By cross correlating two strains of different detectors ($s_I, s_J$) we get rid of the noise as seen in Eq.~\eqref{Eq:E[sI*sJ]}. Thus, the 
expectation of the Fourier transform of the two strains is:
\begin{align}
\mathbb{E}[{\tilde{h}_I}^*(f)\tilde{h}_J(f')] &= \tilde{h}_A^*(f)\tilde{h}_{A'}(f') e^{-\frac{2\pi 
i}{c}\hat{\Omega}_0\cdot(f\vec{x}_I-f'\vec{x}_J)} \notag \\ &\quad \cdot F_I^A(\hat{\Omega}_0)F_J^{A'}(\hat{\Omega}_0).
\end{align}

To maximize the signal to noise ratio, we filter this cross correlated strain with a filter function $\tilde{Q}$:
\begin{align}
\mu &\coloneq \mathbb{E}[Y] = \int \delta_T(f-f')\mathbb{E}[\tilde{h}_I^*(f)\tilde{h}_J(f')]\tilde{Q}(f') df'df \notag \\
&= \int \tilde{h}_A^*(f)\tilde{h}_{A'}(f)  e^{-\frac{2\pi i f}{c}\hat{\Omega}_0\cdot(\vec{x}_I-\vec{x}_J)} \notag\\
&\quad \cdot F_I^A(\hat{\Omega}_0)F_J^{A'}(\hat{\Omega}_0)  \tilde{Q}(f) df,
\end{align}
where $Y$ is the cross correlated signal:
\begin{equation}
Y \coloneq \int_{-\infty}^{\infty} \int_{-\infty}^{\infty} \delta_T(f-f') \tilde{s}_I^*(f)\tilde{s}_J(f')\tilde{Q}(f-f') df' df.
\end{equation}

To find the optimal filter function $\tilde{Q}$ we define a scalar product on the space of smooth complex valued functions $C^\infty(\mathbb{C})$:
\begin{equation}
(A,B) \coloneq \int A^*(f)B(f)P_I(|f|)P_J(|f|) df.
\end{equation}
Since the noise power spectra diverge algebraically at the origin and at infinity, we have to restrict our functions $A$ and $B$ to the 
Schwartz-space $\mathcal{S}(\mathbb{C})$.

We can express the expectation of the correlated signal and its variance in terms of this scalar product:
\begin{align}
\mu &= \left( \tilde{Q},\frac{\tilde{h}_A^*\tilde{h}_{A'} e^{-\frac{2\pi i f}{c}\hat{\Omega}_0\cdot\Delta\vec{x}} F_I^AF_J^{A'}}{P_IP_J} \right), 
\\
\sigma^2 &\coloneq \mathbb{V}[Y] = \mathbb{E}[Y^2] - \mathbb{E}[Y]^2 \approx \mathbb{E}[Y^2] = \frac{T}{4}(\tilde{Q},\tilde{Q}),
\end{align}
where $\Delta\vec{x} \coloneq \vec{x}_I - \vec{x}_J$ is the distance vector between the detectors $I$ and $J$.\\
The signal to noise ratio is therefore given by:
\begin{equation}
SNR = \frac{\mu}{\sigma} = \frac{\left( \tilde{Q},\frac{\tilde{h}_A^*\tilde{h}_{A'} e^{-\frac{2\pi i f}{c}\hat{\Omega}_0\cdot\Delta\vec{x}} 
F_I^AF_J^{A'}}{P_IP_J} \right)}{\sqrt{\frac{T}{4}(\tilde{Q},\tilde{Q})}}.
\end{equation}
This can be maximized, by choosing the filter function $\tilde{Q}$ parallell to the correlated signal with respect to our scalar product.
\begin{align}
\tilde{Q} \propto \frac{\tilde{h}_A^*\tilde{h}_{A'} e^{-\frac{2\pi i f}{c}\hat{\Omega}_0\cdot\Delta\vec{x}} F_I^AF_J^{A'}}{P_IP_J} \eqcolon 
\langle h_Ih_J\rangle.
\end{align}
With a proportionality constant $K$ we get:
\begin{align}
SNR &= \sqrt{\frac{4}{T}} \frac{(K\langle h_Ih_J\rangle,\langle h_Ih_J\rangle)}{\sqrt{(K\langle h_Ih_J\rangle,K \langle h_Ih_J\rangle)}} \notag\\
&= 2\sqrt{T(\langle h_Ih_J\rangle,\langle h_Ih_J\rangle)}.
\end{align}

Without loss of generality we can therefore choose $\tilde{Q} = \langle h_Ih_J\rangle$. Finally, we can calculate the maximal possible signal to 
noise ratio, with this choice of optimal filter function.
\begin{align}
SNR &= 2\sqrt{\frac{1}{T}(\tilde{Q},\tilde{Q})} \notag \\
&= 2\sqrt{\frac{1}{T}\int \frac{(\tilde{h}_A(f)\tilde{h}_{A'}(f) F_I^A(\hat{\Omega}_0)F_J^{A'}(\hat{\Omega}_0))^2}{P_I(|f|)P_J(|f|)} df}.
\end{align}
\\
To extract the frequency dependence and the polarization we insert a harmonic wave with amplitude $h_0$, frequency $f$ and polarization $A$.
\begin{equation}
\tilde{h}(f) \overset{!}{=} h_0\delta_T(f-f_0)\delta_{A'A};
\end{equation}
\begin{align}
SNR &= 2\sqrt{T^2\frac{(|h_0|^2F_I^A(\hat{\Omega}_0)F_J^A\hat{\Omega}_0)^2}{P_I(|f|)P_J(|f|)}} \notag \\
&= 2T\frac{|h_0|^2F_I^A(\hat{\Omega}_0)F_J^A(\hat{\Omega}_0)}{\sqrt{P_I(|f|)P_J(|f|)}}.
\end{align}
\\
So, we get the minimal amplitude required to detect a gravitational wave with polarization $A$ and at a $SNR$ of at least 8.
\begin{equation}
|h_0^A(f)|_{min} = \sqrt{32}\frac{\sqrt[4]{P_I(|f|)P_J(|f|)}}{\sqrt{TF_I^A(\hat{\Omega}_0)F_J^A(\hat{\Omega}_0)}}
\end{equation}

\subsection{Determination of Location and Polarizations of Point Sources}
If one has more than 8 detector pairs $(I,J)$ (\mbox{DECIGO}	 would do for example), one can solve for the direction $\hat{\Omega} \simeq 
(\theta,\phi)$ 
and all 6 possible polarizations $A \in \{+,\times,x,y,b,l\}$ of an incoming gravitational wave from a point source. We determine the $SNR$ for 
each quantity under the assumption that the maximum likelihood method is used to calculate them from the at least 8 cross-correlated signals 
$\mu_{IJ}$. The derivation is analogue to the one given in Nishizawa et al. \cite{Nishizawa2010}.\\
\\
The true parameters are denoted by $\vec{\theta}_{true} = (Y,s_A,\hat{\omega})$:
\begin{equation}
Y_{IJ}(f) = T^{3/2} \left| \sum_A \tilde{s}_A(f)F_I^A(\hat{\omega}) \sum_{A'}\tilde{s}_{A'}(f)F_J^{A'}(\hat{\omega}) \right|.
\end{equation}
The estimated values are $\mu = \langle Y \rangle, h_A = \langle s_A \rangle, \hat{\Omega} = \langle\hat{\omega}\rangle$.
\begin{align}
\mu_{IJ}(f) &= Y_{IJ}(f) + n_{IJ}(f) \notag \\
&= T^{3/2} \left| \sum_A \tilde{h}_A(f)F_I^A(\hat{\Omega}) \sum_{A'}\tilde{h}_{A'}(f)F_J^{A'}(\hat{\Omega}) \right|,
\end{align}
where the noise $n_{IJ}(f)$ satisfies:
\begin{subequations}
\begin{align}
\mathbb{E}[n_{IJ}(f)] &= 0 \textit{ , }\\
\mathbb{V}[n_{IJ}(f)] &= \frac{T}{4}P_I(f)P_J(f) \eqcolon \mathcal{N}_{IJ}(f)
\end{align}
\end{subequations}
Our likelihood function is given by:
\begin{equation}
L(\mu_{IJ}|\vec{\theta}) = \exp\left[ -\sum_{(I,J)} \frac{(Y_{IJ} - \mu_{IJ})^2}{2\mathcal{N}_{IJ}} \right],
\end{equation}
with the parameters $\vec{\theta} = (\theta,\phi,+,\times,x,y,b,l)$.

The Fisher information matrix can then be calculated as follows:
\begin{equation}
F_{ij} = \mathbb{E}\left[\left( \partial_{\theta_i}\ln L(\mu_{IJ}|\vec{\theta}) \right)\left( \partial_{\theta_j}\ln L(\mu_{IJ}|\vec{\theta}) 
\right)\right],
\end{equation}
\begin{equation}
\textbf{F} = \begin{pmatrix}
F_{\theta\theta} & F_{\theta\phi} & F_{\theta A'} \\
F_{\phi\theta} & F_{\phi\phi} & F_{\phi A'} \\
F_{A\theta} & F_{A\phi} & F_{AA'}
\end{pmatrix}.
\end{equation}
\\
To simplify the notation, we define: $\alpha' \coloneq \frac{2\pi f}{c}$. We now calculate the $\theta\theta$- and the $\theta A$-components for a 
GW with polarization $A_0$:
\begin{widetext}
\begin{align}
\left.F_{\theta\theta}\right|_{h=h_{A_0}} &= \left.\mathbb{E}\left[\left( \partial_{\theta}\ln L \right)^2\right]\right|_{h=h_{A_0}} = 
\mathbb{E}\left[\left( \sum_{(I,J)}\frac{1}{\mathcal{N}_{IJ}}(Y_{IJ} - \mu_{IJ}) T^3 |\tilde{h}_{A_0}|^2 \partial_{\theta} F^{A_0}_I F^{A_0}_J 
\right)^2 \right] \notag \\
&= \sum_{(I,J)}\frac{1}{\mathcal{N}_{IJ}^2} \underset{= \mathcal{N}_{IJ}}{\underbrace{\mathbb{E}\left[(Y_{IJ} - \mu_{IJ})^2\right]}} T^3\left( 
|\tilde{h}_{A_0}|^2 \partial_{\theta} F^{A_0}_I F^{A_0}_J \right)^2 \notag \\
&\textit{ $ $ $ $ } + \sum_{(I,J) \neq (I',J')}\frac{1}{\mathcal{N}_{IJ}\mathcal{N}_{I'J'}} \underset{= 0}{\underbrace{\mathbb{E}\left[(Y_{IJ} - 
\mu_{IJ})(Y_{I'J'} - \mu_{I'J'})\right]}} T^3\left( |\tilde{h}_{A_0}|^2 \partial_{\theta} F^{A_0}_I F^{A_0}_J \right)\left(  |\tilde{h}_{A_0}|^2 
\partial_{\theta} F^{A_0}_{I'} F^{A_0}_{J'} \right) \notag \\
&= \sum_{(I,J)}\frac{T^3}{\mathcal{N}_{IJ}} \left( |\tilde{h}_{A_0}|^2 \partial_{\theta} F^{A_0}_I F^{A_0}_J \right)^2,
\end{align}
\begin{align}
\left.F_{\theta A}\right|&_{h=h_{A_0}} = \left.\mathbb{E}\left[\left( \partial_{\theta}\ln L \right)\left( \partial_{|\tilde{h}_A|^2}\ln L 
\right)\right]\right|_{h=h_{A_0}} \notag \\
&= \mathbb{E}\left[ \left( \sum_{(I,J)}\frac{1}{\mathcal{N}_{IJ}}(Y_{IJ} - \mu_{IJ}) T^{\frac{3}{2}} |\tilde{h}_{A_0}|^2 \partial_{\theta} 
F^{A_0}_I F^{A_0}_J \right)\left.\left( \sum_{(I,J)}\frac{1}{\mathcal{N}_{IJ}}(Y_{IJ} - \mu_{IJ}) 
T^{\frac{3}{2}}\partial_{|h_A|^2}|\tilde{h}_I\tilde{h}_J| \right)\right|_{h=h_{A_0}} \right] \notag \\
&= \sum_{(I,J)}\frac{T^3}{\mathcal{N}_{IJ}} \left( F_I^AF_J^A + (1-\delta_{AA_0})\frac{1}{2}\left[ \frac{(F_I^A)^2F_J^{A_0}}{F_I^{A_0}} + 
\frac{(F_J^A)^2F_I{A_0}}{F_J^{A_0}} \right]\right) |\tilde{h}_{A_0}|^2 \partial_{\theta} F^{A_0}_I F^{A_0}_J.
\end{align}
\end{widetext}
A detailed calculation of the matrix elements for the more general case, where we have different integration times for different detectors, can be 
found in Appendix~\ref{ET_appendix:fishermatrix}. We list here the rest of the components again for an $A_0$ polarized wave:
\begin{widetext}
\begin{align}
F_{\phi\phi} = \sum_{(I,J)}\frac{T^3}{\mathcal{N}_{IJ}} &\left( |\tilde{h}_{A_0}|^2 \partial_{\phi} F^{A_0}_I F^{A_0}_J \right)^2, \notag \\
F_{\theta\phi} = \sum_{(I,J)}\frac{T^3}{\mathcal{N}_{IJ}} &(|\tilde{h}_{A_0}|^2)^2 \left( \partial_{\theta} F^{A_0}_I F^{A_0}_J \right)\left( 
\partial_{\phi} F^{A_0}_I F^{A_0}_J \right), \notag \\
F_{\phi A} = \sum_{(I,J)}\frac{T^3}{\mathcal{N}_{IJ}} &\left( F_I^AF_J^A + (1-\delta_{AA_0})\frac{1}{2}\left[ \frac{(F_I^A)^2F_J^{A_0}}{F_I^{A_0}} 
+ \frac{(F_J^A)^2F_I^{A_0}}{F_J^{A_0}} \right]\right) |\tilde{h}_{A_0}|^2 \partial_{\phi} F^{A_0}_I F^{A_0}_J, \notag \\
F_{AA'} = \sum_{(I,J)}\frac{T^3}{\mathcal{N}_{IJ}} &\left( F_I^AF_J^A + (1-\delta_{AA_0})\frac{1}{2}\left[ \frac{(F_I^A)^2F_J^{A_0}}{F_I^{A_0}} + 
\frac{(F_J^A)^2F_I^{A_0}}{F_J^{A_0}} \right]\right) \notag \\
\cdot &\left( F_I^{A'}F_J^{A'} + (1-\delta_{A'A_0})\frac{1}{2}\left[ \frac{(F_I^{A'})^2F_J^{A_0}}{F_I^{A_0}} + 
\frac{(F_J^{A'})^2F_I^{A_0}}{F_J^{A_0}} \right]\right).
\end{align}
\end{widetext}
The inverse of the Fisher matrix is the covariance matrix, which has the variance of $\theta_i$ in the $i$-th diagonal entry. So, the square of 
the $SNR$ for measuring polarization $A$ is given by:
\begin{align}
SNR_{A}^2 &\coloneq \left.\frac{(|\tilde{h}_A|^2)^2}{\sigma_A^2}\right|_{h=h_A} = 
\left.\frac{(|\tilde{h}_A|^2)^2}{(\textbf{F}^{-1})_{AA}}\right|_{h=h_A} \notag \\
&= \left.\frac{(|\tilde{h}_A|^2)^2 \det\textbf{F}}{\mathcal{F}_A}\right|_{h=h_A},
\end{align}
where $\mathcal{F}_{\theta_i}$ is the determinant of the minor one gets from removing the $i$-th row and column from the Fisher matrix 
$\mathbf{F}$. The $SNR$ of the cross correlation is related to the one of amplitude by:
\begin{equation}
SNR[\mu]=SNR[h^2]=SNR^2[h].
\end{equation}
Again, demanding an $SNR[h]$ of at least 8 gives us the minimal amplitude. We can read off the prefactors by comparing with the result for one 
detector pair above:
\begin{equation}\label{Eq:h_min(Omega)}
|\tilde{h}_A|_{min} = \sqrt{\frac{32}{T}}\sqrt[4]{\left|\frac{\mathcal{F}_A}{\det\mathbf{F}}\right|_{h=h_A}}.
\end{equation}

The variance of the position in the sky is given by:
\begin{subequations}
\begin{align}
\label{Eq:V(Omega)1}
\mathbb{V}[\theta] &= (\mathbf{F}^{-1})_{\theta\theta} = \frac{\mathcal{F}_\theta}{\det\mathbf{F}},\\
\label{Eq:V(Omega)2}
\mathbb{V}[\phi] &= (\mathbf{F}^{-1})_{\phi\phi} = \frac{\mathcal{F}_\phi}{\det\mathbf{F}}.
\end{align}
\end{subequations}

It turns out that the angular pattern functions of the breathing and the longitudinal modes are proportional to each other: $F_I^l = 
-\sqrt{2}F_I^b$. Therefore it is impossible to distinguish these two with laser interferometry, and we thus focus on the distinction between 
the 4 tensor and vector polarizations and the scalar mode. From now on we use the polarization $A$ as:
\begin{equation}
A \in \lbrace +, \times, x, y, S \rbrace.
\end{equation}

\begin{figure}[h!]
\includegraphics[width=1.0\linewidth]{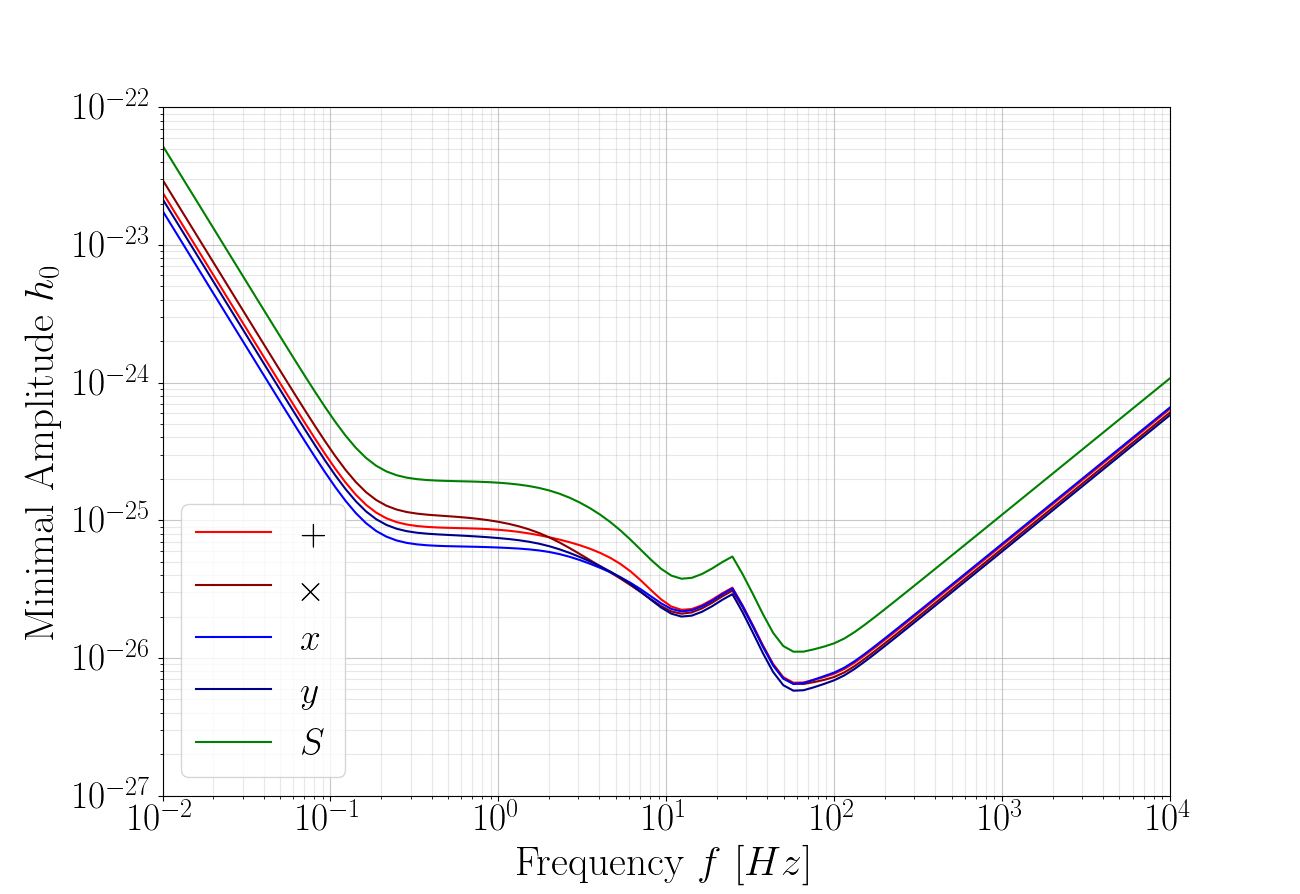}
\caption{\label{ET_fig:h(f)DECIGO} Frequency dependent sensitivity of ET, LIGO and DECIGO towards the polarizations $A \in \lbrace +, \times, x, 
y, S \rbrace$.}
\end{figure}

\begin{figure*}
\includegraphics[width=0.49\linewidth]{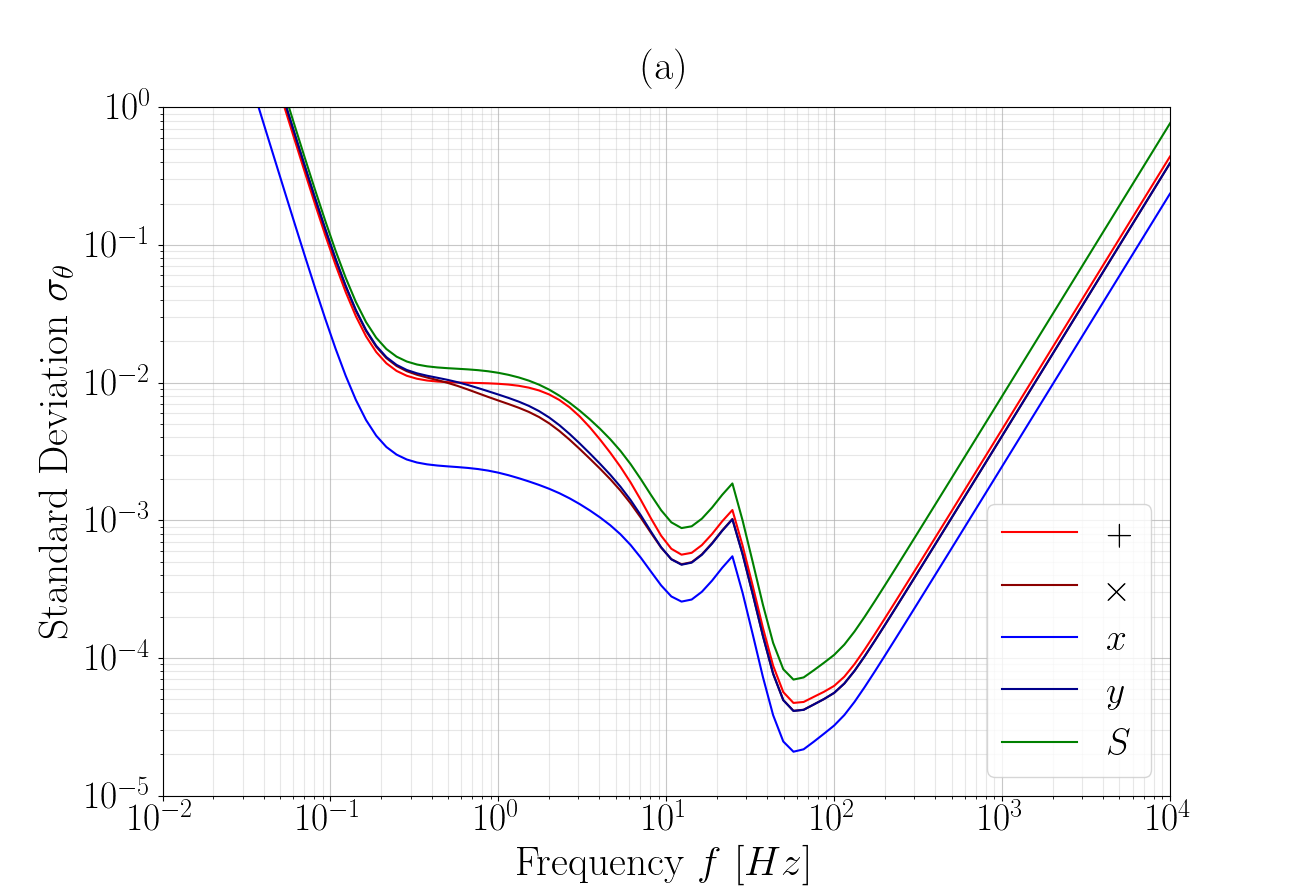}
\includegraphics[width=0.49\linewidth]{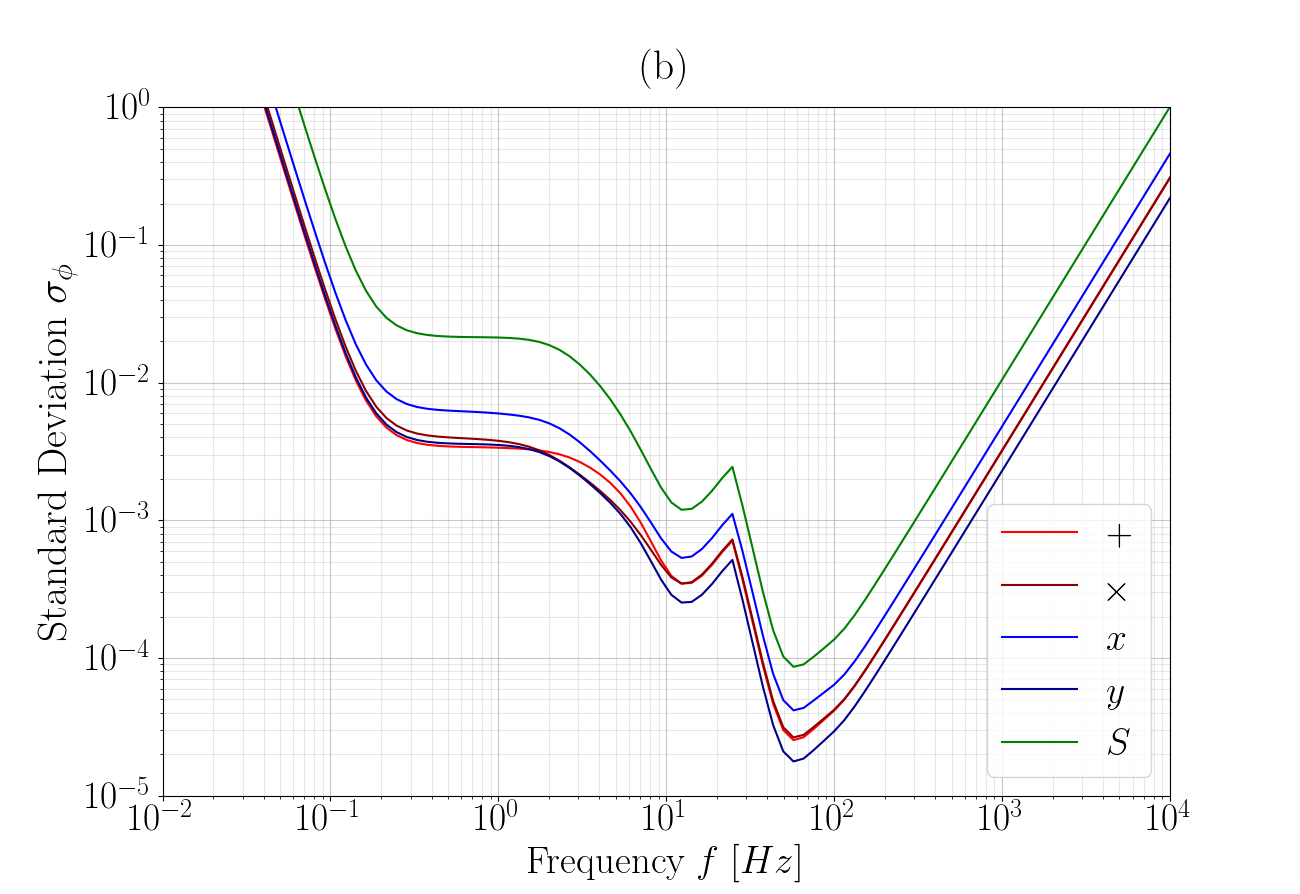}\\
\includegraphics[width=0.49\linewidth]{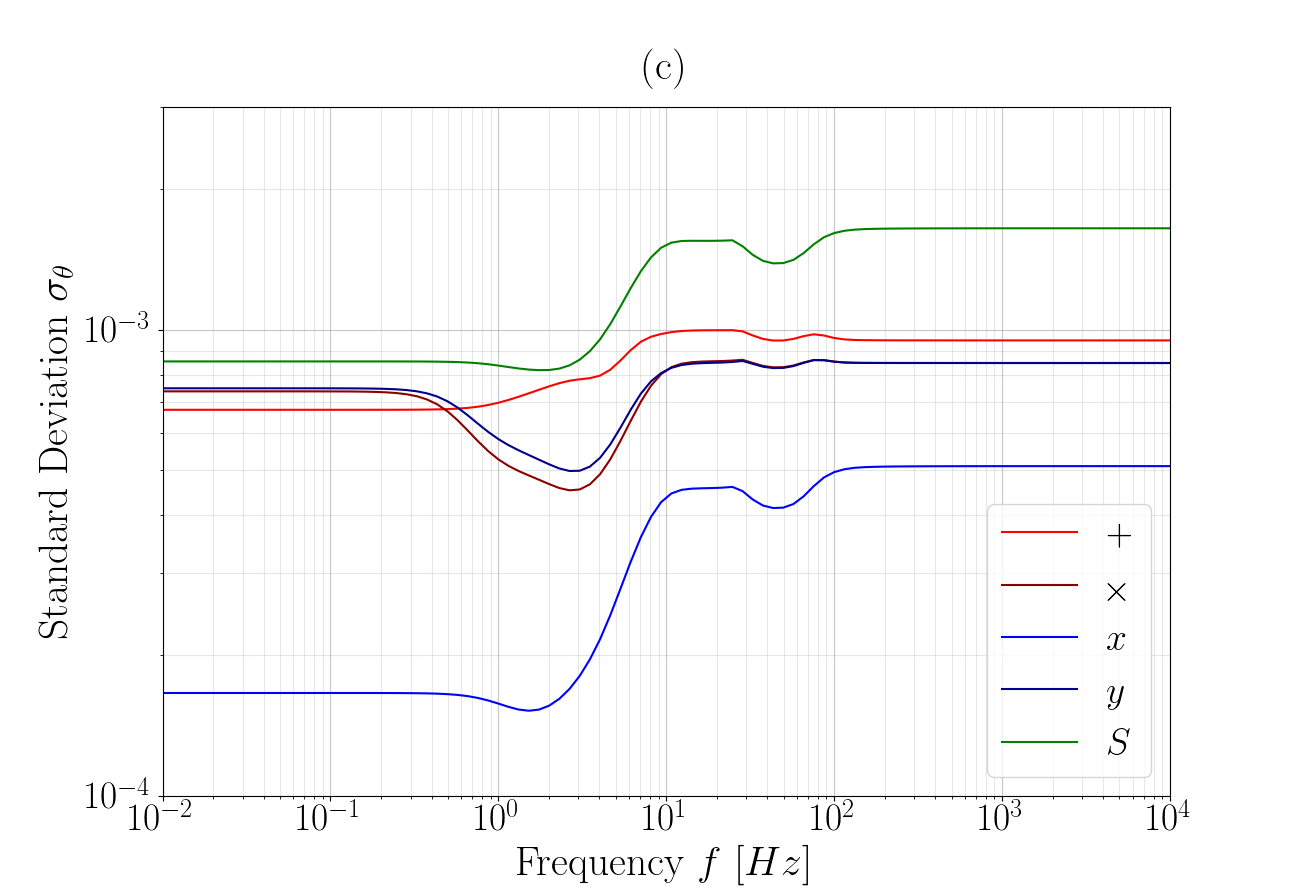}
\includegraphics[width=0.49\linewidth]{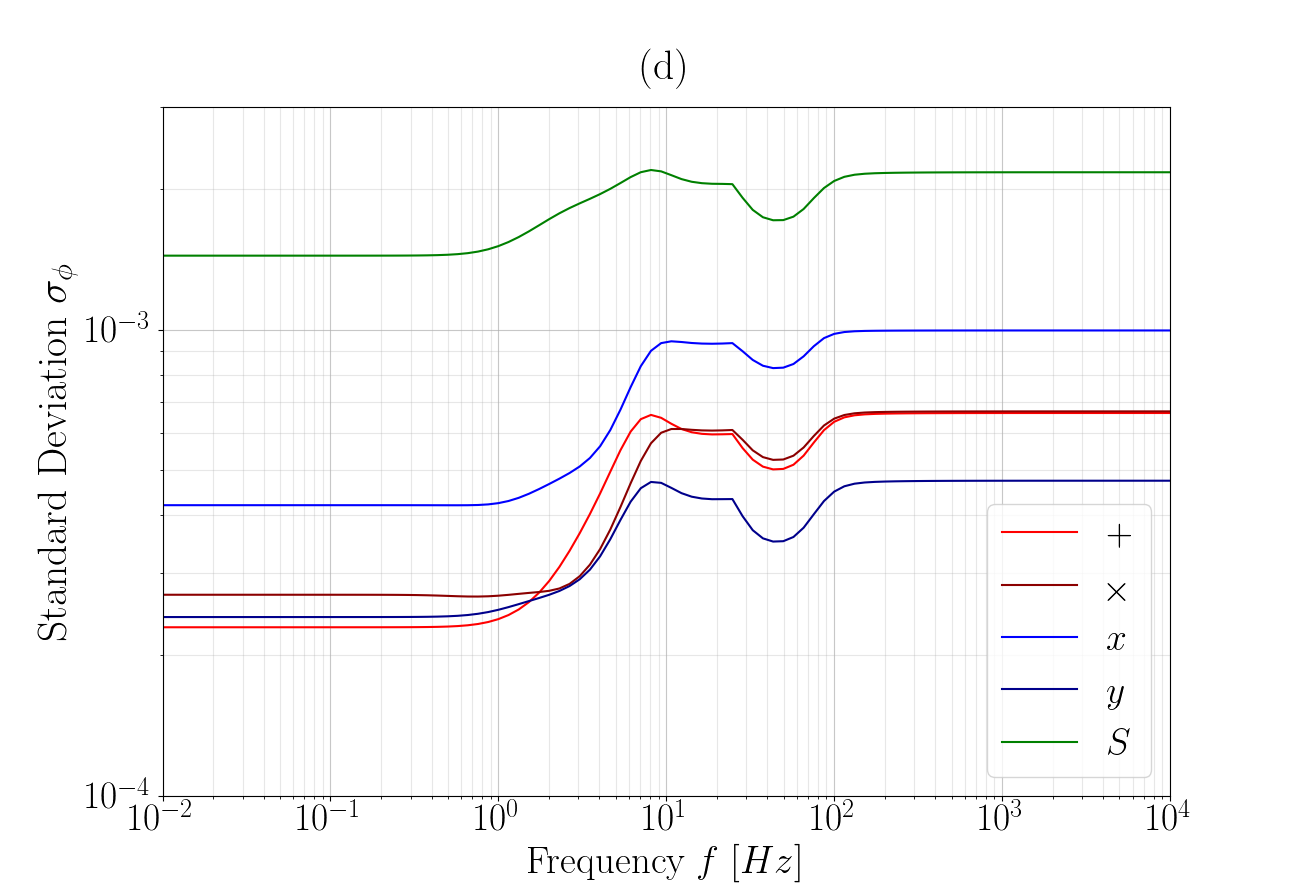}
\caption{\label{ET_fig:o_tp(f)DECIGO} Frequency dependent standard deviation of $\theta$ and $\phi$ for (a)-(b): a GW with amplitude $h = 
10^{-25}$ and (c)-(d): for twice the minimal amplitude of the scalar mode $h = 2|h_S|_{min}$, see Fig.~\ref{ET_fig:h(f)DECIGO}.}
\includegraphics[width=0.7\linewidth]{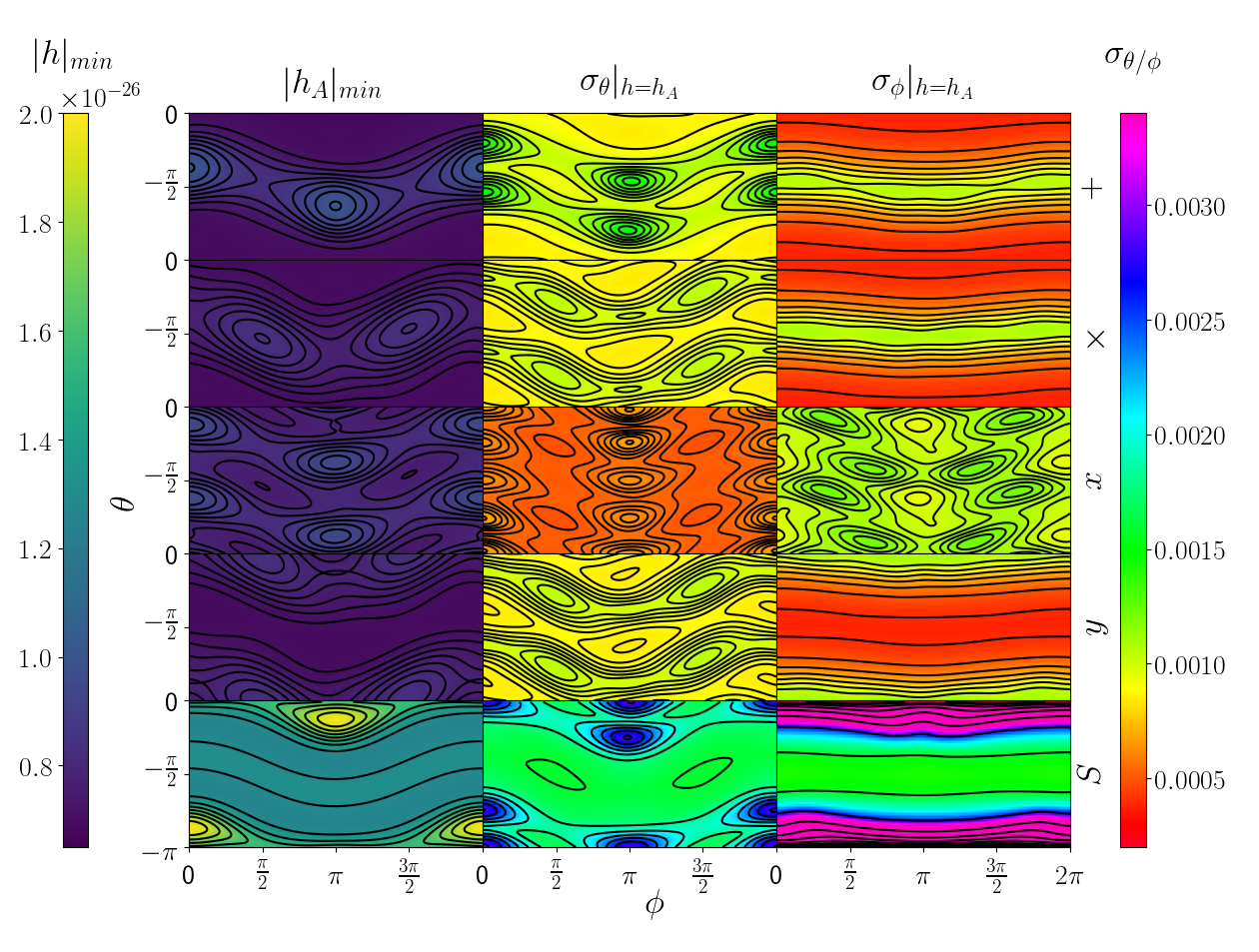}
\caption{\label{ET_fig:Multi-Image} Sensitivity of ET, LIGO and DECIGO at \SI{100}{Hz} towards the polarizations $A \in \lbrace +, \times, x, y, S 
\rbrace$ in the left column. Standard deviation of the $\theta$ and $\phi$ angle for a GW with polarization $A$ and amplitude of $h_A = 
2.6\cdot10^{-26}$ in the middle and right column respectively.}
\end{figure*}

\begin{figure*}
\includegraphics[width=0.7\linewidth]{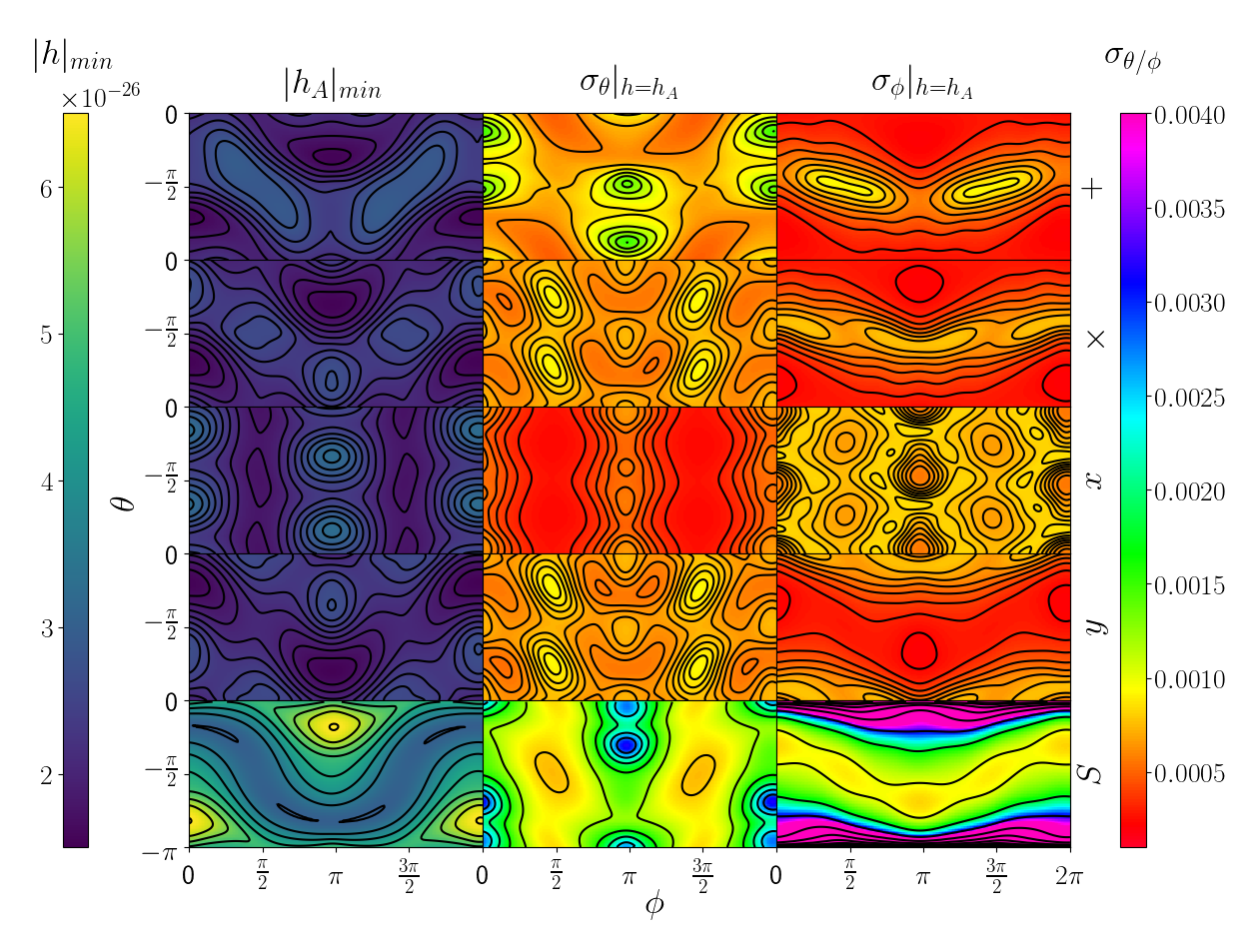}
\caption{\label{ET_fig:Multi-Image10Hz} Sensitivity of ET, LIGO and DECIGO at \SI{10}{Hz} towards the polarizations $A \in \lbrace +, \times, x, 
y, S \rbrace$ in the left column. Standard deviation of the $\theta$ and $\phi$ angle for a GW with polarization $A$ and amplitude of $h_A = 
8.9\cdot10^{-26}$ in the middle and right column respectively.}
\includegraphics[width=0.7\linewidth]{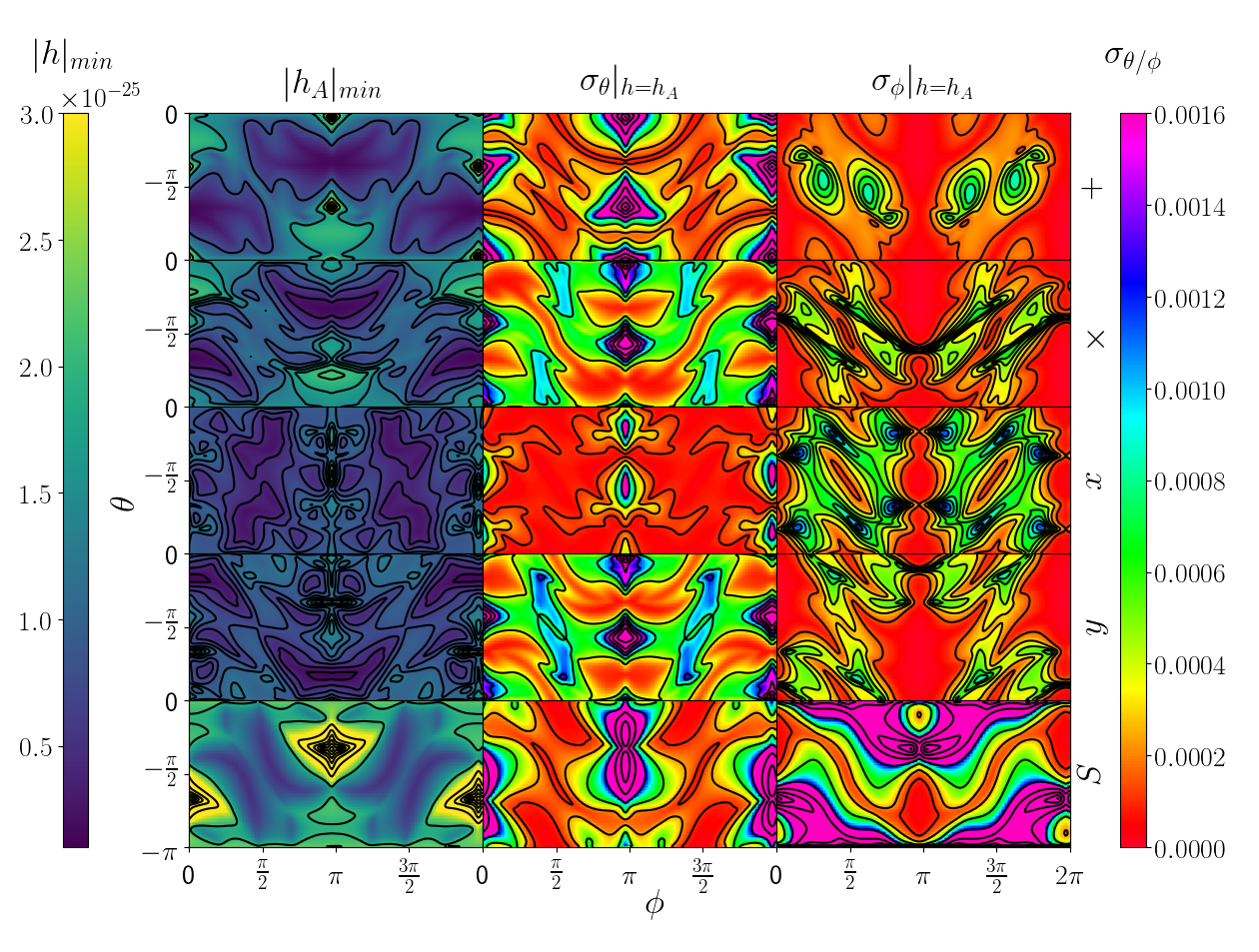}
\caption{\label{ET_fig:Multi-Image1Hz} Sensitivity of ET, LIGO and DECIGO at \SI{1}{Hz} towards the polarizations $A \in \lbrace +, \times, x, y, 
S \rbrace$ in the left column. Standard deviation of the $\theta$ and $\phi$ angle for a GW with polarization $A$ and amplitude of $h_A = 
3.8\cdot10^{-25}$ in the middle and right column respectively.}
\end{figure*}

In our calculations, we only use one cluster out of all the DECIGO clusters, namely the one closest to Earth ($\phi = -20^\circ$ from the Earth 
position, on its orbit around the Sun). We keep including ET and the already existing LIGO detectors. We adopt the HEALPix pixelization scheme to 
evenly distribute $n$ points on the sky (we used $n=48$ and $N_\text{side}=2$ to generate Fig.~\ref{ET_fig:h(f)DECIGO}-\ref{ET_fig:Multi-Image1Hz}; more detailed explanations about the HEALPix scheme can be found in \cite{HEALPix})
and then average over the $h_{min}$ values for each $f$ and obtain the frequency dependant behaviour of 
the average sensitivity of ET, LIGO and DECIGO in Fig.~\ref{ET_fig:h(f)DECIGO}.

There are two aspects of the frequency dependent standard deviation of $\theta$ and $\phi$. One is that if one measures a signal with a certain 
amplitude, then we can measure the position of the source more precisely if it emits GW in frequencies in which we are more sensitive. We plot 
this in Fig.~\ref{ET_fig:o_tp(f)DECIGO}.
The other aspect is that the standard deviations vary with the frequency, relative to the sensitivity at 
that frequency. This means that if we consider for instance a wave which has twice the minimal amplitude for each frequency, we still get 
frequency dependence. In Fig.~\ref{ET_fig:o_tp(f)DECIGO} (below) we take $h = 2|h_S|_{min}(f)$ since it is always higher than the other 
polarizations and we can therefore detect it, no matter which polarization we choose and for any fixed frequency we have the same amplitude for 
all polarizations. This allows us to compare the polarizations with each other.

In Fig.\ref{ET_fig:Multi-Image}, we give the direction dependent sensitivities using Eq.~\eqref{Eq:h_min(Omega)}, along with the angular 
resolution for waves with these polarizations by taking the square root of Eqs.\eqref{Eq:V(Omega)1}-\eqref{Eq:V(Omega)2} at a frequency of 
\SI{100}{Hz}, where our set of detectors is most sensitive. We do a similar procedure at \SI{10}{Hz} and \SI{1}{Hz}, and the results are given in 
Fig.~\ref{ET_fig:Multi-Image10Hz} and \ref{ET_fig:Multi-Image1Hz}. Note that the purple zones correspond to true poles.

\section{Conclusion}
The Einstein Telescope alone cannot be used to distinguish between GW polarization modes, and small changes in its geometry lead to no almost no 
difference. However, by combining ET with second generation detectors such as LIGO, VIRGO and KAGRA, one can detect a GW background with a strain 
amplitude down to $10^{-26}$ and distinguish its polarization modes around a frequency of \SI{100}{Hz}. One can enhance the sensitivity for lower 
frequencies by cross correlating the network with the DECIGO detector, especially in its C3 configuration. In that case, the observation 
window is enhanced and allow measurement of strains below $10^{-23}$, down to $10^{-26}$, in a frequency band from \SI{0.1}{Hz} to \SI{10}{kHz}.\\

It is possible to use an alternative method to distinguish the modes by using the time dependence of the signals. Using the right detector pairs, 
one can then clearly distinguish the scalar and vector modes by the blind spots. The effect is quite significant over a time period of one 
year for a ground-space network of detectors involving DECIGO, ET and LIGO. In the case of a scaled-down B-DECIGO detector, the time 
dependence is however very chaotic, due to the angular frequency around Earth which is then an irrational fraction of Earth rotation, and the 
distance to Earth detectors, which is varying significantly. The sensitivities of the different modes are too close together, whenever the 
detectors are too close to Earth. This can be resolved however if one chooses an orbit on a higher altitude such that it circles the Earth once a 
day. In that case the method becomes more complicated than with the original DECIGO, but is still feasible.

We have thus showed that second and third generation detectors, combined with space-detectors, can provide two different methods to test GR or 
contrain alternative theories by measuring the polarization of a GW background.

In a future project, one could investigate the possibilities of detecting inhomogeneities in the GW background, analogue to the 
ones in the cosmic microwave background. Up to now we only calculated the minimal strain of a GW to be detected, but to find out how large the 
deviation from the mean would have to be to detect them, we would have to deal with the variance of the parameter estimation. It would be 
interesting to find out what angular resolution one could get with various detector combinations.

Gravitational waves should travel undisturbed since Planck time, which would make it possible to measure properties of the early quantum 
gravitational universe directly. This could give us valuable hints on the search of a unifying theory. This advantage of GW over the 
electromagnetic ones also has its disadvantages. Due to the enormous density of the early universe, many emissions of GW would be expected from 
different epochs after the Big Bang, and the difficulty would be to distinguish a signal of an earlier epoch from a later one. Overcoming that 
problem could however allow to establish a complete gravitational map of the beginning of our universe.

\begin{acknowledgments}
L.P. was supported by the UZH research grant FK-17-097. P.J. thanks the Swiss National Foundation for support. We also thank the referee for the 
useful comments.
\end{acknowledgments}

\appendix

\section{Symmetry of ET}

\subsection{ET Perturbations}\label{ET_appendix:ET_perturbation}
We have seen in section~\ref{ET_section:ET} that one cannot distinguish between the three polarization modes with the ET alone, although we 
have three signals, because of the symmetric arrangement of the three interferometers composing ET. One has thus to break the symmetry in order to 
make the three rows in the detector correlation matrix independent. We are going to consider two ways of doing that perturbatively and use the 
framework of the previous section to determine their impact on the sensitivity, which will allow us to compare both methods.

\subsubsection{Irregular Triangle}
We make one opening angle smaller by a small angle $\epsilon_\phi$ and make another angle bigger by the same amount, which leaves the third angle 
unchanged. We now have a completely irregular triangle with three different angles and therefore the ORFs of all three detector pairs become 
different, and the detector correlation matrix becomes invertible.

Changing the angles will also change the arm lengths and therefore the distance between a detector pair. We use the sine-law to determine the 
impact of a change in the angles on the change in the distance $\delta$. To estimate the order of magnitude of effect of the perturbation on the 
detector correlation matrix, we calculate the change in the ORF of the detector pair $(1,2)$, when we shrink the angle $\phi_3$ and enlarge 
$\phi_1$ by $\epsilon_\phi$, which leaves $\phi_2$ unchanged but shortens $d_{12}$:
\begin{equation}
\frac{d-\delta}{\sin(\phi_3)} = \frac{d}{\sin(\phi_1)} \textit{ , $ $ } \phi_3 = \phi-\epsilon_\phi \textit{ , $ $ } \phi_1 = \phi + \epsilon_\phi
\end{equation}
for $\phi=\pi/3$. To first order we get:
\begin{equation}
\delta \approx \sqrt{3}d\epsilon_\phi.
\end{equation}
With that expression, we can relate the effects of a change in the distance to the change in the angles:
\begin{align}
\rho_i^M &\mapsto (\rho_i^M)^{(0)} - \sqrt{3}\alpha(\rho_i^M)'\epsilon_\phi, \\
\cos\beta &\mapsto \cos\beta^{(0)} + \sqrt{3}\frac{d^2}{R_E^2}\epsilon_\phi.
\end{align}
The only coordinate angle that changes is $\sigma_{1+}$. Therefore, we insert the values of the other angles:
\begin{align}
&\sigma_{1-} = 0 \ , \quad \sigma_{1+} \mapsto \sigma_{1+}^{(0)} + \epsilon_\phi \ , \quad \sigma_{1+}^{(0)} = 
\frac{\pi}{3},\\ \label{Eq:ET_angles}
&\sigma_{2-} = \frac{2\pi}{3} \ , \quad \sigma_{2+} = \pi;
\end{align}
in Eq.~\eqref{Eq:ORFIJ} to get the new ORF:
\begin{widetext}
\begin{align}
\gamma_{12}^M &= \frac{1}{16}\left\lbrace -3\rho_1^M\sin^2\sigma_{1+} + \frac{\sqrt{3}}{2}\left[ 2\rho_1^M\cos\beta + 
\rho_2^M\frac{1+\cos\beta}{2} \right]\sin(2\sigma_{1+}) \right. \notag \\
&\left.\textit{ $ $ }+ \frac{3}{4}\left[ 4\rho_1^M\cos^2\beta + 2\rho_2^M(1+\cos\beta)\cos\beta + \rho_3^M(1+\cos\beta)^2 
\right](\cos^2\sigma_{1+}-1) \right\rbrace \notag \\
&\mapsto \frac{1}{16}\left\lbrace -3\left((\rho_1^M)^{(0)} - \sqrt{3}\alpha(\rho_1^M)'\epsilon_\phi\right)\left(\sin^2\sigma_{1+}^{(0)} + 
2\sin\sigma_{1+}^{(0)}\cos\sigma_{1+}^{(0)}\epsilon_\phi\right) \right. \notag \\
&\textit{ $ $ $ $ }+ \frac{\sqrt{3}}{2}\left[ 2\left((\rho_1^M)^{(0)} - \sqrt{3}\alpha(\rho_1^M)'\epsilon_\phi\right)\left(\cos\beta^{(0)} + 
\sqrt{3}\frac{d^2}{R_E^2}\epsilon_\phi\right) \right. \notag \\
&\textit{ $ $ $ $ }\left.+ \frac{1}{2}\left((\rho_2^M)^{(0)} - \sqrt{3}\alpha(\rho_2^M)'\epsilon_\phi\right)\left(1+\cos\beta^{(0)} + 
\sqrt{3}\frac{d^2}{R_E^2}\epsilon_\phi\right) \right]\left(\sin(2\sigma_{1+}^{(0)}) + 2\cos(2\sigma_{1+}^{(0)})\epsilon_\phi\right) \notag \\
&\textit{ $ $ $ $ }+ \frac{3}{4}\left[ 4\left((\rho_1^M)^{(0)} - \sqrt{3}\alpha(\rho_1^M)'\epsilon_\phi\right)\left(\cos\beta^{(0)} + 
\sqrt{3}\frac{d^2}{R_E^2}\epsilon_\phi\right)^2 \right. \notag \\
&\textit{ $ $ $ $ }+ 2\left((\rho_2^M)^{(0)} - \sqrt{3}\alpha(\rho_2^M)'\epsilon_\phi\right)\left(1 + \cos\beta^{(0)} + 
\sqrt{3}\frac{d^2}{R_E^2}\epsilon_\phi\right)\left(\cos\beta^{(0)} + \sqrt{3}\frac{d^2}{R_E^2}\epsilon_\phi\right) \notag \\
&\textit{ $ $ $ $ }\left.\left.+ \left((\rho_3^M)^{(0)} - \sqrt{3}\alpha(\rho_3^M)'\epsilon_\phi\right)\left(1 + \cos\beta^{(0)} + 
\sqrt{3}\frac{d^2}{R_E^2}\epsilon_\phi\right)^2\right]\left(\cos^2\sigma_{1+}^{(0)} - 2\cos\sigma_{1+}^{(0)}\sin\sigma_{1+}^{(0)}\epsilon_\phi - 
1\right)\right\rbrace.
\end{align}
\end{widetext}
We can simplify this expression, by plugging in the values for $\sigma_{1+}^{(0)}$ and approximating $\cos\beta^{(0)}$, using the fact that 
$\frac{d^2}{R_E^2} = 2.5 \cdot 10^{-6} \ll 1$:
\begin{equation}
\cos\beta^{(0)} = 1 - \frac{d^2}{2R_E^2} \approx 1.
\end{equation}
To first order in $\epsilon_\phi$ we get:
\begin{widetext}
\begin{align}
\epsilon_M \approx \frac{1}{16}&\left\lbrace -\frac{3\sqrt{3}}{2}\left[ (\rho_1^M)^{(0)} - \frac{3}{2}\alpha(\rho_1^M)' \right] - 
\frac{\sqrt{3}}{2}\left[ 2(\rho_1^M)^{(0)} + 3\alpha(\rho_1^M)' + (\rho_2^M)^{(0)} + \frac{3}{2}\alpha(\rho_2^M)' \right]\right. \notag \\
&\textit{ $ $ }\left.- \frac{3\sqrt{3}}{4}\left[ 2\left((\rho_1^M)^{(0)}+(\rho_2^M)^{(0)}+(\rho_3^M)^{(0)}\right) - 
3\alpha\left((\rho_1^M)'+(\rho_2^M)'+(\rho_3^M)'\right) \right] \right\rbrace\epsilon_\phi \notag \\
\approx -\frac{\sqrt{3}}{32}&\left\lbrace 8(\rho_1^M)^{(0)} + 4(\rho_2^M)^{(0)} + 3(\rho_3^M)^{(0)} - 3\alpha\left( 2(\rho_1^M)' + (\rho_2^M)' + 
\frac{3}{2}(\rho_3^M)' \right) \right\rbrace\epsilon_\phi.
\end{align}
\end{widetext}

\begin{figure}[h!]
\includegraphics[width=1.0\linewidth]{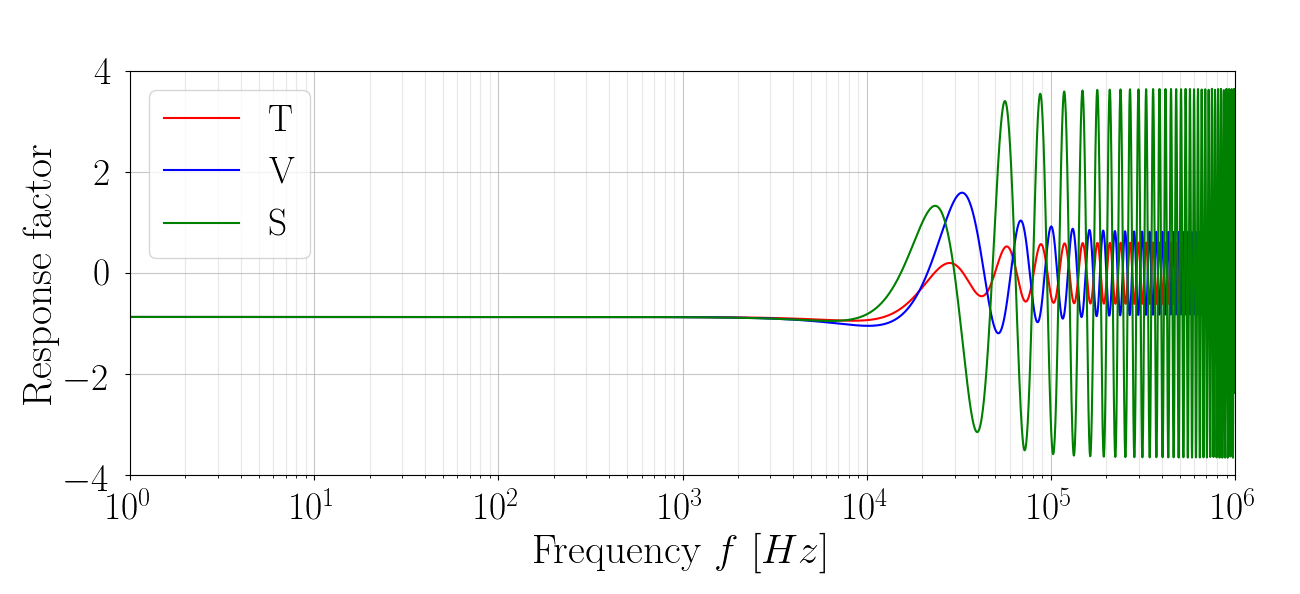}
\caption{The factor with which the ORF responds to a small change in the detector angles.}
\label{IrrTplot}
\end{figure}

In Fig.~\ref{IrrTplot} we plot the response factor, which multiplies to $\epsilon_\phi$ to get the change in the ORF 
$\epsilon_M$.

The response factor stays almost constant at a value of -0.87, since $\alpha \ll 1$ until we get close to the critical frequency $f_{crit} =$ 
\SI{3e4}{Hz}, defined over $\frac{f_{crit}d}{c} \coloneq 1$.

\subsubsection{Tilted Detector Planes}
Now we leave the angles and the arm-lengths of the three Michelson-interferometers invariant but tilt the plane in which one of the three 
detectors lies. We tilt the plane of detector 1, such that $\hat{u}_1$ gets tilted in negative $z$-direction. The other detector arms stay 
unchanged ($\hat{u}_J = \hat{u}_J^{(0)}$ for $J \neq 1$ and $\hat{v}_J = \hat{v}_J^{(0)}$ for all $J$), and we can write the perturbation as:
\begin{equation}
\hat{u}_1 \mapsto \hat{u}_1^{(0)} + \delta\hat{u}_1 \textit{ , $ $ } \delta\hat{u}_1 = \begin{pmatrix}
0 \\ 0 \\ -\delta u
\end{pmatrix}.
\end{equation}
The angle $\alpha$ by which $\hat{u}_1$ is rotated can be approximated by:
\begin{equation}
\alpha \approx \sin\alpha \approx \delta u.
\end{equation}
\\
We calculate the contractions of the perturbed detector tensors, analogue to Eqs.~\eqref{Eq:DD}-\eqref{Eq:DDdddd} to first order in $\delta u$:
\begin{widetext}
\begin{align}
D_1^{ij}D^2_{ij} &= \frac{1}{4}\left[ (\hat{u}_1^{(0)}\cdot\hat{u}_2 + \delta\hat{u}_1\cdot\hat{u}_2)^2 - (\hat{v}_1\cdot\hat{u}_2)^2 - 
(\hat{u}_1^{(0)}\cdot\hat{v}_2 + \delta\hat{u}_1\cdot\hat{v}_2)^2 + (\hat{v}_1\cdot\hat{v}_2)^2 \right] \notag \\
&\approx \frac{1}{4}(D_1^{ij}D^2_{ij})^{(0)} + \frac{1}{2}\left[ (\hat{u}_1^{(0)}\cdot\hat{u}_2)(\delta\hat{u}_1\cdot\hat{u}_2) - 
(\hat{u}_1^{(0)}\cdot\hat{v}_2)(\delta\hat{u}_1\cdot\hat{v}_2) \right]. \notag
\end{align}
% \end{widetext}
Using the angles for ET as in Eq.~\eqref{Eq:ET_angles} we get:
\begin{align}
\delta(D_1^{ij}D^2_{ij}) &= \frac{1}{2}\left[ (\cos\beta\cos\sigma_{1-}\cos\sigma_{2-} + \sin\sigma_{1-}\sin\sigma_{2-})\delta 
u\sin\beta\cos\sigma_{2-} \right. \notag \\
&\left.\textit{ $ $ $ $ } - (\cos\beta\cos\sigma_{1-}\cos\sigma_{2+} + \sin\sigma_{1-}\sin\sigma_{2+})\delta u\sin\beta\cos\sigma_{2+} \right] 
\notag \\
&= \frac{1}{2}\cos\beta\sin\beta(\cos^2\sigma_{2-}-1)\delta u = -\frac{3}{8}\cos\beta\sin\beta\delta u,\\
D^i_{1,k}D_2^{kj}\hat{d}_i\hat{d}_j &= \frac{1}{4}\left( (\hat{u}_1^{(0)}\cdot\hat{d} + \delta\hat{u}_1\cdot\hat{d})(\hat{u}_1^{(0)} + 
\delta\hat{u}_1) - (\hat{v}_1\cdot\hat{d})\hat{v}_1 \right)\cdot\left( (\hat{u}_2\cdot\hat{d})\hat{u}_2 - (\hat{v}_2\cdot\hat{d})\hat{v}_2 \right) 
\notag \\
&\approx (D^i_{1,k}D_2^{kj}\hat{d}_i\hat{d}_j)^{(0)} + \frac{1}{4}\left( (\delta\hat{u}_1\cdot\hat{d})\hat{u}_1^{(0)} + 
(\hat{u}_1^{(0)}\cdot\hat{d})\delta\hat{u}_1 \right)\cdot\left( (\hat{u}_2\cdot\hat{d})\hat{u}_2 - (\hat{v}_2\cdot\hat{d})\cdot\hat{v}_2 \right), 
\notag \\
\delta(D^i_{1,k} D_2^{kj}\hat{d}_i\hat{d}_j) &= \frac{1}{8(1-\cos\beta)}\left( \delta u(1-\cos\beta)\hat{u}_1^{(0)} + 
\cos\sigma_{1-}\sin\beta\delta\hat{u}_1 \right) \cdot\left( \sin\beta\cos\sigma_{2-}\hat{u}_2 - \sin\beta\cos\sigma_{2+}\hat{v}_2 \right) 
\notag \\
&= \frac{\sin\beta}{8}\left( \cos\sigma_{2-}\underset{\hat{u}_1^{(0)}\cdot\hat{u}_2}{\underbrace{\cos\beta\cos\sigma_{2-}}} - 
\underset{\hat{u}_1^{(0)}\cdot\hat{v}_2}{\underbrace{\cos\beta}} \right)\delta u + \frac{1+\cos\beta}{8}\left( 
\cos\sigma_{2-}\underset{\delta\hat{u}_1\cdot\hat{u}_2}{\underbrace{\delta u\sin\beta\cos\sigma_{2-}}} - 
\underset{\delta\hat{u}_1\cdot\hat{v}_2}{\underbrace{\delta u\sin\beta}} \right) \notag \\
&= \frac{\sin\beta}{8}\left[ \cos\beta(\cos^2\sigma_{2-}-1) + (1+\cos\beta)(\cos^2\sigma_{2-}-1) \right]\delta u \notag \\
&= -\frac{\sin\beta}{8}(1+2\cos\beta)\sin^2\sigma_{2-}\delta u = -\frac{3}{4}\frac{\sin\beta}{8}(1+2\cos\beta)\delta u, \\
D_1^{ij}D_2^{kl} \hat{d}_i\hat{d}_j\hat{d}_k\hat{d}_l &= \frac{1}{4}\left( (\hat{u}_1^{(0)}\cdot\hat{d} + \delta\hat{u}_1\cdot\hat{d})^2 - 
(\hat{v}_1\cdot\hat{d})^2 \right)\left( (\hat{u}_2\cdot\hat{d})^2 - (\hat{v}_2\cdot\hat{d})^2 \right) \notag \\
&= (D_1^{ij}D_2^{kl}\hat{d}_i\hat{d}_j\hat{d}_k\hat{d}_l)^{(0)} + \frac{1}{2}\left( (\hat{u}_1^{(0)}\cdot\hat{d})(\delta\hat{u}_1\cdot\hat{d}) 
\right)\left( (\hat{u}_2\cdot\hat{d})^2 - (\hat{v}_2\cdot\hat{d})^2 \right), \notag \\
\delta(D_1^{ij}D_2^{kl} \hat{d}_i\hat{d}_j\hat{d}_k\hat{d}_l) &= \frac{1}{8(1-\cos\beta)^2}\cos\sigma_{1-}\sin\beta\delta u(1-\cos\beta) \left( 
\sin^2\beta\cos^2\sigma_{2-} - \sin^2\beta\cos^2\sigma_{2+} \right) \notag \\
&= \frac{1+\cos\beta}{8}\sin\beta\left( \cos^2\sigma_{2-} - 1 \right)\delta u = -\frac{3}{4}\frac{1+\cos\beta}{8}\sin\beta\delta u.
\end{align}
\end{widetext}
Finally, we can patch all terms together in order to calculate the perturbation $\epsilon_M$:
\begin{widetext}
\begin{align}
\epsilon_M &= \rho_1^M\delta(D_1^{ij}D^2_{ij}) + \rho_2^M\delta(D^i_{1,k}D_2^{kj}\hat{d}_i\hat{d}_j)  + 
\rho_3^M\delta(D_1^{ij}D_2^{kl}\hat{d}_i\hat{d}_j\hat{d}_k\hat{d}_l) \notag \\
&= -\frac{3}{8}\sin\beta\left\lbrace \rho_1^M\cos\beta + \frac{1}{4}\rho_2^M(1+2\cos\beta) + \frac{1}{4}\rho_3^M(1+\cos\beta) \right\rbrace\delta 
u.
\end{align}
\end{widetext}

Again, we find that $\epsilon_M$ is almost independent of $f$, but the effect is three orders of magnitude smaller if we tilt one plane, instead 
of deforming the equilateral triangle.
\begin{equation}
\epsilon_M = 1.2 \cdot 10^{-3} \delta u
\end{equation}

\begin{figure}[h!]
\includegraphics[width=1.0\linewidth]{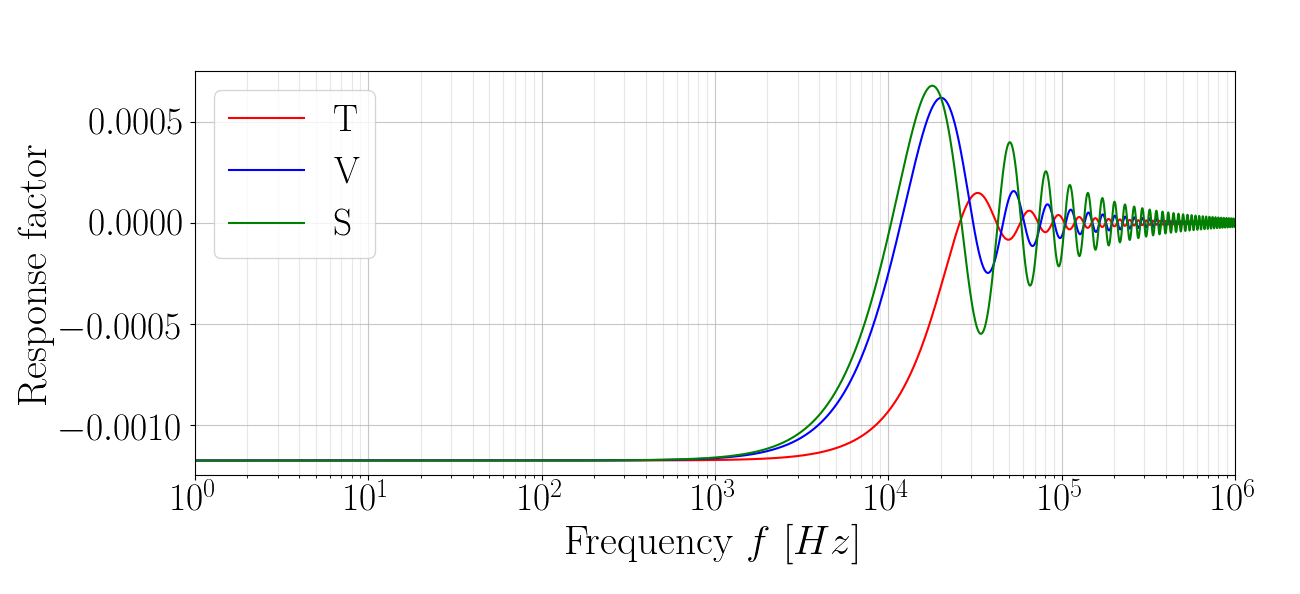}
\caption{Factor with which the ORF responds to a small tilt of one of the detector planes.}
\label{TPplot}
\end{figure}

The response factor for the tilted plane, shown in Fig.~\ref{TPplot}, stays at about -0.001 for frequencies far below $f_{crit}$ and oscillates 
ever closer around zero for increasing frequencies above \SI{10^5}{Hz}. Since ET is designed to measure in a frequency range from \SI{1.5}{Hz} to 
\SI{10}{kHz} the oscillations are not relevant. We find that the response to the same small change in the tilt angle is three orders of magnitude 
smaller than that of the change in the opening angle.\\
\\
The effect of a perturbation is at best as small as the angle by which we change ET's geometry, in the case of the irregular triangle. As we will 
argue in the next section, the problem is resolved if one adds additional detectors, for example LIGO, which exists already anyway, and changing 
the geometry of ET is therefore not worth the effort.

\section{B-DECIGO}\label{ET_appendix:B-DECIGO}
The scaled-down detector B-DECIGO \cite{BDECIGO2018} orbits around the Earth on an altitude of \SI{2000}{km} which is on the same order of 
magnitude as the radius of the Earth (\SI{6371}{km}). If we replace DECIGO by this smalle version, we can see in Fig.~\ref{ET_fig:ET,LIGO,B-DECIGO} 
that the sensitivity gets worse below \SI{10}{Hz} for all polarizations, as compared to DECIGO.
\begin{figure}[h!]
\includegraphics[width=0.9\linewidth]{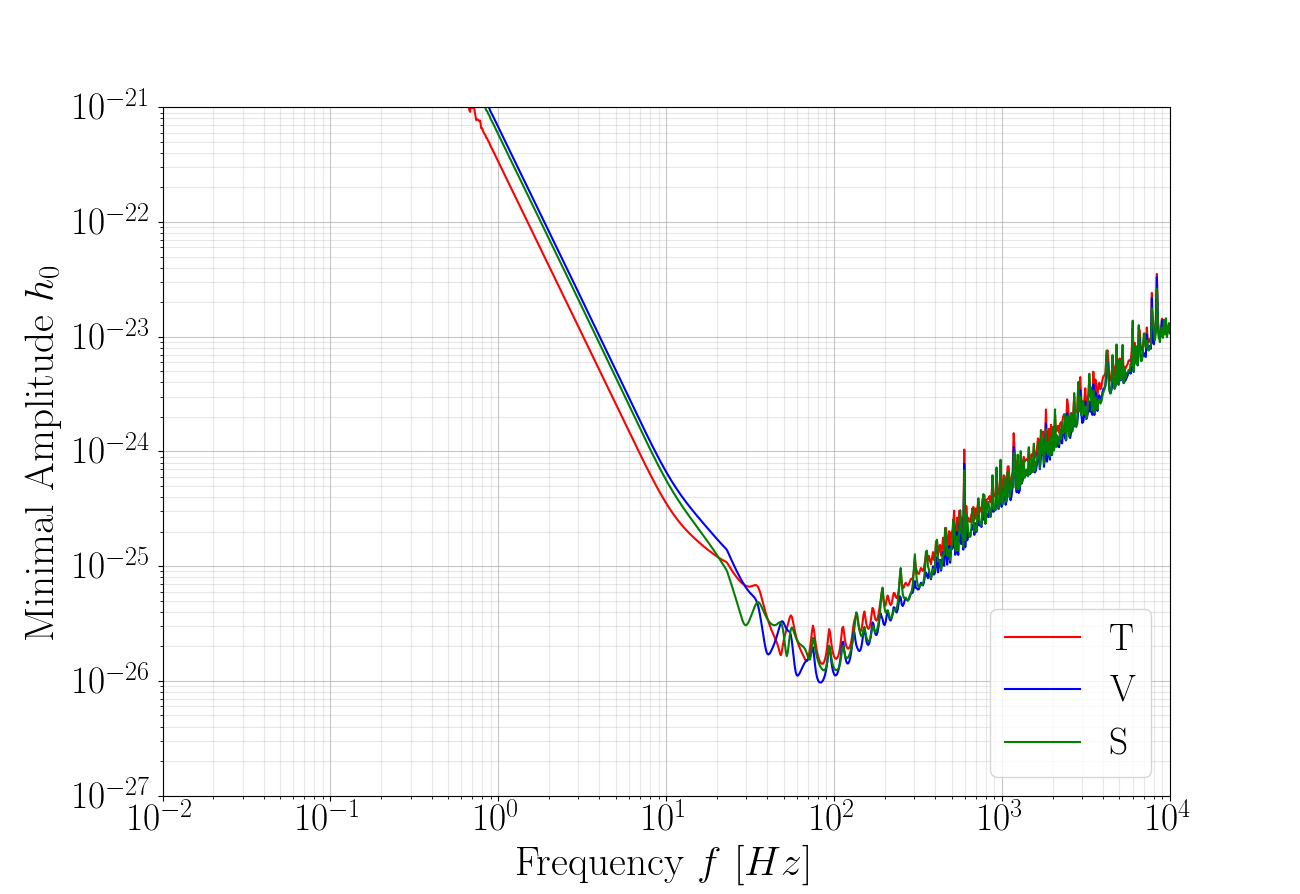}
\caption{\label{ET_fig:ET,LIGO,B-DECIGO} Combined sensitivity of B-DECIGO, ET and both advanced LIGO detectors.}
\end{figure}

B-DECIGO circles the Earth in a Sun-synchronous dusk-dawn orbit with an angular frequency of about \SI{8.2 e-4}{s^{-1}} while 
Earth rotation corresponds to \SI{7.3e-5}{s^{-1}}. This leads to rapidly varying distances and directions of the detector arms and the irrational 
ratio between the two angular velocities leads to a chaotic behaviour, which makes the use of the time dependent sensitivity very complicated. 
Additionally one can observe that the sensitivities for the different modes get closer together as one moves a space detector closer to Earth.

If one would instead let B-DECIGO take the same type of orbit but on a higher altitude (\SI{35867}{km}), such that it would circle Earth in one 
day, one would get almost the same signal every day over a period of about a week, because the change would now be on the time-scale of a year. 
The detectors would also be far enough from Earth to get relevantly different sensitivity curves for the different modes. The procedure would be 
more complicated than in the case of DECIGO, but one could still use certain blind spots or other characteristics that only one mode shows. A 
large disadvantage to DECIGO would also be that one would have to spot those characteristics in a model in advance, since the sensitivities are 
not periodic.

\begin{figure}[h!]
\includegraphics[width=1\linewidth]{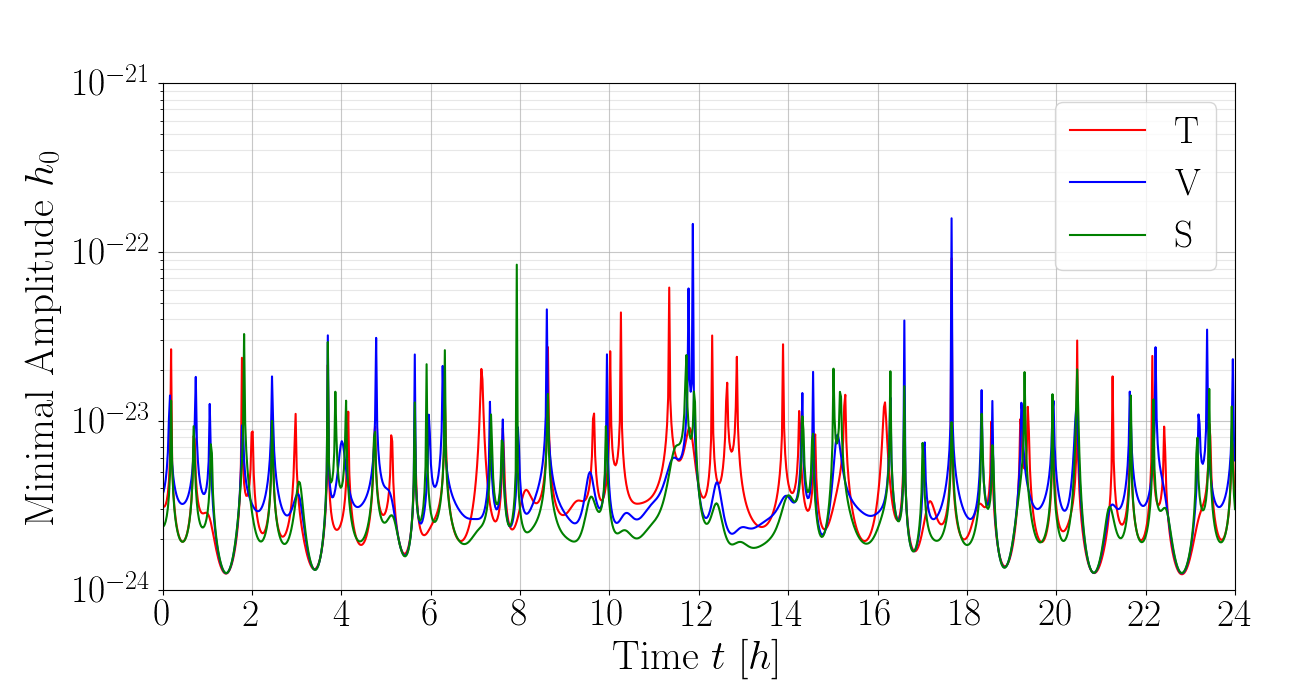}\\
\includegraphics[width=1\linewidth]{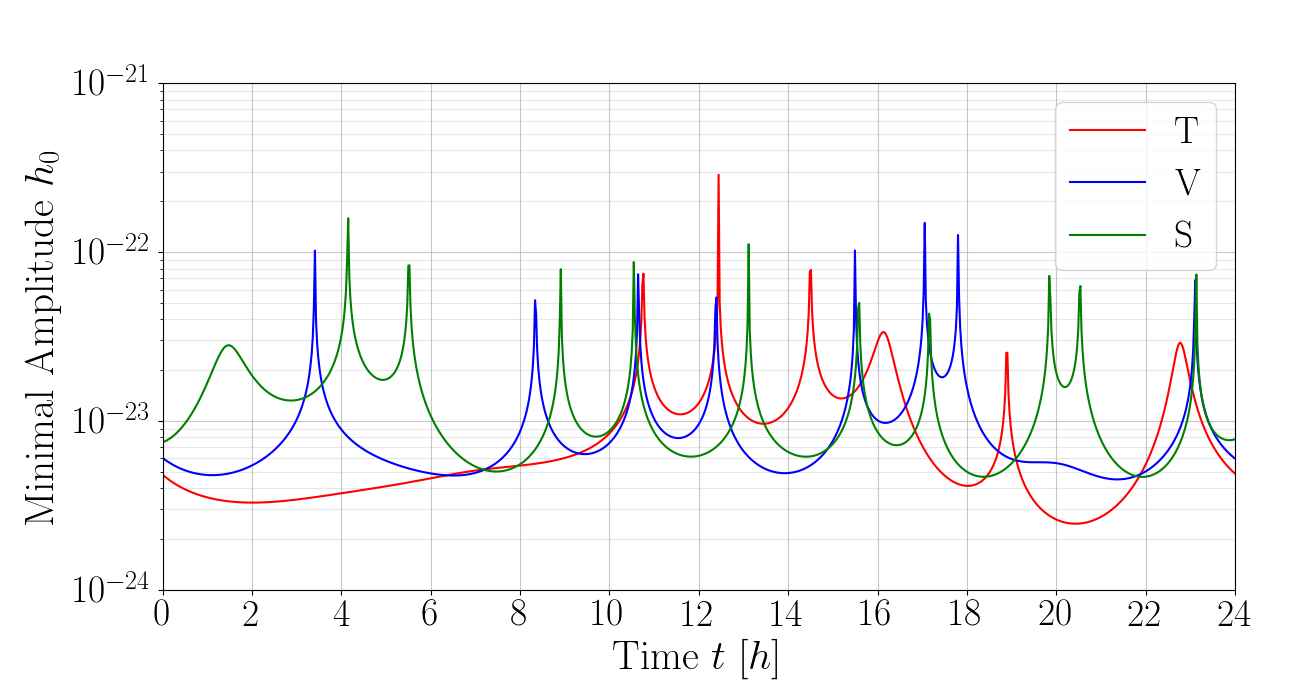}
\caption{\label{ET_fig:tdepB-DECIGO} Time dependence of the sensitivity for B-DECIGO for one day (above) and for a higher altitude of 
\SI{35867}{km} (below) for a frequency of \SI{100}{Hz}.}
\end{figure}

We compare the time dependent sensitivities of both versions (original B-DECIGO and higher altitude) for time-span of one day in 
Fig.~\ref{ET_fig:tdepB-DECIGO}.

\begin{figure}[h!]
\includegraphics[width=1\linewidth]{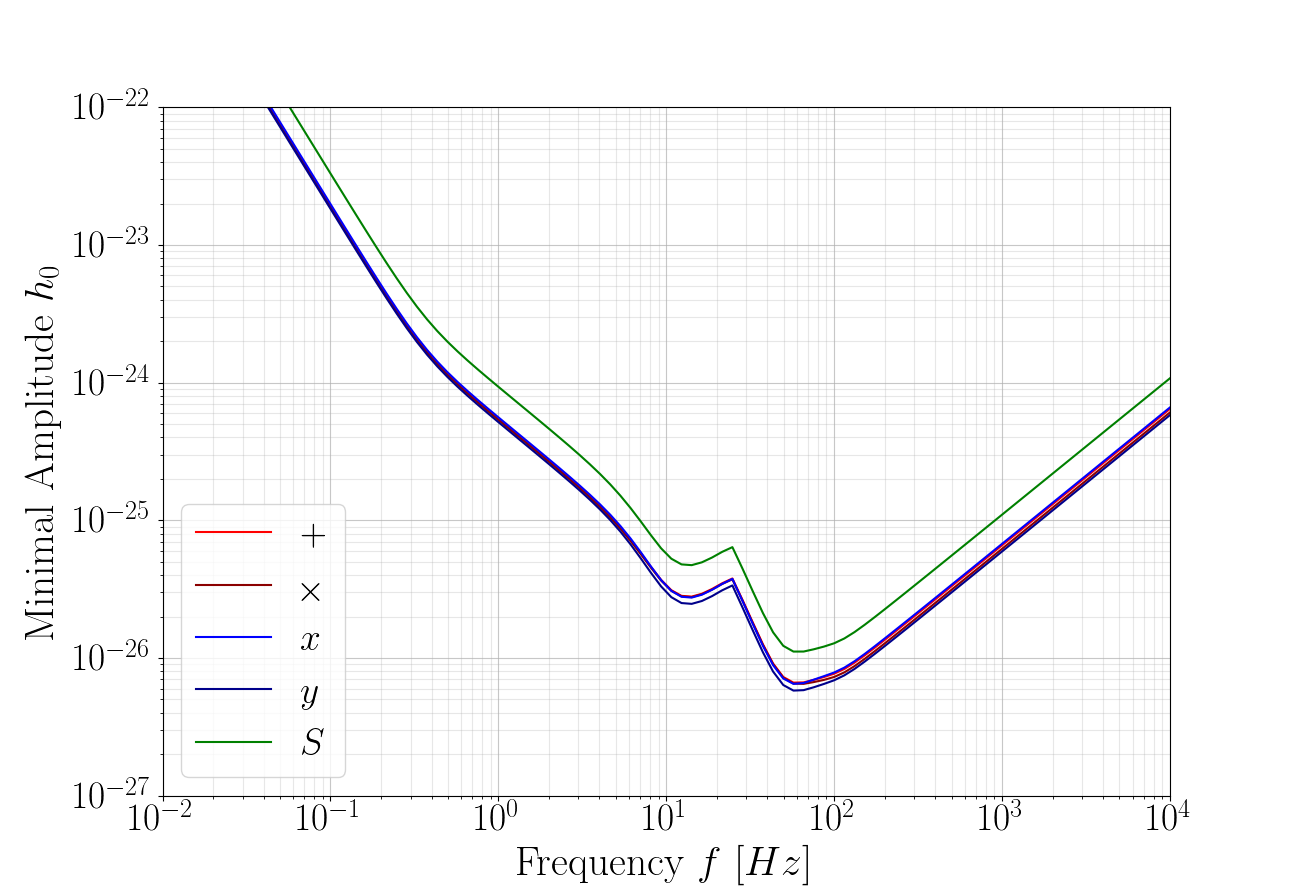}
\caption{\label{ET_fig:h(f)B-DECIGO} Frequency dependent sensitivity of ET, LIGO and B-DECIGO.}
\end{figure}

In Fig.~\ref{ET_fig:h(f)B-DECIGO} we plot the frequency dependent sensitivity of B-DECIGO together with ET and LIGO in the case of point sources. 
The behaviour is very similar to the one with DECIGO, except that the plateau around \SI{1}{Hz} is missing. Since B-DECIGO is not as sensitive as 
DECIGO, it can only increase the sensitivity there a bit.

\section{Delta Distribution Approximation}\label{ET_appendix:delta}
In this Appendix we give a detailed derivation of the signal to noise ratio for a merger by focusing on the approximations of the Dirac delta 
distribution and the Fourier transforms. We first use a scalar signal, measured by two detectors, to simplify the calculation and then generalize 
to a wave with arbitrary polarizations measured by multiple detectors.\\
In future all GW detectors together could be sensitive enough to measure the in spiral of a binary Black Hole or neutron star merger, months 
before the merger event happens. In this case detectors with different distances from the source would have different observation times. This 
would help to measure the position of the source in the sky:
\begin{align}\label{Eq:DT=Omega}
c\Delta T = \hat{\Omega}\cdot\Delta\vec{x}_{IJ}, && \Delta T = T_I - T_J, && \Delta\vec{x}_{IJ} = \vec{x}_I - \vec{x}_J,
\end{align}
where $T_I$ and $T_J$ are the observation times of the detectors $I$ and $J$, $\vec{x}_I$ and $\vec{x}_J$ their position vectors and 
$\hat{\Omega}$ is the direction of travel of the GW.\\
\\
We define the cross correlated and filtered strain amplitude of the detector pair $(I,J)$ by:
\begin{equation}
Y \coloneq \int_{-T_I/2}^{T_I/2}\int_{-T_J/2}^{T_J/2} s_I(t)s_J(t')Q(t-t') dt'dt,
\end{equation}
where $Q$ is the filter function and $s_I$, $s_J$ are the strains measured by the detectors  $I$,$J$, which are the sum of the signal $h_I$ and 
the noise $n_I$ in detector $I$:
\begin{equation}
s_I(t) = h_I(t) + n_I(t).
\end{equation}

By taking the ensemble average we get rid of the noise terms:
\begin{widetext}
\begin{align}
\mu &\coloneq \mathbb{E}[Y] \notag \\
&= \int_{-T_I/2}^{T_I/2}\int_{-T_J/2}^{T_J/2} \left\lbrace \mathbb{E}[h_I(t)h_J(t')] + \mathbb{E}[h_I(t)n_J(t')] + \mathbb{E}[n_I(t)h_J(t')] + 
\mathbb{E}[n_I(t)n_J(t')] \right\rbrace Q(t-t') dt'dt \notag \\
&= \int_{-T_I/2}^{T_I/2}\int_{-T_J/2}^{T_J/2} \int\tilde{h}_I^*(f)e^{2\pi i f\left(t-\frac{\hat{\Omega}\cdot\vec{x}_I}{c}\right)}df 
\int\tilde{h}_J(f')e^{-2\pi i f'\left(t'-\frac{\hat{\Omega}\cdot\vec{x}_J}{c}\right)}df' Q(t-t') dt'dt \notag \\
&= \int \tilde{h}_I^*(f)\tilde{h}_J(f')e^{-\frac{2\pi i}{c}\hat{\Omega}\cdot(f\vec{x}_I-f'\vec{x}_J)} \int_{-T_I/2}^{T_I/2}\int_{-T_J/2}^{T_J/2} 
Q(t-t')e^{-2\pi i(f't'-ft)} dt'dt df'df,
\end{align}
\end{widetext}
where we replaced the signal by its Fourier transform: $h_I(t) = \int\tilde{h}_I(f)e^{-2\pi i 
f\left(t-\frac{\hat{\Omega}\cdot\vec{x}_I}{c}\right)}df$.\\

We apply the following substitution to the integral over $t'$: $\tau = t-t'$, $d\tau = dt$,
\begin{align}
\mu &= \int \tilde{h}_I^*(f)\tilde{h}_J(f')e^{-\frac{2\pi i}{c}\hat{\Omega}\cdot(f\vec{x}_I-f'\vec{x}_J)} \notag\\
& \cdot \int_{-T_J/2}^{T_J/2} \int_{-T_I/2-t'}^{T_I/2-t'}Q(\tau)e^{2\pi if\tau}d\tau e^{-2\pi i(f'-f)t'} dt' df'df.
\end{align}
Then we approximate the integral over $\tau$ with the Fourier transform of the filter function $Q$:
\begin{equation}
\int_{-T_I/2-t'}^{T_I/2-t'}Q(\tau)e^{2\pi if\tau}d\tau \approx \tilde{Q}(f).
\end{equation}
If we shift a wave packed in time, it is still composed of the same frequencies. Therefore, we can ignore the time shift in the integration volume 
by $-t$.\\
\\
We pull this out of the $t$ integral and get:
\begin{align}
 \int_{-T_J/2}^{T_J/2} e^{-2\pi i(f'-f)t'} dt' &= -\frac{1}{\pi\Delta f}\frac{1}{2i}\left(e^{-\pi i\Delta f T_J} - e^{\pi i\Delta f T_J}\right) 
\notag\\
&=\frac{\sin(\pi\Delta f T_J)}{\pi\Delta f} \eqcolon \delta_{T_J}(f'-f). \label{Eq:deltaT}
\end{align}

If $\Delta f = f' - f$ approaches zero, we get: $\underset{\Delta f \rightarrow 0}{\lim}\delta_{T_J}(\Delta f) = \underset{\Delta f \rightarrow 
0}{\lim} \frac{1}{\pi\Delta f}(0 + \pi T_J\Delta f + \mathcal{O}(\Delta f^2)) = T_J$
and for big $\Delta f$, $\delta_{T_J}$ gets small:
\begin{equation}
\left|\frac{\sin(\pi\Delta f T_J)}{\pi\Delta f}\right| \leqslant \frac{1}{\pi\Delta f} \overset{\Delta f \rightarrow \infty}{\longrightarrow} 0.
\end{equation}
\\
By approximating $\delta_{T_J}(f'-f) \approx \delta(f'-f)$ with the Dirac delta distribution we can evaluate the integral over $f'$.
\begin{align}
\mu &\approx \int \tilde{h}_I^*(f)\tilde{h}_J(f') e^{-\frac{2\pi i}{c}\hat{\Omega}\cdot(f\vec{x}_I-f'\vec{x}_J)} \tilde{Q}(f')\delta(f'-f) df'df 
\notag\\
&= \int \tilde{h}_I^*(f)\tilde{h}_J(f)\tilde{Q}(f) e^{-2\pi if\frac{\hat{\Omega}\cdot\Delta\vec{x}_{IJ}}{c}} df.
\end{align}

We now have an expression for the signal. To calculate the signal to noise ratio we need to deal with noise which is the square root of the 
variance in absence of a signal:
\begin{widetext}
\begin{align}
\sigma^2 &\coloneq \mathbb{V}[Y]|_{h=0} = \mathbb{E}[Y^2] - \mathbb{E}[Y]^2 |_{h=0} = \mathbb{E}[Y^2]|_{h=0} \notag \\
&= \int_{-T_I/2}^{T_I/2}\int_{-T_I/2}^{T_I/2}\int_{-T_J/2}^{T_J/2}\int_{-T_J/2}^{T_J/2} \mathbb{E}[s_I(t)s_I(t')s_J(\tau)s_J(\tau')] 
Q(t-\tau)Q(t'-\tau') d\tau'd\tau dt'dt|_{h=0} \notag \\
&= \int_{-T_I/2}^{T_I/2}\int_{-T_I/2}^{T_I/2}\int_{-T_J/2}^{T_J/2}\int_{-T_J/2}^{T_J/2} \mathbb{E}[n_I(t)n_I(t')]\mathbb{E}[n_J(\tau)n_J(\tau')] 
Q(t-\tau)Q(t'-\tau') d\tau'd\tau dt'dt.
\end{align}
\end{widetext}
Since the noises of the two detectors are independent of each other, we can take their expectation separately. We then insert the Fourier 
transformation (FT) of the noise, in the time interval in which the measurement is taken:
\begin{equation}
n_I(t) = \int \tilde{n}_I(f)e^{-2\pi ift} df,
\end{equation}
And then swap the time and frequency integrals and approximate the FT of the filter function and the delta distribution as before:
\begin{align}
\sigma^2 &= \int \mathbb{E}[\tilde{n}_I^*(f)\tilde{n}_I(f')]\mathbb{E}[\tilde{n}_J(\nu)\tilde{n}_J^*(\nu')] \notag\\
&\cdot \tilde{Q}(\nu)\delta(\nu-f)\tilde{Q}^*(\nu')\delta(\nu'-f') d\nu'd\nu df'df \notag \\
&= \int \mathbb{E}[\tilde{n}_I^*(f)\tilde{n}_I(f')]\mathbb{E}[\tilde{n}_J(f)\tilde{n}_J^*(f')] \tilde{Q}(f)\tilde{Q}^*(f') df'df.
\end{align}

Now we use that different frequencies in the noise are not correlated to each other and the definition of the two sided noise power spectral 
density:
\begin{equation}
\mathbb{E}[\tilde{n}_I^*(f)\tilde{n}_I(f')] \eqcolon \frac{1}{2}P_I(|f'|)\delta(f'-f).
\end{equation}
If we would carelessly plug in this identity, we would get a multiplication of two delta distributions, which is not definable. But we cannot take 
the expectation of the noise squared over an infinite time integral anyway. So, the delta distribution is actually a $\delta_{T_I}$. This is a 
smooth function and not a distribution and can therefore be multiplied with another $\delta_{T_J}$.
\begin{align}
\sigma^2 &= \frac{1}{4}\int P_I(|f'|)P_J(|f'|)\tilde{Q}(f)\tilde{Q}^*(f') \notag\\ &\cdot \int_{-T_I/2}^{T_I/2}e^{-2\pi i(f'-f)t}dt 
\int_{-T_J/2}^{T_J/2}e^{2\pi i(f'-f)t'}dt' df'df \notag \\
&= \frac{1}{4}\int P_I(|f'|)P_J(|f'|)\tilde{Q}(f)\tilde{Q}^*(f') \notag\\ &\cdot \int_{-T_I/2}^{T_I/2}\int_{-T_J/2}^{T_J/2} e^{2\pi i(f'-f)(t'-t)} 
dt'dt df'df.
\end{align}

\begin{figure}[h!]
\includegraphics[width=0.9\linewidth]{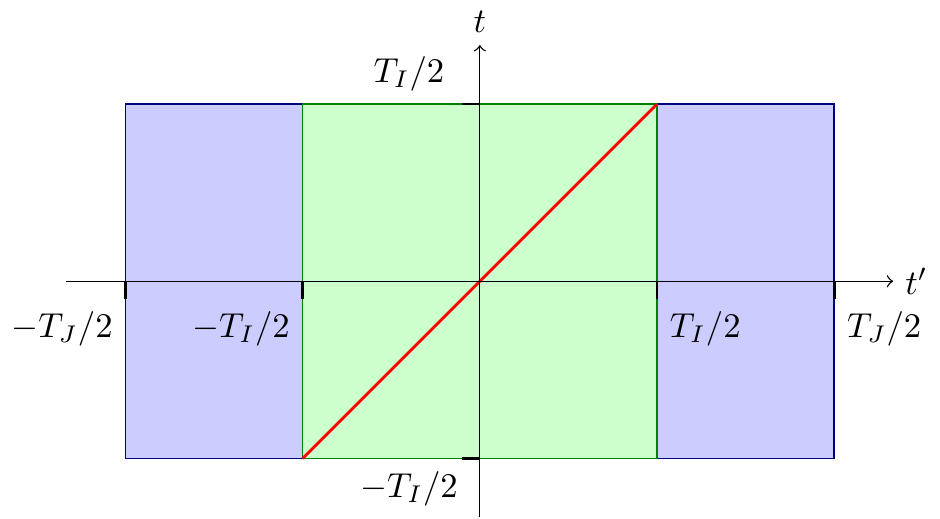}
\caption{\label{Fig:IntVol} The green region is symmetric around $t'-t = 0$ (red line). The blue rectangle marks the entire integration region.}
\end{figure}

To evaluate the time integrals we have to split the integration domain into three regions as depicted in Fig.~\ref{Fig:IntVol}, since we need an 
integration region which is symmetric around $t'-t=0$, where we can use Eq.~\eqref{Eq:deltaT}. The rest can be evaluated separately.

Let $T_I < T_J$, $\Delta T = T_J - T_I$ and $\Delta f = f'-f$, then the time integrals read:
\begin{align}
\delta_{T_I}\delta_{T_J} &= \int_{-T_I/2}^{T_I/2} \int_{-T_J/2}^{-T_I/2}e^{-2\pi i\Delta f(t'-t)}dt' \notag\\
&\quad + \int_{-T_I/2}^{T_I/2}e^{-2\pi i\Delta f(t'-t)}dt' \notag\\
&\quad + \int_{T_I/2}^{T_J/2}e^{-2\pi i\Delta f(t'-t)}dt' dt.
\end{align}
We substitute $\eta = -t'$ in the first integral over $t'$, to bring it into the same form as the third one.
\begin{widetext}
\begin{align}
\delta_{T_I}\delta_{T_J} &= \int_{-T_I/2}^{T_I/2} -\int_{T_J/2}^{T_I/2}e^{2\pi i\Delta f(\eta+t)}d\eta + \int_{T_I/2}^{T_J/2}e^{-2\pi i\Delta 
f(t'-t)}dt' dt + \int_{-T_I/2}^{T_I/2}\int_{-T_I/2}^{T_I/2}e^{-2\pi i\Delta f(t'-t)}dt'dt \notag \\
&= \int_{-T_I/2}^{T_I/2}e^{2\pi i\Delta ft} \int_{T_I/2}^{T_J/2} e^{2\pi i\Delta ft'} + e^{-2\pi i\Delta ft'}dt' dt + \delta_{T_I}^2(\Delta f) 
\notag \\
&= \int_{-T_I/2}^{T_I/2}e^{2\pi i\Delta ft}dt \int_{T_I/2}^{T_J/2} 2\cos(2\pi\Delta ft') dt' + \delta_{T_I}^2(\Delta f) \notag \\
&= \delta_{T_I}(\Delta f)\left( \frac{\sin(\pi\Delta f T_J)-\sin(\pi\Delta f T_I)}{\pi\Delta f} + \delta_{T_I}(\Delta f) \right) \notag \\
&\approx \delta(f'-f)\left( \frac{\sin(\pi\Delta fT_I) + \pi\Delta f\Delta T + \mathcal{O}((\pi\Delta f\Delta T)^2) - \sin(\pi\Delta 
fT_I)}{\pi\Delta f} + \delta_{T_I}(\Delta f) \right),
\end{align}
where we assumed, that $\Delta T \ll T_I$.

This approximated distribution acts on functions as:
\begin{align}
\int g(f') \delta_{T_I}\delta_{T_J}(\Delta f) df' %\notag\\
&\approx \int g(f')\left( \Delta T + \mathcal{O}(f'-f) + \delta_{T_I}(f'-f) \right)\delta(f'-f) df \notag\\
&= g(f)(T_I + \Delta T), \ \forall g \in C^\infty(\mathbb{C}).
\end{align}
\end{widetext}

Inserting this into the variance and integrating over $f'$ we get:
\begin{align}\label{Eq:sigma^2}
\sigma^2 &= \frac{1}{4}\int P_I(|f'|)P_J(|f'|)\tilde{Q}(f)\tilde{Q}^*(f') \delta_{T_I}\delta_{T_J}(f'-f) df'df \notag \\
&= \frac{T_I+\Delta T}{4}\int P_I(|f|)P_J(|f|)|\tilde{Q}(f)|^2 df.
\end{align}

Using matched filtering with the scalar product: $(\tilde{A},\tilde{B}) \coloneq \int \tilde{A}^*(f)\tilde{B}(f)P_I(|f|)P_J(|f|)df$,\\
leads us to a filter function:
\begin{equation}
\tilde{Q}(f) = \frac{\tilde{h}_I^*(f)\tilde{h}_J(f)e^{2\pi if\frac{\hat{\Omega}\cdot\Delta\vec{x}_{IJ}}{c}}}{P_I(|f|)P_J(|f|)}.
\end{equation}
We can write the signal and noise in terms of the filter function and arrive at the signal to noise ratio:
\begin{align}
SNR &= \frac{\mu}{\sigma} = \frac{(\tilde{Q},\tilde{Q})}{\sqrt{\frac{T_I+\Delta T}{4}(\tilde{Q},\tilde{Q})}} \notag\\
&= 2\sqrt{\frac{1}{T_I+\Delta T}\int \frac{|\tilde{h}_I(f)\tilde{h}_J(f)|^2}{P_I(|f|)P_J(|f|)} df}.
\end{align}

We now model the merger as a periodic source, which stops radiating at the end of the merging event at its time coordinate $t_0$. Under the 
assumption that the detectors are far away from the source, we can model the incoming wave as a plane wave with amplitude $h_0$ and frequency 
$f_0$, traveling in direction $\hat{\Omega}$:
\begin{equation}
h(t) = h_0 e^{2\pi if_0(t-\frac{\hat{\Omega}\cdot\vec{x}}{c})} \theta\left(t_0 - \frac{\hat{\Omega}\cdot\vec{x}}{c} - t\right).
\end{equation}
The detector $I$ will measure the signal over a time period $T_I$ and the Fourier transform of the measured signal is therefore:
\begin{align}
\tilde{h}_I(f) &= \int_{-T_I/2}^{T_I/2} h_0 e^{2\pi if_0(t-\frac{\hat{\Omega}\cdot\vec{x}}{c})} e^{-2\pi ift} dt \notag\\
&= h_0 e^{-2\pi if_0\frac{\hat{\Omega}\cdot\vec{x}}{c}} \int_{-T_I/2}^{T_I/2} e^{-2\pi i(f-f_0)t} dt
\end{align}

Again, we cannot approximate this with a delta distribution, otherwise we would get a $\delta^4$ for the $|\tilde{h}_I\tilde{h}_J|^2$ term.
\begin{align}
& |\tilde{h}_I(f)\tilde{h}_J(f)|^2 \notag\\ &= h_0^4 \int_{-T_I/2}^{T_I/2}\int_{-T_I/2}^{T_I/2}\int_{-T_J/2}^{T_J/2}\int_{-T_J/2}^{T_J/2} 
\notag \\ &\quad e^{-2\pi i(f-f_0)(t'-t+\tau'-\tau)} d\tau'd\tau dt'dt.
\end{align}

We do the same splitting of the $T_J$ interval as above, under the assumption $T_I < T_J$ and using the short hand $\rho = t'-t+\tau'-\tau$:
\begin{widetext}
\begin{align}
|\tilde{h}_I\tilde{h}_J|^2 &\propto \delta_{T_I}^2\delta_{T_J}^2 = \int_{-T_I/2}^{T_I/2}\int_{-T_J/2}^{T_J/2} e^{-2\pi i(f'-f)(t'-t+\tau'-\tau)} 
d\tau'd\tau dt'dt \notag \\
&= \int_{-T_I/2}^{T_I/2} \int_{-T_J/2}^{-T_I/2} e^{-2\pi i\Delta f\rho} d\tau'd\tau + \int_{-T_I/2}^{T_I/2} e^{-2\pi i\Delta f\rho} d\tau'd\tau + 
\int_{T_I/2}^{T_J/2} e^{-2\pi i\Delta f\rho} d\tau'd\tau dt'dt \notag \\
&= \int_{-T_I/2}^{T_I/2}e^{-2\pi i\Delta f(t'-t)}dt'dt \int_{T_I/2}^{T_J/2} e^{2\pi i\Delta f(\tau'-\tau)} + e^{-2\pi i\Delta f(\tau'-\tau)} 
d\tau'd\tau + \int_{-T_I/2}^{T_I/2}e^{-2\pi i\Delta f\rho}d\tau'd\tau dt'dt \notag \\
&= \delta_{T_I}^2(\Delta f)\int_{T_I/2}^{T_J/2} 2\cos(2\pi\Delta f(\tau'-\tau))d\tau'd\tau + \delta_{T_I}^4(\Delta f) \notag \\
&= \delta_{T_I}^2(\Delta f)\left( \frac{1}{\pi\Delta f} \int_{T_I/2}^{T_J/2} \sin(\pi\Delta f(T_J-2\tau))-\sin(\pi\Delta f(T_I-2\tau)) d\tau + 
\delta_{T_I}^2(\Delta f) \right) \notag \\
&= \delta_{T_I}^2(\Delta f)\left( -\frac{1}{2(\pi\Delta f)^2}\lbrace \cos(\pi\Delta f(T_J-T_J)) - \cos(\pi\Delta f(T_J-T_I)) \right. \notag \\
&\textit{ $ $ $ $ $ $ $ $ $ $ $ $ $ $ $ $ $ $ $ $ $ $ $ $ $ $ $ $ $ $ }\left. - \cos(\pi\Delta f(T_I-T_J)) + \cos(\pi\Delta f(T_I-T_I)) \rbrace + 
\delta_{T_I}^2(\Delta f) \right) \notag \\
&= \delta_{T_I}^2(\Delta f)\left( \frac{1}{(\pi\Delta f)^2}\lbrace \cos(\pi\Delta f\Delta T) -1 \rbrace + \delta_{T_I}^2(\Delta f) \right) \notag 
\\
&= \delta(f'-f)\delta_{T_I}(\Delta f)\left( \frac{1}{(\pi\Delta f)^2}\left\lbrace 1 - \frac{1}{2}(\pi\Delta f\Delta T)^2 + \mathcal{O}(\Delta f^4) 
- 1 \right\rbrace + \delta_{T_I}^2(\Delta f) \right).
\end{align}

The action on a function $g \in C^\infty(\mathbb{C})$ is:
\begin{align}
\int g(f')\delta_{T_I}^2\delta_{T_J}^2(f'-f) df' &= \int g(f') \delta_{T_I}(\Delta f)\left( \frac{\Delta T^2}{2} + \mathcal{O}(\Delta 
f^2) + \delta_{T_I}(\Delta f)^2 \right) \delta(f'-f) df' \notag \\
&= g(f)\delta_{T_I}(0)\left( \frac{\Delta T^2}{2} + \delta_{T_I}(0)^2 \right) \notag\\
&= g(f)T_I\left( \frac{\Delta T^2}{2} + T_I^2 \right).
\end{align}

\end{widetext}

When we plug this into the $SNR$ we get:
\begin{align}\label{Eq:SNR_PtSrcDT}
SNR &= 2\sqrt{\frac{T_I}{T_I+\Delta T}\left( \frac{\Delta T^2}{2} + T_I^2 \right) \frac{h_0^4}{P_I(|f_0|)P_J(|f_0|)}} \notag\\
&\approx 2\left( T_I + \frac{\hat{\Omega}\cdot\Delta\vec{x}_{IJ}}{2c} \right)\frac{h_0^2}{\sqrt{P_I(f_0)P_J(f_0)}},
\end{align}
where we used the identification of the integration time with the direction of the source in Eq.~\eqref{Eq:DT=Omega}.\\

The minimal amplitude is then given by:
\begin{align}
h_{min} &= \sqrt{32}\left( \frac{(T_I+\Delta T)P_I(f)P_J(f)}{T_I\left(\frac{\Delta T^2}{2}+T_I^2\right)} \right)^{1/4} \notag\\
&\approx \sqrt{32}\left( \frac{1}{\sqrt{T_I}} + \frac{\hat{\Omega}\cdot\Delta\vec{x}_{IJ}}{4c\sqrt{T_I}^3} \right)\sqrt[4]{P_I(f)P_J(f)}.
\end{align}

Including polarizations, we have a gravitational wave $h_{ij}(t)$, which induces the signal $h_I(t)$ in detector $I$:
\begin{align}
h_{ij}(t) &= \sum_A h_A e^{2\pi if_0(t-\frac{\hat{\Omega}\cdot\vec{x}}{c}) + \varphi_A} \theta\left(t_0 - \frac{\hat{\Omega}\cdot\vec{x}}{c} - 
t\right) e_{ij}^A \\
h_I(t) &= \sum_A h_A F_I^A(\hat{\Omega}) e^{2\pi if_0(t-\frac{\hat{\Omega}\cdot\vec{x}_I}{c}) + \varphi_A} \theta\left(t_0 - 
\frac{\hat{\Omega}\cdot\vec{x}_I}{c} - t\right),
\end{align}
where $h_A$ is the amplitude of the wave in polarization $A$ and $\varphi_A$ accounts for the fact that the polarizations could be phase shifted.\\
For the absolute value squared of the cross correlated signals of two detectors we get:
\begin{widetext}
\begin{align}
&|\tilde{h}_I(f)\tilde{h}_J(f)|^2 = |\tilde{h}_I(f)|^2|\tilde{h}_J(f)|^2 \notag \\
&= \left| \sum_A h_A F_I^A(\hat{\Omega})e^{-\frac{2\pi if_0}{c}\hat{\Omega}\cdot\vec{x}_I + i\varphi_A} \right|^2 \left| \sum_A h_A 
F_J^A(\hat{\Omega})e^{-\frac{2\pi if_0}{c}\hat{\Omega}\cdot\vec{x}_J + i\varphi_A} \right|^2 \int_{-T_I/2}^{T_I/2}\int_{-T_J/2}^{T_J/2} e^{-2\pi 
i(f-f_0)\rho} d^4\rho \notag \\
&= \left| \sum_A h_A F_I^A(\hat{\Omega})e^{i\varphi_A} \right|^2 \left| \sum_A h_A F_J^A(\hat{\Omega})e^{i\varphi_A} \right|^2 
\delta_{T_I}(f-f_0)\left( \frac{\Delta T^2}{2} + \delta_{T_I}(f-f_0)^2 \right).
\end{align}
\end{widetext}
We make the assumption, that the gravitational wave has only one of the polarizations $h = \sum_{A'} h_{A'} \delta_{A'A}$, to get the signal to 
noise ratio for that polarization.
\begin{align}
& SNR_A \coloneq \left.\frac{\mu}{\sigma}\right|_{h=h_A} \notag\\
& = 2\sqrt{\frac{T_I}{T_I+\Delta T}\left( \frac{\Delta T^2}{2} + T_I^2 \right) 
\frac{(|h_A|^2F_I^A(\hat{\Omega})F_J^A(\hat{\Omega}))^2}{P_I(|f_0|)P_J(|f_0|)}} \notag \\
&\approx 2\left( T_I + \frac{\hat{\Omega}\cdot\Delta\vec{x}_{IJ}}{2c} 
\right)\frac{|h_A|^2F_I^A(\hat{\Omega})F_J^A(\hat{\Omega})}{\sqrt{P_I(f_0)P_J(f_0)}},
\end{align}
which we get by replacing $h_0^4 \mapsto (|h_A|^2F_I^A(\hat{\Omega})F_J^A(\hat{\Omega}))^2$ in Eq.~\eqref{Eq:SNR_PtSrcDT}.

For multiple detectors we use the maximum likelihood method and calculate the Fisher matrix. The likelihood function is given by:
\begin{equation}
L(\mu_{IJ},\vec{\theta}) = e^{-\sum_{(I,J)}\frac{(Y_{IJ}-\mu_{IJ})^2}{2\sigma_{IJ}^2}},
\end{equation}
where $\mu_{IJ} = \mathbb{E}[Y_{IJ}]$ is the ensemble average of the correlated signals of the detectors $I$ and $J$. Its variance $\sigma_{IJ}^2 
= \mathbb{V}[Y_{IJ}]$ is given by Eq.~\eqref{Eq:sigma^2} without the filtering. Multiplying the $SNR$ Eq.~\eqref{Eq:SNR_PtSrcDT} with the noise we 
get:
\begin{equation}
\mu_{IJ} = \sqrt{T_I\left( \frac{\Delta T^2}{2} + T_I^2 \right)}|\tilde{h}_I\tilde{h}_J|.
\end{equation}
The matrix element $F_{AA'}$ of the Fisher matrix is then given by:
\begin{widetext}
\begin{align}
F_{AA'} &= \mathbb{E}\left[ \left(\partial_{|h_A|^2}\ln L\right)\left(\partial_{|h_{A'}|^2}\ln L\right) \right] \notag \\
&= \mathbb{E}\left[ \left( \sum_{(I,J)}\frac{1}{\sigma_{IJ}^2}(Y_{IJ}-\mu_{IJ})\sqrt{T_I\left(\frac{\Delta T^2}{2}+T_I^2\right)} \right)^2 
\left(\partial_{|h_A|^2}|\tilde{h}_I\tilde{h}_J|\right) \left(\partial_{|h_{A'}|^2}|\tilde{h}_I\tilde{h}_J|\right) \right] \notag \\
&= \sum_{(I,J)} \frac{1}{\sigma_{IJ}^4} \underset{\mathbb{V}[Y_{IJ}]=\sigma_{IJ}^2}{\underbrace{\mathbb{E}[(Y_{IJ}-\mu_{IJ})^2]}} 
T_I\left(\frac{\Delta T^2}{2}+T_I^2\right) \left(\partial_{|h_A|^2}|\tilde{h}_I\tilde{h}_J|\right) 
\left(\partial_{|h_{A'}|^2}|\tilde{h}_I\tilde{h}_J|\right) \notag \\
&\textit{ $ $ $ $ } + \sum_{(I,J)\neq (I',J')} \frac{1}{\sigma_{IJ}^2\sigma_{I'J'}^2} 
\underset{Cov(Y_{IJ},Y_{I'J'})=0}{\underbrace{\mathbb{E}[(Y_{IJ}-\mu_{IJ})(Y_{I'J'}-\mu_{I'J'})]}} T_I\left(\frac{\Delta T^2}{2}+T_I^2\right) 
\left(\partial_{|h_A|^2}|\tilde{h}_I\tilde{h}_J|\right) \left(\partial_{|h_{A'}|^2}|\tilde{h}_{I'}\tilde{h}_{J'}|\right) \notag \\
&= \sum_{(I,J)} \frac{4T_I}{(T_I+\Delta T)P_IP_J}\left(\frac{\Delta T^2}{2}+T_I^2\right) \left(\partial_{|h_A|^2}|\tilde{h}_I\tilde{h}_J|\right) 
\left(\partial_{|h_{A'}|^2}|\tilde{h}_I\tilde{h}_J|\right)
\end{align}
\end{widetext}

The $SNR$ squared of a specific polarization $A$ is defined by dividing the square of the quantity we are looking for $|h_A|^2$ by its variance 
$\sigma_A$, under the condition that the incoming wave has only that polarization.
\begin{align}
SNR_A^2 &\coloneq \left.\frac{(|h_A|^2)^2}{\sigma_A^2}\right|_{h=h_A} = \left.\frac{(|h_A|^2)^2}{(F^{-1})_{AA}}\right|_{h=h_A} \notag \\
&= \left.\frac{(|h_A|^2)^2 \det\mathbf{F}}{\mathcal{F}_A}\right|_{h=h_A}
\end{align}

\section{Fisher Matrix Entries}\label{ET_appendix:fishermatrix}
As can be seen in Appendix~\ref{ET_appendix:delta}, the Fisher matrix can be written as a sum of Fisher matrices of single detector pairs, 
which consist of a pre-factor and two derivative terms for row and column of the entry. If $\theta_{i,j}$ are polarizations we have:
\begin{equation}
F_{ij} \propto \left(\partial_{\theta_i}|\tilde{h}_I\tilde{h}_J|\right) \left(\partial_{\theta_j}|\tilde{h}_I\tilde{h}_J|\right),
\end{equation}
where $\vec{\theta} = (\theta, \phi, +, \times, x, y, b, l)$ are the parameters we are looking for.\\

Here we calculate those derivative terms.\\
We start by writing out the absolute value squared of the correlation signal.
\begin{align}
|\tilde{h}_I\tilde{h}_J|^2 = \left| \sum_Ah_AF_I^A \sum_{A'}h_{A'}F_J^{A'} \right|^2,
\end{align}
where the phase $\varphi_A$ of the polarization $A$ is integrated in the complex valued amplitude $h_A \in \mathbb{C}$.\\
We split the multiplied signals up into sums over terms where the polarizations coincide and where they are different:
\begin{widetext}
\begin{align}
|\tilde{h}_I\tilde{h}_J|^2 &= \left| \sum_Ah_A^2F_I^AF_J^A + \sum_{A\neq A'}h_Ah_{A'}F_I^AF_J^{A'} \right|^2 \notag \\
&= \left(\sum_Ah_A^2F_I^AF_J^A\right)\left(\sum_Bh_B^2F_I^BF_J^B\right)^* + \left(\sum_Ah_A^2F_I^AF_J^A\right)\left(\sum_{B\neq 
B'}h_Bh_{B'}F_I^BF_J^{B'}\right)^* \notag \\
&\textit{ $ $ $ $ } + \left(\sum_{A\neq A'}h_Ah_{A'}F_I^AF_J^{A'}\right)\left(\sum_Bh_B^2F_I^BF_J^B\right)^* + \left(\sum_{A\neq 
A'}h_Ah_{A'}F_I^AF_J^{A'}\right)\left(\sum_{B\neq B'}h_Bh_{B'}F_I^BF_J^{B'}\right)^* \notag \\
\notag \\
&= \sum_A\left(|h_A|^2F_I^AF_J^A\right)^2 + \sum_{A\neq B}(h_Ah_B^*)^2F_I^AF_J^AF_I^BF_J^B \notag \\
&\textit{ $ $ $ $ } + \sum_Ah_A|h_A|^2F_I^AF_J^A\sum_{B\neq A}h_B^*(F_I^AF_J^B+F_I^BF_J^A) + \sum_{A\neq B\neq 
B'}h_A^2F_I^AF_J^Ah_B^*h_{B'}^*F_I^BF_J^{B'} \notag \\
&\textit{ $ $ $ $ } + \sum_Bh_B^*|h_B|^2F_I^BF_J^B\sum_{A\neq B}h_A(F_I^BF_J^A+F_I^AF_J^B) + \sum_{A\neq A'\neq 
B}h_Ah_{A'}F_I^AF_J^{A'}(h_B^*)^2F_I^BF_J^B \notag \\
&\textit{ $ $ $ $ } + \sum_{A\neq A'}|h_A|^2|h_{A'}|^2\left[(F_I^AF_J^{A'})^2+F_I^AF_J^{A'}F_I^{A'}F_J^A\right] \notag \\
&\textit{ $ $ $ $ } + \sum_A|h_A|^2 \sum_{\underset{B,B'\neq A}{B\neq B'}}h_Bh_{B'}^*\left[(F_I^A)^2F_J^BF_J^{B'} + 
F_I^AF_J^A(F_I^BF_J^{B'}+F_I^{B'}F_J^B) + (F_J^A)^2F_I^BF_I^{B'}\right] \notag \\
&\textit{ $ $ $ $ } + \sum_{A\neq A'\neq B\neq B'}h_Ah_{A'}h_B^*h_{B'}^*F_I^AF_J^{A'}F_I^BF_J^{B'}.
\end{align}
\end{widetext}
When we take the derivative after $|h_A|^2$ all sums which do not contain such a term vanish.
\begin{widetext}
\begin{align}
\partial&_{|h_A|^2}|\tilde{h}_I\tilde{h}_J| = \frac{1}{2\sqrt{|\tilde{h}_I\tilde{h}_J|}}\left\lbrace 2|h_A|^2(F_I^AF_J^A)^2 + 
h_AF_I^AF_J^A\sum_{B\neq A}h_B^*(F_I^AF_J^B+F_I^BF_J^A) \right.\notag \\
&\left. + h_A^*F_I^AF_J^A\sum_{B\neq A}h_B(F_I^AF_J^B+F_I^BF_J^A) + \sum_{A'\neq 
A}|h_{A'}|^2\left[(F_I^AF_J^{A'})^2+2F_I^AF_J^{A'}F_I^{A'}F_J^A+(F_I^{A'}F_J^A)^2\right] \right.\notag \\
&\left. + \sum_{\underset{B,B'\neq A}{B\neq B'}}h_Bh_{B'}^*\left[(F_I^A)^2F_J^BF_J^{B'} + F_I^AF_J^A(F_I^BF_J^{B'}+F_I^{B'}F_J^B) + 
(F_J^A)^2F_I^BF_I^{B'}\right] \right\rbrace.
\end{align}
\end{widetext}
We add the condition, that we have an incoming wave with polarization $A_0$ and therefore all terms proportional to two different polarizations 
are zero.
\begin{widetext}
\begin{align}
\left.\partial_{|h_A|^2}|\tilde{h}_I\tilde{h}_J|\right|_{h=h_{A_0}} &= 
\frac{1}{2\sqrt{\left(|h_{A_0}|^2F_I^{A_0}F_J^{A_0}\right)^2+0}}\left\lbrace 2\delta_{AA_0}|h_A|^2(F_I^AF_J^A)^2 \right.
\notag \\ &\left. + (1-\delta_{AA_0})|h_{A_0}|^2\left[(F_I^AF_J^{A_0})^2+2F_I^AF_J^{A_0}F_I^{A_0}F_J^A+(F_I^{A_0}F_J^A)^2\right] \right\rbrace 
\notag \\
&= F_I^AF_J^A + (1-\delta_{AA_0})\frac{1}{2}\left[ \frac{(F_I^A)^2F_J^{A_0}}{F_I^{A_0}} + \frac{(F_J^A)^2F_I^{A_0}}{F_J^{A_0}} \right]
\end{align}
\end{widetext}

If we calculate a matrix element in the $\theta$ or $\phi$ row or column, we cannot pull the term $\sqrt{T_I\left(\frac{\Delta 
T^2}{2}+T_I^2\right)}$ out in front, so the general Fisher matrix element looks like:
\begin{align}
F_{ij} &= \mathbb{E}\left[(\partial_{\theta_i}\ln L)(\partial_{\theta_j}\ln L)\right] \notag \\
&= \sum_{(I,J)}\frac{1}{\sigma_{IJ}^2} \left( \partial_{\theta_i}\sqrt{T_I\left(\frac{\Delta T^2}{2}+T_I^2\right)}|\tilde{h}_I\tilde{h}_J| \right) 
\notag\\
& \cdot \left( \partial_{\theta_j}\sqrt{T_I\left(\frac{\Delta T^2}{2}+T_I^2\right)}|\tilde{h}_I\tilde{h}_J| \right)
\end{align}
The variance of the true signal $Y_{IJ}$ is dependent on the true time difference and we can treat it as a parameter when we take the derivative 
after the estimated $\theta$-value.
\begin{align}
\mathbb{V}[Y_{IJ}] = \sigma_{IJ}^2 = \frac{T_I+\Delta T}{4}P_IP_J,	&&	\Delta T = \frac{\hat{\omega}\cdot\Delta\vec{x}_{IJ}}{c},
\end{align}
where $\hat{\omega}$ is the true direction of the source.\\

The derivative term for the angle $\theta$ for a wave with polarization $A_0$ is given by:
\begin{align}
& \left.\partial_\theta \sqrt{T_I\left(\frac{\Delta T^2}{2}+T_I^2\right)} |\tilde{h}_I\tilde{h}_J| \right|_{h=h_{A_0}} \notag \\ &= 
\frac{T_I\Delta T}{2\sqrt{T_I\left(\frac{\Delta T^2}{2}+T_I^2\right)}}\frac{\hat{\Omega}_{,\theta}\cdot\Delta\vec{x}_{IJ}}{c} 
|h_{A_0}|^2F_I^{A_0}F_J^{A_0} \notag \\
&+ \sqrt{T_I\left(\frac{\Delta T^2}{2}+T_I^2\right)}|h_{A_0}|^2\left( F_{I,\theta}^{A_0}F_J^{A_0} + F_I^{A_0}F_{J,\theta}^{A_0} \right)
\end{align}

\bibliographystyle{apsrev4-1}
\bibliography{paper_et_bibliography}

\end{document}